\documentclass[twocolumn]{aa}  
\usepackage{graphicx,natbib}
\usepackage[figuresright]{rotating}
\usepackage{txfonts}
\bibpunct{(}{)}{;}{a}{}{,}
\begin{document}
   \title{Analysis of 26 Barium Stars 
   \thanks{Based on spectroscopic observations collected at the 
           European Southern Observatory (ESO), within the 
           Observat\'orio Nacional ON/ESO and ON/IAG agreements, 
           under FAPESP project n$^{\circ}$ 1998/10138-8. Photometric 
           observations collected at the Observat\'orio do Pico dos Dias (LNA/MCT).}
           \fnmsep\thanks{Tables \ref{ferger} and \ref{abun} are only available in elecronic at the
           CDS via anonymous ftp to cdsarc.u-strasbg.fr/Abstract.html}}

   \subtitle{I. Abundances 
            }

   \author{D.M. Allen
          \inst{}\thanks{Present address: Observat\'orio do Valongo/UFRJ,
              Ladeira do Pedro Antonio 43, 20080-090 Rio de Janeiro, RJ, Brazil}
          \and
          B. Barbuy
	  \inst{}
          }

   \offprints{D.M. Allen}

   \institute{Instituto de Astronomia, Geof\'\i sica e Ci\^encias
    Atmosf\'ericas, Universidade de S\~ao Paulo,
              Rua do Mat\~ao 1226, 05508-900 S\~ao Paulo, Brazil,
              \email{dinah@astro.iag.usp.br, barbuy@astro.iag.usp.br}
              }

   \date{Received, 2006; accepted, 2006}

 
  \abstract
   {We present a detailed analysis of 26 barium stars, including dwarf barium stars, 
   providing their atmospheric parameters 
   (T$_{\rm eff}$, log g, [Fe/H], v$_t$) and elemental abundances.}
   {We aim at deriving gravities and luminosity classes of the sample stars, 
   in particular to confirm the existence of dwarf barium stars. Accurate abundances
   of chemical elements were derived. Abundance ratios between nucleosynthetic processes, 
   by using Eu and Ba as representatives of the r- and s-processes are presented.}
   {High-resolution spectra with the FEROS spectrograph at the ESO-1.5m Telescope, 
    and photometric data with Fotrap at the Zeiss telescope at the LNA were obtained.
    The atmospheric parameters were derived in an iterative way, with temperatures
    obtained from colour-temperature calibrations.
    The abundances were derived using spectrum synthesis for Li, Na, Al, $\alpha$-, 
    iron peak, s- and r-elements atomic lines, and C and N molecular lines.}
   {Atmospheric parameters in the range 4300 $<$ T$\sb {eff}$ $<$ 6500, 
    -1.2 $<$ [Fe/H] $<$ 0.0 and 1.4 $\leq$ log g $<$ 4.6 were derived, confirming that 
    our sample contains giants, subgiants and dwarfs. The abundance results obtained for 
    Li, Al, Na, $\alpha$- and iron peak elements for the sample stars show that 
    they are compatible with the values found in the literature for 
    normal disk stars in the same range of metallicities. Enhancements of C, N and 
    heavy elements relative to Fe, that characterise barium stars, were derived and
    showed that [X/Ba] vs. [Ba/H] and [X/Ba] vs. [Fe/H] present different behaviour as
    compared to [X/Eu] vs. [Eu/H] and [X/Eu] vs. [Fe/H], reflecting the different 
    nucleosynthetic sites for the s- and r-processes.}
   {}


   \maketitle
%

\section{Introduction}

Barium stars were recognized as a distinct group of peculiar
stars by \citet{bk51}. Initially, the objects included in this group were 
only G and K giants which showed strong lines of s-process elements, particularly 
\ion{Ba}{II} at $\lambda$4554 $\rm \AA$ and \ion{Sr}{II} at $\lambda$4077 $\rm \AA$, 
as well as enhanced CH, CN and C$\sb 2$ molecular bands. 
However, the discovery by \citet{tomk89} that the dwarf star HR 107 showed 
chemical composition similar to that of a mild barium giant, has pushed the 
search for less evolved barium stars \citep{gomez97}. Some studies
have proposed that these stars could be ancestors of the barium giants 
\citep{north94,bjra92}.

\citet{mclu80} revealed that most barium stars, maybe all of them,
show variations in radial-velocity suggesting the presence of companions. This
has been confirmed by \citet{mclu83,mclu84} and 
\citet{udry98a,udry98b}.
\citet{bovit80} and \citet{bovit85} observed an ultraviolet excess in the
barium stars $\zeta$ CAP and $\xi$ Ceti, which could be 
explained by white dwarf companions.

The binarity hypothesis for all barium stars has provided an
interesting explanation for their peculiarity. In this context, the
more massive of them evolves through the 
thermal pulse - asymptotic giant branch (TP-AGB)
phase, when s-process occurs, and afterwards the third dredge-up
brings to the surface carbon and s-process elements. The
enriched material is then transferred to the companion, which 
becomes a barium star. \citet{bond03} observed a
planetary nebula (PN) in Cassiopeia, with a late type star, 
showing overabundance in carbon and s-process elements typical of 
a barium star. This discovery confirmed
the hypothesis about their origin. In the PNs Abell 35
\citep{jacoby81,thev97} and LoTr5 \citep{jasniewicz96,thev97}
barium and s-elements rich stars were also observed. There is a central
hot star in each of these nebulae detected with IUE 
({\it International Ultraviolet Explorer}). \citet{jeffries96}
observed the binary system 2REJ0357+283 consisting of a white dwarf and
an s-element rich K dwarf star, where the high rotation can be
attributed to the mass transfer from the white dwarf progenitor, during 
the AGB phase.

However, several IUE spectra of barium stars analysed by \citet{DL83} showed no
UV excess, putting in check the hypothesis that all barium stars have a
white dwarf companion. \citet{bovit00} observed UV excess for most
of their barium stars, but the estimated cooling time for some of the 
companion white dwarfs, was too long or larger than the evolution time of 
the barium star. If the binarity hypothesis is not confirmed for all
stars of the group, another explanation for their origin would be needed.

In Sect. 2 the observations are reported. In Sect. 3 the atmospheric parameters 
are derived. In Sect.4 the abundances derivation is described. In Sect. 5
conclusions are drawn.


\section{Observations}

The sample stars were selected from \citet{gomez97} and \citet{north94} where
the authors suggested that there were less evolved barium stars among the giants of their
sample. The star HR 107 from \citet{tomk89} was also included in the sample.
\citet{menn97} identified HD 5424, HD 13551, HD 116869 and HD 123396 as halo stars.

Optical spectra were obtained at the 1.52m telescope at ESO, La Silla, 
using the Fiber Fed Extended Range Optical Spectrograph (FEROS) 
(Kaufer et al. 2000), on February and October/2000, January and October/2001, 
January and July/2002. The total spectrum coverage 
is  356-920 nm with a resolving power of 48,000. 
Two fibers, with entrance aperture of 2.7 arcsec, recorded 
simultaneously star light and sky background. The detector is a 
back-illuminated CCD with 2948$\times$4096 pixels of 15 $\mu$m size.
Reductions were carried out through a special pipeline package for
reductions (DRS) of FEROS data, in MIDAS environment. 
The data reduction proceeded with subtraction
of bias and scattered light in the CCD, orders extraction, flatfielding, 
and wavelength calibration with a ThAr calibration frame. 
Radial velocities were taken into account by using IRAF tasks RVIDLINE and DOPCOR. 
The spectra were cut in parts of 100 $\rm \AA$ each using the SCOPY task, and  
the normalization was carried out with the CONTINUUM task.

The photometric observations were obtained using the FOTRAP (Fot\^ometro R\'apido)
at the ZEISS 60cm telescope at LNA (Laborat\'orio Nacional de Astrof\'\i sica)
in June, August and September/2001 and May and July/2002. Data reductions were
done using the appropriate code available at LNA. The colours obtained were
(B-V), (V-R), (R-I) and (V-I).
The log of observations is presented in Table \ref{logs}.

\begin{table}
\caption{Log of spectroscopic ($sp$) and photometric ($pho$) observations.
The S/N ratio was measured in the of $\lambda$5000 $\rm \AA$ region.
Radial velocity $v_r$ (km/s) is shown in column 6.
References used to build this sample: 1 - \citet{gomez97}; 
2 - \citet{north94}; 3 - \citet{tomk89}.}
\label{logs}
\setlength\tabcolsep{2.5pt}
\begin{tabular}{lccrcrl}
\hline
\noalign{\smallskip}
star & date$\sb {pho}$ & date$\sb {sp}$ & Exp. (s) & S/N & $v_r$ & ref \\
\noalign{\smallskip}
\hline
HD749	  & 31/08/2001 & 17/01/2001 & 1800 & 200 &  15.47 & 1 \\
HR107	  & 16/07/2002 & 05/10/2001 &  600 & 200 &   9.76 & 3 \\
HD5424	  & 01/09/2001 & 17/01/2001 & 3000 & 100 &  -1.56 & 1 \\
HD8270	  & 01/09/2001 & 14/02/2000 & 1800 & 150 &  14.97 & 1 \\
HD12392	  & 31/08/2001 & 17/01/2001 & 2700 & 120 & -25.38 & 1 \\
HD13551	  & 01/09/2001 & 14/02/2000 & 2700 & 100 &  34.79 & 1 \\
HD22589	  & 01/09/2001 & 14/02/2000 & 2700 & 200 & -28.90 & 1 \\
HD27271	  & 01/09/2001 & 14/02/2000 & 1800 & 250 & -16.65 & 1 \\
HD48565	  & ...        & 23/01/2002 &  600 & 250 & -33.67 & 2 \\
HD76225	  & 18/05/2002 & 23/01/2002 & 1200 & 150 &  -1.34 & 2 \\
HD87080	  & 18/05/2002 & 14/02/2000 & 2700 & 120 &   8.10 & 1 \\
HD89948	  & 18/05/2002 & 14/02/2000 & 1800 & 250 &   6.01 & 1 \\
HD92545	  & 18/05/2002 & 23/01/2002 & 1200 & 150 &  -0.16 & 2 \\
HD106191  & 19/05/2002 & 03/07/2002 & 3600 & 100 &   0.68 & 2 \\
HD107574  & 18/05/2002 & 23/01/2002 & 1320 & 200 &   0.78 & 2 \\
HD116869  & 18/05/2002 & 14/02/2000 & 3600 & 150 &  -9.84 & 1 \\
HD123396  & 19/05/2002 & 15/02/2000 & 3600 & 150 &  26.64 & 1 \\
HD123585  & 18/05/2002 & 04/07/2002 & 7200 & 100 &  27.87 & 1,2 \\
HD147609  & 18/05/2002 & 04/07/2002 & 3600 & 120 & -19.39 & 2 \\
HD150862  & 18/05/2002 & 04/07/2002 & 3600 & 150 &  50.38 & 2 \\
HD188985  & 19/05/2002 & 04/10/2001 & 1800 & 130 &   8.17 & 1,2 \\
HD210709  & 26/06/2001 & 18/10/2000 & 2700 & 100 &  27.01 & 1 \\
HD210910  & 27/06/2001 & 19/10/2000 & 2400 & 200 &  -9.86 & 1 \\
HD222349  & 15/07/2002 & 02/10/2001 & 2400 & 150 &  27.65 & 2 \\
BD+185215 & ...        & 05/10/2001 & 3600 & 100 & -33.98 & 2 \\
HD223938  & 15/07/2002 & 18/10/2000 & 2700 & 250 &  -0.85 & 1 \\
\noalign{\smallskip}
\hline
\end{tabular}
\\
\end{table}

%

\section{Atmospheric Parameters}

\subsection{Extinction}

Reddening values shown in Table \ref{averm} were derived according to 
\citet{card89}, using effective
wavelengths by \citet{bb88} and \citet{bess79}. 
The visual extinction A$\sb v$ was considered null for stars nearer than 
70 pc according to \citet{verg98}, and for the other stars, A$\sb v$ was 
determinated according to \citet{chen98}.

Distances were taken from \citet{menn97}. Including Hipparcos parallaxes as 
input, these authors used a Maximum Likelihood method, which they considered 
to provide better results for the distances than the ones obtained directly 
from the parallaxes.
The distances missing in \citeauthor{menn97}, were obtained directly from 
Hipparcos parallaxes. In cases where no parallax values were available in the
Hipparcos Catalogue, initial distances were estimated
assuming that these stars are subgiants of absolute magnitude $M\sb v$ = 3.3
from \citet{gomez97}, or dwarfs of $M\sb v$ = 4.5. 
The V magnitudes in these cases were taken from SIMBAD available at the web address 
http://simbad.u-strasbg.fr/Simbad.
Distances are shown in Table \ref{coordBa}.

\begin{table*}
\centering
\caption{Equatorial and galactic coordinates of the sample barium stars. 
The errors on last decimals are given in parenthesis.
Distances D$\sb M$ are from \citet{menn97}; distances D$\sb H$ were determined from
Hipparcos parallaxes (D$_H$=1/$\pi_H$); D$\sb g$ are the distances from spectroscopy
for stars with Hipparcos parallax not available.}
\label{coordBa}
\begin{tabular}{lccccccrr}
\hline
\noalign{\smallskip}
star & $\alpha$ (2000) & $\delta$ (2000) & $\pi_H$ (mas) &
D$_H$(kpc) & D$_M$(kpc) & D$_g$(kpc) & l & b \\
\noalign{\smallskip}
\hline
\noalign{\smallskip}
HD 749     & 00:11:38 & -49:39:20 &  7.11(1.08) & 0.141(20) & 0.161(23)  & ... & 319.03 & -66.21 \\
HR 107     & 00:28:20 & +10:11:23 & 27.51(0.86) & 0.036(1)  & ...    & ... & 113.62 & -52.26 \\
HD 5424    & 00:55:44 & -27:53:36 &  0.22(1.42) & 4.5(20.0) & 0.979(344) & ... & 251.97 & -88.78 \\
HD 8270    & 01:21:10 & -47:31:48 & 13.43(1.16) & 0.074(6)  & 0.078(7)   & ... & 288.97 & -68.78 \\
HD 12392   & 02:01:23 & -04:48:10 & ...     & ...   & ...    & 0.219(30) & 162.77 & -62.14 \\
HD 13551   & 02:10:09 & -60:45:31 &  8.91(0.88) & 0.112(10) & 0.118(12)  & ... & 286.75 & -53.83 \\
HD 22589   & 03:37:55 & -06:58:25 & ...     & ...   & ...    & 0.249(20) & 193.68 & -45.71 \\
HD 27271   & 04:18:34 & +02:28:17 &  6.01(1.13) & 0.166(30) & 0.168(15)  & ... & 190.74 & -32.02 \\
HD 48565   & 06:44:55 & +20:51:38 & 21.77(1.07) & 0.046(2)  & 0.046(2)   & ... & 193.53 & + 7.97 \\
HD 76225   & 08:54:01 & -26:54:56 &  3.37(1.11) & 0.297(98) & 0.208(32)  & ... & 251.43 & +11.42 \\
HD 87080   & 10:02:01 & -33:41:11 &  7.90(1.39) & 0.127(22) & 0.160(27)  & ... & 267.02 & +17.12 \\
HD 89948   & 10:22:22 & -29:33:23 & 23.42(0.93) & 0.043(2)  & 0.043(2)   & ... & 268.02 & +23.03 \\
HD 92545   & 10:40:58 & -12:11:44 &  7.82(1.05) & 0.128(17) & 0.120(13)  & ... & 259.85 & +39.52 \\
HD 106191  & 12:13:11 & -15:13:56 & ...     & ...   & ...    & 0.145(20) & 289.43 & +46.63 \\
HD 107574  & 12:21:52 & -18:24:00 &  5.02(1.06) & 0.199(42) & ...    & ... & 293.17 & +43.91 \\
HD 116869  & 13:26:38 & -04:26:46 &  0.23(1.34) & 4.3(25.3) & 0.703(172) & ... & 319.33 & +57.30 \\
HD 123396  & 14:17:33 & -83:32:52 &  1.73(0.86) & 0.578(287)& 0.834(295) & ... & 305.46 & -21.10 \\
HD 123585  & 14:09:36 & -44:22:01 &  8.75(1.39) & 0.114(18) & 0.121(11)  & ... & 317.36 & +16.30 \\
HD 147609  & 16:21:52 & +27:22:27 & 16.64(0.99) & 0.060(4)  & ...    & ... &  45.97 & +43.60 \\
HD 150862  & 16:44:44 & -25:12:59 & ...     & ...   & ...    & 0.74(10) & 355.20 & +13.21 \\
HD 188985  & 19:59:58 & -48:58:32 & 14.06(1.18) & 0.071(6)  & 0.074(6)   & ... & 350.01 & -31.07 \\
HD 210709  & 22:12:54 & -35:25:51 & -0.10(1.37) & ...	& 0.197(29)  & ... &   9.08 & -55.37 \\
HD 210910  & 22:13:42 & -03:46:31 &  6.13(1.16) & 0.163(31) & 0.195(32)  & ... &  57.73 & -45.72 \\
HD 222349  & 23:40:01 & -56:44:26 & ...     & ...   & ...    & 0.168(20) & 321.31 & -57.77 \\
BD+18 5215 & 23:46:56 & +19:28:22 & ...     & ...   & ...    & 0.153(20) & 102.68 & -40.85 \\
HD 223938  & 23:53:50 & -50:00:00 &  4.42(1.21) & 0.226(62) & 0.218(36)  & ... & 324.82 & -64.61 \\
\noalign{\smallskip}
\hline
\end{tabular}
\\
\end{table*}

\begin{table}
\caption{Reddening values for sample stars (see Sect. 3.1).}
{\scriptsize
\label{averm}
\begin{tabular}{lcccccc}
\hline
\noalign{\smallskip}
star & A$\sb v$ & E(B-V) & E(V-I) & E(V-R) & E(R-I) & E(V-K) \\
\noalign{\smallskip}
\hline
\noalign{\smallskip}
HD 749     & 0.0(1)	& ...   & ...   & ...   & ...   & ...   \\
HR 107     & 0.0(1)	& ...   & ...   & ...   & ...   & ...   \\
HD 5424    & 0.0(1)	& ...   & ...   & ...   & ...   & ...   \\
HD 8270    & 0.0(1)	& ...   & ...   & ...   & ...   & ...   \\
HD 12392   & 0.0(1)	& ...   & ...   & ...   & ...   & ...   \\
HD 13551   & 0.0(1)	& ...   & ...   & ...   & ...   & ...   \\
HD 22589   & 0.030(108) & 0.010 & 0.012 & 0.005 & 0.007 & 0.027 \\
HD 27271   & 0.092(123) & 0.029 & 0.036 & 0.015 & 0.021 & 0.082 \\
HD 48565   & 0.0(1)	& ...   & ...   & ...   & ...   & ...   \\
HD 76225   & 0.038(109) & 0.012 & 0.015 & 0.006 & 0.009 & 0.034 \\
HD 87080   & 0.161(140) & 0.052 & 0.063 & 0.027 & 0.037 & 0.143 \\
HD 89948   & 0.0(1)	& ...   & ...   & ...   & ...   & ...   \\
HD 92545   & 0.045(111) & 0.014 & 0.018 & 0.007 & 0.010 & 0.040 \\
HD 106191  & 0.018(105) & 0.006 & 0.007 & 0.003 & 0.005 & 0.016 \\
HD 107574  & 0.037(109) & 0.012 & 0.015 & 0.006 & 0.009 & 0.033 \\
HD 116869  & 0.0(1)	& ...   & ...   & ...   & ...   & ...   \\
HD 123396  & 0.061(115) & 0.020 & 0.024 & 0.010 & 0.014 & 0.054 \\
HD 123585  & 0.130(132) & 0.042 & 0.051 & 0.022 & 0.030 & 0.115 \\
HD 147609  & 0.0(1)	& ...   & ...   & ...   & ...   & ...   \\
HD 150862  & 0.090(123) & 0.029 & 0.035 & 0.015 & 0.021 & 0.080 \\
HD 188985  & 0.051(113) & 0.016 & 0.020 & 0.008 & 0.012 & 0.045 \\
HD 210709  & 0.0(1)	& ...   & ...   & ...   & ...   & ...   \\
HD 210910  & 0.026(106) & 0.008 & 0.010 & 0.004 & 0.006 & 0.023 \\
HD 222349  & 0.0(1)	& ...   & ...   & ...   & ...   & ...   \\
BD+18 5215 & 0.047(112) & 0.015 & 0.019 & 0.008 & 0.011 & 0.042 \\
HD 223938  & 0.0(1)	& ...   & ...   & ...   & ...   & ...   \\
\noalign{\smallskip}
\hline
\end{tabular}
\\
}
\end{table}

\subsection{Temperatures}

Literature photometric data were taken from the 2MASS Point Source Catalog 
\citep[][K$\sb s$ magnitude]{cutri03}, 
The General Catalogue Photometric Data by J.C. Mermilliod, B. Hauck and 
M. Mermilliod available at the web address http://obswww.unige.ch/gcpd/gcpd.html 
(B$_1$, B$_2$, G and V$_1$ magnitudes from 
Geneva system), Hipparcos Catalogue \citep[][(B-V)$\sb H$ and V, shown in the columns 3 and 8 of Table \ref{colors}, respectively]{perryman97}. 
For sample stars 
missing in the Hipparcos Catalogue, (B-V) values were taken 
from SIMBAD. Photometric data used are shown in Table \ref{colors}.

The colour-temperature calibrations were adopted from \citet{alon96} for stars of
log g $>$ 3.6, 
\citet{alon99} for log g $\leq$ 3.6 and \citet{lcb98} for all stars.
\citeauthor{alon99} calibrations use the Johnson system for UBVRI and TCS
(Telesc\'opio Carlos S\'anchez) for JHKLHMN. \citeauthor{lcb98} used the Cousins system
for UBVRI and MSO for JHKLHMN. The calibrations used for the Geneva colours were
adopted from
\citet{melendez03} for dwarfs and \citet{ram04} for giants.
The transformations between photometric systems were obtained from
\citet{carp01}, \citet{bb88}, \citet{alon96} and \citet{alon99}.

The G band affects the spectrum in the $\lambda < 4320 \rm \AA$ region. It is 
possible to 
see some alteration in the black body distribution in the far infrared,
probably due to dust in the binary system \citep{cat77,hakk89}. The colour 
(B-V) is affected by the CN and C$\sb 2$ bands, causing in the spectrum the depression of 
\citet{bn69}, making the stars redder and as a consequence the 
temperatures derived from this colour can be lower. Colours (R-I) and (V-I) 
are more affected by molecular bands and blanketing effects in K and M 
stars than (V-K). 
The calibrations resulted in slightly different temperatures. In general, 
\citet{lcb98} provide higher temperatures than \citet{alon96} and \citet{alon99}.
The resulting temperatures of the Geneva system are generally higher 
than those from the (B-V) index (see Tables \ref{Tlcb98} and \ref{Talonso}).
Given that some indicators tend to increase and others to reduce the temperature,
effective temperatures adopted were the mean values
obtained from the (B-V) Hipparcos, (B-V), (V-I), (V-R) and (R-I) from LNA, and (V-K)
2MASS indices and the colours of the Geneva system (B$\sb 2$-V$\sb 1$), (B$\sb 2$-G)
and (B$\sb 1$-B$\sb 2$). This is shown in column 2 of Table
\ref{result1}. The uncertainties were estimated to be $\pm$100 K, taking into account
the errors from the colour-temperature calibrations.

\begin{figure}[ht]
\sidecaption
\includegraphics[width=9.0cm]{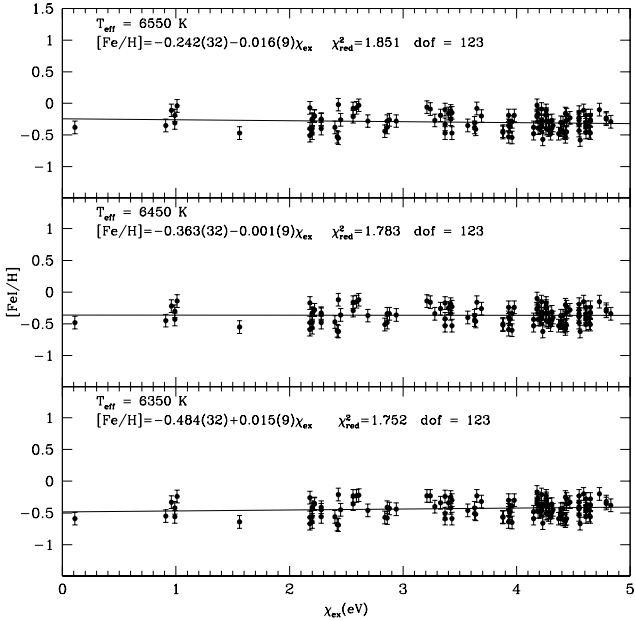}
\caption{[FeI/H] vs. $\chi_{ex}$ (eV) for the star HD 123585.
Least-squares fits for 3 values of temperatures were used to derive T$_{exc}$.
Uncertainties on [FeI/H] and $\chi_{ex}$ were estimated to be $\pm$0.10 and $\pm$0.01,
respectively.}
\label{gratex}
\end{figure}

In order to check photometric temperatures, the excitation temperatures 
T$\sb {exc}$ were 
derived by imposing excitation equilibrium. Lines of 
\ion{Fe}{i} and \ion{Fe}{ii} that give metallicities differing 
between the average and median of $\Delta$[Fe/H] $>$ 0.01 dex were eliminated.
This procedure prevents blended lines to affect the results. Resulting 
T$\sb {exc}$ are given in column 3 of Table \ref{result1}, and illustrated
in Figure \ref{gratex}.

\begin{figure}[ht]
\sidecaption
\includegraphics[width=9.0cm]{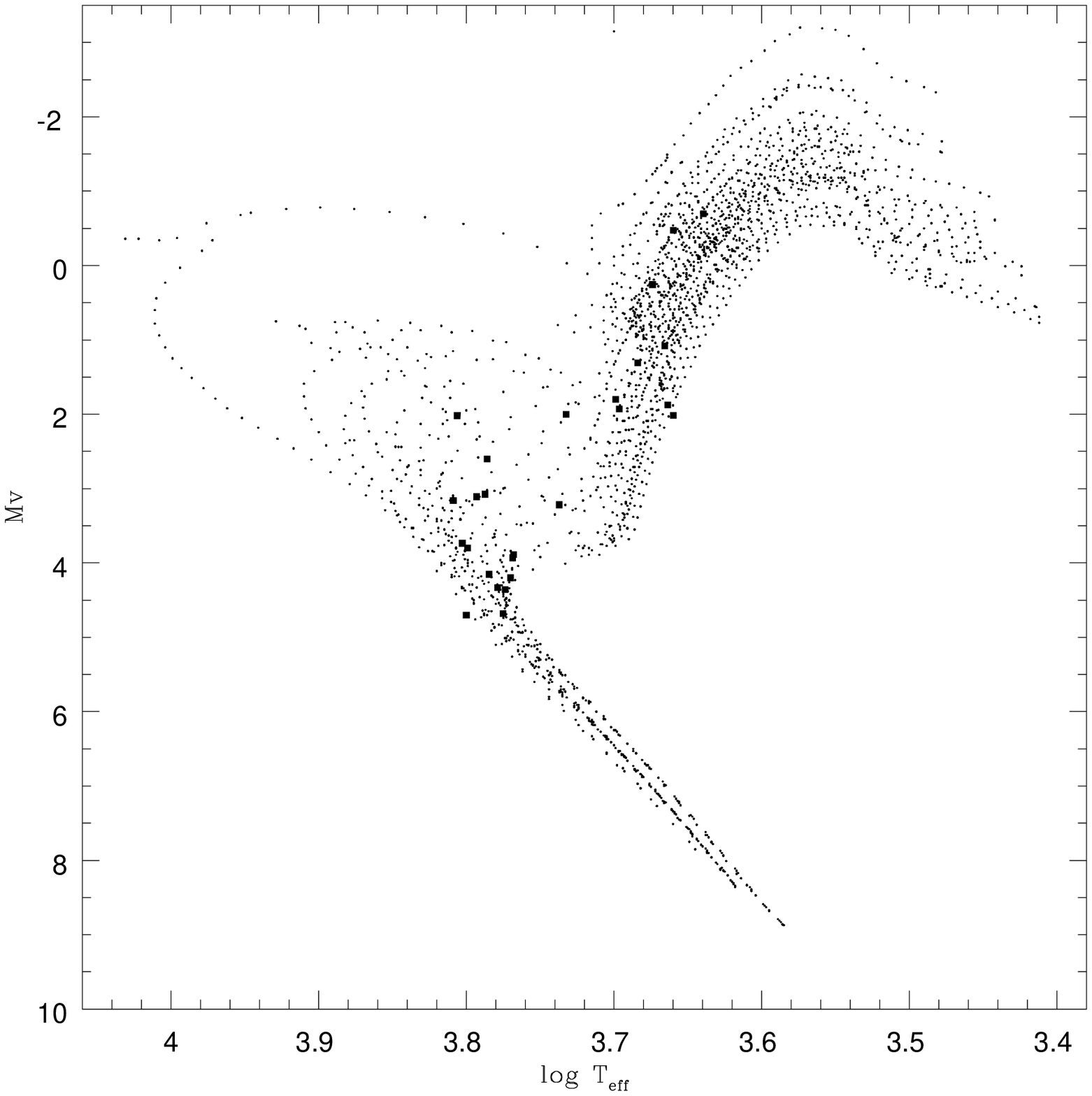}
\caption{Isochrones from \protect\citet{bert94} with sample stars overplotted (black squares).}
\label{isocr}
\end{figure}

\begin{figure*}
\sidecaption
\includegraphics[width=9cm,angle=0]{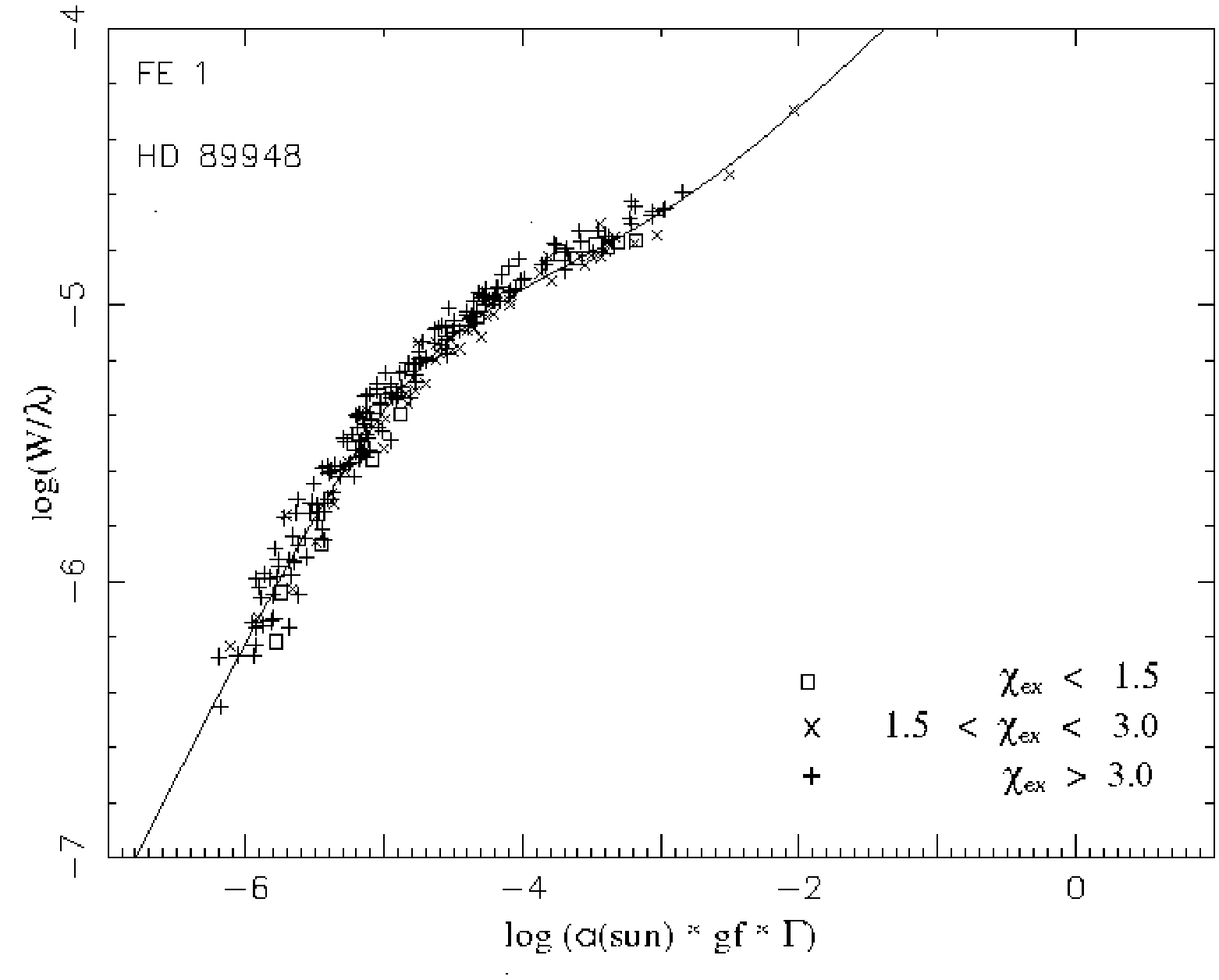}
\includegraphics[width=9cm,angle=0]{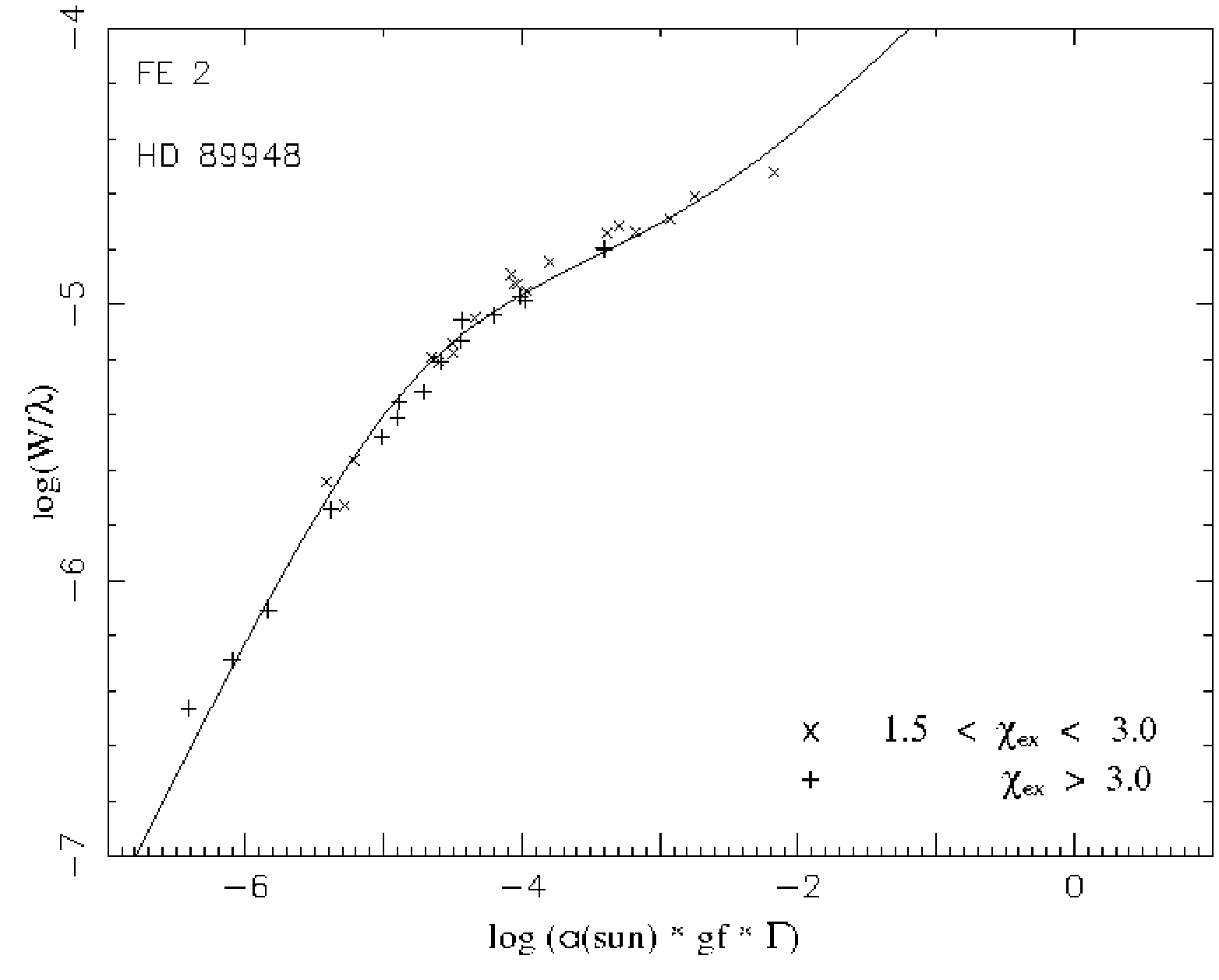}
\caption{\ion{Fe}{i} and \ion{Fe}{ii} curves of growth for the star HD 89948.
'W' are the equivalent widths.
$\log(a(sun))$ = $\log (n_x/n_H)$, where $n_x$ and $n_H$ are the numerical
densities of an element and hydrogen, respectively;
$\log gf$: oscillator strength; 
$\Gamma$ = $(W/\lambda)(n_H/n_x)(gf)^{-1}$, as defined in \protect\citet{cayrel63} and \protect\citet{spispi75}.
The iron abundance is given by the horizontal distance between the linear part of 
the curve of growth and the 45$^\circ$ line through the origin.}
\label{renoir}
\end{figure*}

\begin{table*}
\caption{Colours and magnitudes. V magnitudes are from the Hipparcos database, 
otherwise from SIMBAD (marked with *); K$_s$ from 2MASS Point Source Catalog; 
B$_1$, B$_2$, V$_1$ and G are Geneva magnitudes; ``H'': Hipparcos; ``S'': SIMBAD;
``L'': LNA. Uncertainties for Hipparcos database are of $\pm$ 0.01 whereas for 
SIMBAD they were estimated in $\pm$ 0.05.}
\label{colors}
\begin{tabular}{lcccccccccccc}
\hline
\noalign{\smallskip}
star & (B-V)$\sb S$ & (B-V)$\sb H$ & (B-V)$\sb L$ & (V-I)$\sb L$ & (V-R)$\sb L$ & (R-I)$\sb L$ & V & K$\sb s$ & 
B$_1$ & B$_2$ & V$_1$ & G \\
\noalign{\smallskip}
\hline
\noalign{\smallskip}
HD 749     & 1.130 & 1.126(12) & 1.126 & 1.096 & 0.581 & 0.515 &  7.91  & 5.385(9)  & ... & ... & ... &  \\
HR 107     & 0.430 & 0.447(5)  & 0.408 & 0.506 & 0.245 & 0.261 &  6.05  & 4.942(13) & 0.992 & 1.401 & 1.172 & 1.525 \\
HD 5424    & 1.130 & 1.141(2)  & 1.169 & 1.003 & 0.521 & 0.482 &  9.48  & 7.015(15) & 1.372 & 1.117 & 0.283 & 0.466 \\
HD 8270    & 0.490 & 0.537(17) & 0.539 & 0.664 & 0.334 & 0.330 &  8.82  & 7.524(15) & 1.013 & 1.352 & 1.063 & 1.390 \\
HD 12392   & 1.060 &	...    & 1.126 & 0.842 & 0.460 & 0.382 &  8.49* & 6.231(31) & ... & ... & ... \\
HD 13551   & 0.550 & 0.599(24) & 0.522 & 0.682 & 0.351 & 0.331 &  9.32  &    ...    & 1.014 & 1.363 & 1.076 & 1.420 \\
HD 22589   & 0.710 &	...    & 0.752 & 0.806 & 0.419 & 0.387 &  8.97* & 7.265(11) & ... & ... & ... \\
HD 27271   & 1.000 & 1.011(17) & 1.028 & 1.051 & 0.542 & 0.509 &  7.53  & 5.270(11) & ... & ... & ... \\
HD 48565   & 0.528 & 0.562(22) &    ...& ...   & ...   & ...   &  7.20  & 5.806(17) & 1.007 & 1.355 & 1.028 & 1.352 \\
HD 76225   & 0.510 & 0.567(22) & 0.522 & 0.594 & 0.300 & 0.294 &  9.23  & 7.979(39) & 1.017 & 1.359 & 1.058 & 1.390 \\
HD 87080   & 0.780 & 0.775(12) & 0.761 & 0.797 & 0.405 & 0.392 &  9.40  & 7.567(19) & 1.120 & 1.275 & 0.769 & 1.050 \\
HD 89948   & 0.500 & 0.545(11) & 0.543 & 0.612 & 0.309 & 0.303 &  7.50  & 6.189(19) & 1.036 & 1.338 & 1.021 & 1.347 \\
HD 92545   & 0.470 & 0.503(3)  & 0.500 & 0.592 & 0.299 & 0.293 &  8.55  & 7.282(23) & 1.016 & 1.362 & 1.080 & 1.412 \\
HD 106191  & 0.680 &	...    & 0.600 & 0.643 & 0.323 & 0.320 & 10.00* & 8.586(21) & 1.053 & 1.330 & 0.966 & 1.284 \\
HD 107574  & 0.400 & 0.441(8)  & 0.434 & 0.524 & 0.263 & 0.261 &  8.55  & 7.415(17) & 0.983 & 1.384 & 1.153 & 1.505 \\
HD 116869  & 0.970 & 1.040(15) & 1.041 & 0.997 & 0.520 & 0.477 &  9.49  & 7.143(19) & ... & ... & ...\\
HD 123396  & 1.180 & 1.190(15) & 1.224 & 1.178 & 0.598 & 0.580 &  8.97  & 6.144(17) & ... & ... & ... \\
HD 123585  & 0.520 & 0.519(12) & 0.505 & 0.555 & 0.285 & 0.270 &  9.28  & 8.081(19) & 1.015 & 1.362 & 1.061 & 1.405 \\
HD 147609  & 0.470 & 0.600(15) & 0.584 & 0.655 & 0.332 & 0.323 &  8.57  &    ...    & 1.010 & 1.359 & 1.078 & 1.421 \\
HD 150862  & 0.490 &	...    & 0.515 & 0.578 & 0.296 & 0.282 &  9.17* & 8.009(27) & 1.020 & 1.355 & 1.068 & 1.399 \\
HD 188985  & 0.480 & 0.532(17) & 0.550 & 0.604 & 0.300 & 0.304 &  8.55  & 7.231(23) & 1.034 & 1.345 & 1.019 & 1.347 \\
HD 210709  & 1.060 & 1.103(5)  & 1.117 & 1.051 & 0.559 & 0.492 &  9.23  & 6.711(19) & ... & ... & ... \\
HD 210910  & 1.100 & 1.100(30) & 1.086 & 1.133 & 0.596 & 0.537 &  8.49  & 5.857(19) & 1.307 & 1.144 & 0.344 & 0.513 \\
HD 222349  & 0.500 &	...    & 0.495 & 0.571 & 0.287 & 0.284 &  9.20* & 7.940(25) & 1.001 & 1.365 & 1.082 & 1.421 \\
BD+18 5215 & 0.370 &	...    &   ... & ...   & ...   & ...   &  9.74* & 8.535(11) & 0.991 & 1.368 & 1.110 & 1.452 \\
HD 223938  & 0.890 & 0.905(14) & 0.873 & 0.893 & 0.461 & 0.432 &  8.62  & 6.504(13) & ... & ... & ... \\
\noalign{\smallskip}
\hline
\end{tabular}
\\
\end{table*}

\subsection{Surface Gravity}

Surface gravities (log g) were determined using two methods:

1) Spectroscopic gravities were derived by imposing ionization 
equilibrium of \ion{Fe}{i} and \ion{Fe}{ii} (Figure \ref{renoir}), 
and given in column 4 of Table \ref{result1}. The gravity log g was
varied until the two curves of growth give a same [Fe/H] value, where
\ion{Fe}{ii} lines are more sensitive to gravity variations.
According to \citet{niss97}, 
this method presents a problem because \ion{Fe}{i} lines are affected by
NLTE effects. The uncertainty was estimated as $\pm$0.1 dex.

2) The classical equations of stellar evolution (equation \ref{logtri})
as a function of the distances, according to \citet{niss97} 
and \citet{allende99}:


\begin{eqnarray}
\label{logtri}
\log\biggl({g_\ast\over g_\odot}\biggr) = & \log\biggl({M_\ast\over M_\odot}\biggr) +
4\log\biggl({T_{eff\ast}\over T_{eff\odot}}\biggr)+ 0.4V_\circ + 0.4BC+ \nonumber \\
 & +2\log {1\over D} + 0.1
\end{eqnarray}

\begin{eqnarray}
\sigma_{logg}= & \biggl[\biggl({\sigma_M\over
{M\ln(10)}}\biggr)^2+\biggl({4\sigma_{Teff\ast}\over{T_{eff\ast}\ln(10)}}\biggr)^2+\biggl({4\sigma_{Teff\odot}\over
{T_{eff\odot}\ln(10)}}\biggr)^2+ \nonumber \\
+ & \sigma_{logg\odot}^{2}+(0.4\sigma_{V\circ})^2+(0.4\sigma_{BC})^2+\biggl({2\sigma_\pi\over{\pi\ln(10)}}\biggr)^2\biggr]^{0.5}
\end{eqnarray}
where $M\sb \ast$ is the stellar mass, V$_\circ$ is the V corrected magnitude,
BC is the bolometric correction and D is the distance derived as explained in Sect. 3.1. 
For the Sun, $T\sb {eff\odot}$ = 5781 K 
\citep{bess98}; log g$\sb \odot$ = 4.44; $M\sb {bol\odot}$ = 4.75
\citep{cram99} were adopted.

Stellar masses in equation \ref{logtri} were adopted from \citet{bert94} isochrones,
as shown in Figure \ref{isocr}, and reported in column 9 of Table \ref{result2},
corresponding to metallicities and temperatures as close as possible to those of the
sample (columns 7 (or 6) and 2 of Table \ref{result1}, respectively). 
Uncertainties were estimated as $\pm$ 0.1 M$_\odot$.
These log g values are shown in column 5 of Table \ref{result1}.

According to \citet{niss97}, the gravities resulting from the first method are 
systematically lower than those of the second, but there is no clear explanation.
It could reside in NLTE effects in \ion{Fe}{i} lines, uncertainties
on masses or temperatures.

\subsection {Oscillator Strengths}

The \ion{Fe}{i} line list and respective oscillator strengths from
National Institute
of  Standards \& Technology (NIST) library \citep{mart88,mart02} and \ion{Fe}{ii}
oscillator strengths renormalized by \citet{melendez06}
were adopted. The list of \ion{Fe}{i} and \ion{Fe}{ii} lines is given in 
Table \ref{ferger}. 
Damping constants for neutral lines were computed  using the collisional  broadening
theory of \citet[][and references therein]{barklem98,barklem00},
as described in \citet{zoccali04} and \citet{coelho05}.
The oscillator strengths for the elements other than Fe and respective sources
are shown in Table \ref{abun}. For $\alpha$- and iron peak elements, most values
are from NIST. Laboratory values were
preferred over theoretical ones.

For lines of \ion{Cu}{I}, \ion{Eu}{II}, \ion{La}{II}, \ion{Ba}{II} and \ion{Pb}{I}, 
hyperfine structure (hfs) was taken into account employing a code made available by 
McWilliam, following the calculations described by \citet{proc00}. 
The hfs constants were taken from \citet{rut78} for \ion{Ba}{II}, 
\citet{law01a} for \ion{La}{II}, \citet{law01b} for \ion{Eu}{II} and
\citet{Biemont00} for \ion{Pb}{I}. The final hfs components
were determined by using the solar isotopic mix by \citet{lod03} and total 
log gf values from laboratory measurements, as shown in Table \ref{abun}. For
copper, the hfs from \citet{bi76} was used, with isotopic fractions of 0.69 for 
$\sp {63}$Cu and 0.31 for $\sp {65}$Cu. In this case, small corrections were 
applied such that the total log gf equals the gf-value adopted in this work.
The lines for which hfs were used were checked using the solar spectrum
\citep{kurucz84}.

\begin{table}[ht!]
\caption{Equivalent widths and atomic constants for \ion{Fe}{i} and \ion{Fe}{ii} lines.
e1 - HD 749; e2 - HR 107; e3 - HD 5424; e4 - HD 8270; e5 - HD 12392;
e6 - HD 13551; e7 - HD 22589; e8 - HD 27271; e9 - HD 48565; e10 - HD 76225; e11 - HD 87080; e12 - HD 89948;
e13 - HD 92545; e14 - HD 106191; e15 - HD 107574; e16 - HD 116869; e17 - HD 123396; e18 - HD 123585;
e19 - HD 147609; e20 - HD 150862; e21 - HD 188985; e22 - HD 210709; e23 - HD 210910; e24 - HD 222349;
e25 - BD+18 5215; e26 - HD 223938.  Full table is only available in electronic form.}
{\scriptsize
\label{ferger}
 $$
\setlength\tabcolsep{3pt}
\begin{tabular}{ccccccrrrrcrrr}
\hline\hline
ion & $\lambda$ & $\chi_{ex}$ & log gf & e1 & e2 & e3 & e4 & e5 & e6 & e7 & e8 & e9 & e10 ... \\ 
\noalign{\smallskip}                         
\hline
\noalign{\smallskip}                         
Fe I  & 4514.19 & 3.05 & -2.050 &  89 &  24 &  76 &  34 &  79 &  36 &  67 &  87 &  24 &  32 ... \\
Fe I  & 4551.65 & 3.94 & -2.060 &  48 &   5 &  41 &  12 &  45 & ... &  31 &  50 &   7 & ... \\
Fe I  & 4554.46 & 2.86 & -3.050 & ... &  13 & ... &  16 & ... & ... & ... & ... & ... & ... \\
Fe I  & 4579.82 & 3.07 & -2.830 &  47 & ... &  41 & ... &  36 & ... &  25 & ... & ... &   9 ... \\
Fe I  & 4587.13 & 3.57 & -1.780 &  83 &  26 &  79 &  27 &  80 &  34 &  61 &  86 &  23 &  34 ... \\
Fe I  & 4613.21 & 3.29 & -1.670 &  87 & ... &  95 &  46 &  91 &  62 &  73 &  93 &  33 &  47 ... \\
. & . & . & . & . & . & . & . & . & . & . & . & . & \\
. & . & . & . & . & . & . & . & . & . & . & . & . & \\
\noalign{\smallskip}                         
\hline
\end{tabular} 
$$                                           
}
\end{table}

\subsection{Metallicities and Microturbulent Velocities}

Equivalent widths were measured with IRAF, and \ion{Fe}{i} 
and \ion{Fe}{ii} lines with 10 $<$ W$_{\lambda}$ $<$ 160 m$\rm\AA$ were considered.
Photospheric 1D models were extracted from the NMARCS grid \citep{plez92}, 
originally developed by \citet{bell76} and \citet{gben75} for gravities
log g $<$ 3.3. For less evolved stars, with log g $\geq$ 3.3 the models by
\citet{edv93} were adopted.

Microturbulent velocities v$_{\rm t}$ were determined by
canceling the trend of \ion{Fe}{i} abundance vs. equivalent width.

The uncertainties on the metallicity are due to uncertainties on the
input parameters: temperature, log g and
microturbulent velocity. The variation of these parameters affects the iron 
abundance $\epsilon$(Fe), as follows

\begin{equation} 
\sigma_{\epsilon(Fe)}=\sqrt{(\Delta \epsilon(Fe)_T)^2+(\Delta \epsilon(Fe)_{lg})^2+(\Delta \epsilon(Fe)_v)^2}
\end{equation}
where $\Delta \epsilon(Fe)_T$, $\Delta \epsilon(Fe)_{lg}$ and $\Delta \epsilon(Fe)_v$ 
are the differences on the iron abundances due to variations on temperature, log g
and microturbulent velocity, respectively.

The contributions of the equivalent widths (from 0.5 to 0.9 m$\rm \AA$) to the uncertainty 
in metallicity are negligible \citep{cayrel89}, where uncertainties on continuum placement are not 
taken into account.

For HD 5424, a variation of $\Delta T$ = 100 K in the temperature
results in $\Delta\epsilon(Fe)_T$ = 1.940 $\times$ $10^6$, $\Delta\log g$ = 0.3 dex in log g
results in $\Delta\epsilon(Fe)_{lg}$ = 2.855 $\times$ $10^6$ and $\Delta v_t$ = 0.1 km/s in
v$\sb t$ results in $\Delta\epsilon(Fe)_{v}$ = 0.803 $\times$ $10^6$ in the $\epsilon$(Fe).
For HD 150862, 
$\Delta\epsilon(Fe)_T$ = 5.65 $\times$ $10^5$ 
for $\Delta T$ = 100 K, $\Delta\epsilon(Fe)_{lg}$ = 1.69 $\times$ $10^6$ for $\Delta \log g$ = 0.1 dex 
and $\Delta\epsilon(Fe)_{v}$ = 0.113 $\times$ $10^6$ for $\Delta v_t$ = 0.1 km/s.
The quantity of interest is the logarithm of $\epsilon(Fe)$, such that
$\sigma_{\log\epsilon(Fe)} = (\sigma_{\epsilon(Fe)})/(\epsilon(Fe)\ln{10})$.
Consequently, $\sigma_{[Fe/H]}=\sqrt{\sigma_{\log\epsilon(Fe)}^2+\sigma_{\log\epsilon(Fe)\odot}^2}$,
resulting $\sigma_{[Fe/H]}$ = 0.18 for HD 5424, and this result should be typical 
for all sample stars with $\log g <$ 3.3. For HD 150862, $\sigma_{[Fe/H]}$ = 0.04, 
this uncertainty being adopted for stars with $\log g \geq$ 3.3.

\begin{table}
\caption{Temperatures based on \citet{lcb98} calibrations.}
\label{Tlcb98}
\setlength\tabcolsep{1.5pt}
\begin{tabular}{lccccccc}
\hline
star & T$\sb {(B-V)S}$ & T$\sb {(B-V)H}$ & T$\sb {(B-V)L}$ & 
T$\sb {(V-I)L}$ & T$\sb {(V-R)L}$ & T$\sb {(R-I)L}$ & T$\sb {(V-K)}$ \\
\hline
HD 749     & 4700 & 4710 & 4770 & 4570 & 4590 & 4540 & 4620 \\
HR 107     & 6480 & 6400 & 6630 & 6450 & 6540 & 6370 & 6460 \\
HD 5424    & 4550 & 4520 & 4530 & 4690 & 4660 & 4700 & 4650 \\
HD 8270    & 6160 & 5960 & 5990 & 5870 & 5880 & 5860 & 6200 \\
HD 12392   & 4690 &...   & 4610 & 5230 & 5100 & 5430 & 4870 \\
HD 13551   & 5890 & 5720 & 6030 & 5800 & 5760 & 5850 &...   \\
HD 22589   & 5460 &...   & 5390 & 5420 & 5380 & 5480 & 5600 \\
HD 27271   & 5010 & 4990 & 5010 & 4700 & 4800 & 4630 & 4950 \\
HD 48565   & 5930 & 5760 &...	&...   &...   &...   & 6080 \\
HD 76225   & 6170 & 5930 & 6150 & 6180 & 6170 & 6180 & 6310 \\
HD 87080   & 5360 & 5370 & 5440 & 5630 & 5600 & 5680 & 5400 \\
HD 89948   & 6160 & 5970 & 6010 & 6060 & 6060 & 6050 & 6180 \\
HD 92545   & 6440 & 6290 & 6350 & 6200 & 6220 & 6180 & 6270 \\
HD 106191  & 5570 &...   & 5860 & 5970 & 6000 & 5940 & 6050 \\
HD 107574  & 6550 & 6340 & 6410 & 6420 & 6390 & 6450 & 6460 \\
HD 116869  & 4910 & 4780 & 4820 & 4680 & 4690 & 4660 & 4760 \\
HD 123396  & 4410 & 4350 & 4310 & 4440 & 4450 & 4430 & 4380 \\
HD 123585  & 6210 & 6220 & 6330 & 6460 & 6390 & 6560 & 6500 \\
HD 147609  & 6450 & 5890 & 6000 & 5920 & 5970 & 5870 &...   \\
HD 150862  & 6410 &...   & 6340 & 6330 & 6300 & 6370 & 6500 \\
HD 188985  & 6360 & 6120 & 6080 & 6170 & 6210 & 6140 & 6250 \\
HD 210709  & 4780 & 4690 & 4720 & 4620 & 4620 & 4610 & 4620 \\
HD 210910  & 4740 & 4740 & 4830 & 4520 & 4540 & 4500 & 4550 \\
HD 222349  & 6040 &...   & 6100 & 6220 & 6190 & 6240 & 6270 \\
BD+18 5215 & 6800 &...   &...	&...   &...   &...   & 6400 \\
HD 223938  & 5000 & 4970 & 5080 & 4940 & 4980 & 4890 & 5000 \\
\noalign{\smallskip}
\hline
\end{tabular}
\\
\end{table}

\begin{table*}
\begin{center}
\caption{Temperatures based on calibrations by \citet{alon96} or 
\citet{alon99} and for the Geneva system those by \citet{melendez03} for subgiants or dwarfs 
and \citet{ram04} for giants.}
\label{Talonso}
\begin{tabular}{lcccccccccc}
\hline
star & T$\sb {(B-V)S}$ & T$\sb {(B-V)H}$ & T$\sb {(B-V)L}$ & T$\sb {(V-I)L}$ & 
T$\sb {(V-R)L}$ & 
T$\sb {(R-I)L}$ & T$\sb {(V-K)}$ & T$\sb {B2-V1}$ & T$\sb {B2-G}$ & T$\sb {B1-B2}$ \\
\hline
HD 749     & 4570 & 4570 & 4570 & 4530 & 4570 & 4570 & 4590 &...&...&...\\
HR 107     & 6440 & 6370 & 6540 & 6460 & 6600 & 6180 & 6410 & 6410 & 6430 & 6360 \\
HD 5424    & 4430 & 4410 & 4360 & 4720 & 4740 & 4690 & 4620 & 4400 & 4470 & 4460 \\
HD 8270    & 6130 & 5940 & 5940 & 5760 & 5860 & 5650 & 6080 & 6050 & 6050 & 6050 \\
HD 12392   & 4670 &...   & 4560 & 5110 & 5010 & 5190 & 4840 &...&...&...\\
HD 13551   & 5890 & 5730 & 6000 & 5700 & 5740 & 5640 &...   & 6060 & 6120 & 6130 \\
HD 22589   & 5420 &...   & 5290 & 5320 & 5370 & 5300 & 5470 &...&...&...\\
HD 27271   & 4950 & 4930 & 4890 & 4700 & 4780 & 4690 & 4940 &...&...&...\\
HD 48565   & 5910 & 5790 &...   &...   &...   &...   & 5920 & 5820 & 5860 & 5820 \\
HD 76225   & 6150 & 5930 & 6110 & 6110 & 6180 & 5970 & 6210 & 6070 & 6110 & 6080 \\
HD 87080   & 5280 & 5290 & 5330 & 5510 & 5560 & 5480 & 5450 & 5250 & 5370 & 5370 \\
HD 89948   & 6150 & 5970 & 5980 & 5970 & 6070 & 5820 & 6060 & 5950 & 5990 & 6000 \\
HD 92545   & 6390 & 6250 & 6260 & 6130 & 6240 & 5950 & 6190 & 6220 & 6240 & 6200 \\
HD 106191  & 5550 &...   & 5820 & 5870 & 6000 & 5720 & 5920 & 5780 & 5850 & 5830 \\
HD 107574  & 6520 & 6330 & 6360 & 6450 & 6440 & 6330 & 6430 & 6370 & 6380 & 6370 \\
HD 116869  & 4760 & 4620 & 4620 & 4740 & 4750 & 4710 & 4740 &...&...&...\\
HD 123396  & 4230 & 4220 & 4160 & 4430 & 4480 & 4360 & 4360 &...&...&...\\
HD 123585  & 6160 & 6170 & 6220 & 6470 & 6400 & 6430 & 6460 & 6140 & 6220 & 6240 \\
HD 147609  & 6400 & 5880 & 5940 & 5800 & 5950 & 5620 &...   & 6210 & 6280 & 6210 \\
HD 150862  & 6350 &...   & 6240 & 6280 & 6310 & 6150 & 6460 & 6260 & 6280 & 6270 \\
HD 188985  & 6300 & 6080 & 6010 & 6090 & 6210 & 5910 & 6120 & 5960 & 6020 & 6020 \\
HD 210709  & 4670 & 4600 & 4570 & 4620 & 4630 & 4660 & 4590 &...&...&...\\
HD 210910  & 4660 & 4660 & 4680 & 4480 & 4540 & 4510 & 4520 & 4520 & 4570 & 4430 \\
HD 222349  & 6020 &...   & 6040 & 6150 & 6190 & 6040 & 6140 & 6040 & 6060 & 6070 \\
BD+18 5215 & 6670 &...   &...   &...   &...   &...   & 6310 & 6270 & 6280 & 6280 \\
HD 223938  & 4930 & 4900 & 4970 & 4980 & 4990 & 4920 & 4990 &...&...&...\\
\noalign{\smallskip}
\hline
\end{tabular}
\\
\end{center}
\end{table*}

\subsection {Adopted Atmospheric Parameters}

The atmospheric parameters were derived in an iterative way, adopting 
initial values of log g and [Fe/H] according to \citet{north94} and 
\citet{gomez97}. The lines of \ion{Fe}{i} 
and \ion{Fe}{ii} were used separately in order to test the ionization equilibrium 
(see Sect. 3.3). In this first iteration, one searchs  for values of log g, [Fe/H] and 
v$\sb t$ corresponding to the ionization equilibrium in order to verify the trend of 
the results. Using the average of the temperatures and masses from isochrones, the 
surface gravity was determined from equation \ref{logtri}, in order to be compared with 
that from the ionization equilibrium.

In the first iteration, values different from the ones adopted initially for [Fe/H]
and log g from equation \ref{logtri} were obtained. In this case,
these new values were used as input in the colour-temperature calibrations 
and the procedure was restarted. The procedure was repeated until the input 
parameters matched the output ones, providing a consistent set of 
atmospheric parameters. After several iterations, it was possible to define a set of atmospheric 
parameters for each star, shown in Table \ref{result1}. These results indicate
that our sample contains giants, subgiants and dwarfs. 
Table \ref{lit} shows atmospheric parameters for barium stars found in the literature.

The use of models by \citet{gben75} led to $\Delta$[FeI/H] and $\Delta$[FeII/H] 
$\leq$ 0.03 dex relative to models by \citet{edv93} and \citet{plez92} for
most stars. For two stars, $\Delta$[Fe/H] $\approx$ 0.15 dex and for another 4 ones, 
0.04 dex $<$ $\Delta$[Fe/H] $<$ 0.09 dex.

In Table \ref{result2} are given the bolometric corrections BC(V) determined
according to \citet{alon95} for dwarfs and
subgiants and to \citet{alon99} for giants, with good agreement with
values determined from a linear interpolation on \citet{lcb98} grids. The uncertainties 
on BC(V) were estimated by computing how the uncertainties on temperatures modify its value; 
from equations \ref{magab} to \ref{massa}, one obtains the absolute
magnitude M$\sb v$, the bolometric magnitude M$\sb {bol}$, the luminosity
(L$\sb \ast$/L$\sb \odot$), the radius (R$\sb \ast$/R$\sb \odot$)  and the mass
(M$\sb \ast$/M$\sb \odot$), providing as input the distance D(pc), A$\sb v$, V
and log g:

\begin{equation}
\label{magab}
M_v = V-5\log D+5-A_v
\end{equation}
\begin{equation}
\sigma_{Mv}=\biggl[({\sigma_V)^2+\biggl({5\sigma_D\over D\ln(10)}\biggr)^2+\sigma_{Av}^2}\biggr]^{0.5}
\end{equation}
\begin{equation}
\label{magbol}
M_{bol\ast} = M_v+BC(V)
\end{equation}
\begin{equation}
\sigma_{Mbol\ast}=\biggl({\sigma_{Mv}^2+\sigma_{BC}^2}\biggr)^{0.5}
\end{equation}
\begin{equation}
\label{lumin}
L_\ast = 10^{-0.4(M_{bol\ast}-M_{bol\odot})}L_\odot
\end{equation}
\begin{equation}
\sigma_L=L0.4\ln(10)\biggl({\sigma_{Mbol\ast}^2+\sigma_{Mbol\odot}^2}\biggr)^{0.5}
\end{equation}
\begin{equation}
\label{raio}
R_\ast = \biggl({L_\ast T_{ef\odot}^4\over L_\odot T_{ef\ast}^4}\biggr)^{0.5}R_\odot
\end{equation}
\begin{equation}
\sigma_R=R\biggl[{\biggl({\sigma_L\over 2L}\biggr)^2+\biggl({2\sigma_{Teff\odot}\over T_{eff\odot}}\biggr)^2+\biggl({2\sigma_{Teff\ast}\over T_{eff\ast}}\biggr)^2}\biggr]^{0.5}
\end{equation}
Recalling that $g=10^{\log(g)}$ and $\sigma_g=g\ln(10)\sigma_{\log(g)}$
\begin{equation}
\label{massa}
M_\ast = {g_\ast R_\ast^2\over g_\odot R_\odot^2}M_\odot
\end{equation}
\begin{equation}
\sigma_M=M\biggl[{\biggl({\sigma_{g\ast}\over g\ast}\biggr)^2+\biggl({\sigma_{g\odot}\over g\odot}\biggr)^2+\biggl({2\sigma_{R}\over R}\biggr)^2}\biggr]^{0.5}.
\end{equation}

For masses computed with equation \ref{massa} and log g from
ionization equilibrium, the results are those shown in column 8 of 
Table \ref{result2}. These masses are lower than those 
derived from the isochrones (column 9), resulting in higher values for log g
computed from equation \ref{logtri}.

Our derivations are essentially compatible with \citet{menn97} masses
for barium stars, according to their luminosity classes: 1 - 1.6 M$\sb \odot$ 
for dwarfs, and 1 - 3 M$\sb \odot$ for giants and subgiants.

For stars with no Hipparcos parallaxes, distances also had to be derived iteratively.
In the cases where extinction was null, the determination of the colours was 
straight forward, the temperatures and the other parameters, log g,
[Fe/H] and BC. The distance can be determined by setting the mass. 
In the case of extinction variation on which an average of A$\sb v$ was adopted,
the distance obtained can be different from that used by computing A$\sb v$. In this
case, the new distance was used to restart the procedure, computing new A$\sb v$, 
colours and temperatures. The procedure was repeated until the output distance
matched the input one. In both cases, the calculation was made
for several masses inside the range given by \citet{menn97} and the absolute 
magnitudes were used to derive the masses from isochrones in order to be compared 
with the input masses. The log g was obtained with equation \ref{logtri}. 
The mass chosen was that for which log g from the equation \ref{logtri} was inside
the range of 0.3 dex of the spectroscopic one. Once the mass was set, the distances
were recalculated, as shown in Table \ref{mdsempi}.

\begin{table}[h!]
\caption{Masses and distances adopted for stars with no Hipparcos parallax $\pi \sb H$ values.}
\label{mdsempi}
   $$ 
\begin{tabular}{lccc}
\hline
\noalign{\smallskip}
star & D$\sb {min}$-D$\sb {max}$ & M$\sb {min}$-M$\sb {max}$ & 
adopted \\
\noalign{\smallskip}
\hline
\noalign{\smallskip}
HD 12392    & 150-240pc  & 1.6-2,4M$\sb \odot$ & 219(30)pc  2,0(1)M$\sb \odot$ \\
HD 22589    & 240-288pc  & 1,6-1,8M$\sb \odot$ & 249(20)pc  1,6(1)M$\sb \odot$ \\
HD 106191   & 140-180pc  & 1,0-1,1M$\sb \odot$ & 145(20)pc  1,0(1)M$\sb \odot$ \\
HD 150862   & 70-90pc    & 1,0-1,2M$\sb \odot$ & 74(10)pc   1,1(1)M$\sb \odot$ \\
HD 222349   & 154-182pc  & 1,1-1,2M$\sb \odot$ & 168(20)pc  1,2(1)M$\sb \odot$ \\
BD+18 5215  & 140-190pc  & 1,1-1,2M$\sb \odot$ & 153(20)pc  1,1(1)M$\sb \odot$ \\
\noalign{\smallskip}
\hline
\end{tabular}
   $$ 
\end{table}

According to \citet{ti99}, the NLTE affects \ion{Fe}{i} lines more than 
\ion{Fe}{ii} ones, and it directly affects the log g derived from ionization equilibrium.
For this reason, the gravities adopted were those resulting from expression
\ref{logtri}, whereas for the metallicities, \ion{Fe}{ii} lines were used, as
shown in column 7 of Table \ref{result1}. It is worth noting that using
the gf-values for \ion{Fe}{ii} lines by \citet{melendez06}, the abundances 
of Fe derived from the \ion{Fe}{i} and \ion{Fe}{ii} lines are closer to each other
and the log g from ionization equilibrium increases by 0.2 dex relative to
the case of the gf-values from NIST. 

Table \ref{result1} compared with Table \ref{lit}, shows that differences are stronger for
HD 27271 and HD 147609. Temperatures derived from the colours indices in the present 
work for HD 27271 and HD 147609 are lower than those found in the literature.
Furthermore, the metallicities used were those
derived from \ion{Fe}{ii} lines instead of ionization equilibrium. 
Also the masses derived using log g from ionization equilibrium are
very small (see column 8 of Table \ref{result2}), therefore we preferred
to use log g derived by using masses from isochrones, and \ion{Fe}{ii} lines
for the determination of metallicities.

\begin{table}
\caption{Stellar parameter results. T$_{eff}$: photometric temperature; 
T$_{exc}$: excitation temperature; $\log$g(C): $\log$g related to curve of growth;
$\log$g(D): $\log$g related to masses from isochrones.
Numbers in parenthesis are errors in last decimals.}
\label{result1}
\setlength\tabcolsep{2pt}
\begin{tabular}{lccccccc}
\hline
star & T$_{eff}$ & T$_{exc}$ & $\log$g(C) & $\log$g(D) & 
[\ion{Fe}{i}/H] & [\ion{Fe}{ii}/H] & v$\sb {t}$ \\
\hline
HD 749     & 4610 & 4580 & 2.3 & 2.8(1)  & -0.06 & +0.17 & 0.9 \\
HR 107     & 6440 & 6650 & 4.0 & 4.08(7) & -0.34 & -0.36 & 1.6 \\
HD 5424    & 4570 & 4700 & 1.8 & 2.0(3)  & -0.51 & -0.55 & 1.1 \\
HD 8270    & 5940 & 6070 & 4.2 & 4.2(1)  & -0.44 & -0.42 & 0.9 \\
HD 12392   & 5000 & 5000 & 3.2 & 3.2(1)  & -0.06 & -0.12 & 1.2 \\
HD 13551   & 5870 & 6050 & 3.7 & 4.0(1)  & -0.44 & -0.24 & 1.1 \\
HD 22589   & 5400 & 5630 & 3.3 & 3.3(1)  & -0.12 & -0.27 & 1.1 \\
HD 27271   & 4830 & 4830 & 2.3 & 2.9(1)  & -0.09 & +0.17 & 1.3 \\
HD 48565   & 5860 & 6050 & 3.8 & 4.01(8) & -0.71 & -0.62 & 1.0 \\
HD 76225   & 6110 & 6330 & 3.7 & 3.8(1)  & -0.34 & -0.31 & 1.4 \\
HD 87080   & 5460 & 5550 & 3.7 & 3.7(2)  & -0.49 & -0.44 & 1.0 \\
HD 89948   & 6010 & 6010 & 4.2 & 4.30(8) & -0.28 & -0.30 & 1.2 \\
HD 92545   & 6210 & 6270 & 4.0 & 4.0(1)  & -0.15 & -0.12 & 1.3 \\
HD 106191  & 5890 & 5890 & 4.2 & 4.2(1)  & -0.22 & -0.29 & 1.1 \\
HD 107574  & 6400 & 6400 & 3.6 & 3.6(2)  & -0.56 & -0.55 & 1.6 \\
HD 116869  & 4720 & 4850 & 2.1 & 2.2(2)  & -0.35 & -0.32 & 1.3 \\
HD 123396  & 4360 & 4480 & 1.2 & 1.4(3)  & -1.19 & -0.99 & 1.2 \\
HD 123585  & 6350 & 6450 & 4.2 & 4.2(1)  & -0.44 & -0.48 & 1.7 \\
HD 147609  & 5960 & 5960 & 3.3 & 4.42(9) & -0.45 & +0.08 & 1.5 \\
HD 150862  & 6310 & 6310 & 4.6 & 4.6(1)  & -0.11 & -0.10 & 1.4 \\
HD 188985  & 6090 & 6190 & 4.3 & 4.3(1)  & -0.25 & -0.30 & 1.1 \\
HD 210709  & 4630 & 4680 & 2.3 & 2.4(2)  & -0.07 & -0.04 & 1.1 \\
HD 210910  & 4570 & 4770 & 2.0 & 2.7(2)  & -0.37 & +0.04 & 2.0 \\
HD 222349  & 6130 & 6190 & 3.9 & 3.9(1)  & -0.58 & -0.63 & 1.1 \\
BD+18 5215 & 6300 & 6300 & 4.2 & 4.2(1)  & -0.44 & -0.53 & 1.5 \\
HD 223938  & 4970 & 5150 & 2.7 & 3.1(1)  & -0.35 & -0.13 & 1.0 \\
\noalign{\smallskip}
\hline
\end{tabular}
\\
\end{table}

\begin{table}
\caption{Bolometric corrections, absolute magnitudes, liminosities, radii and 
masses for the sample stars. BC$_a$(V): bolometric corrections using 
\citet{alon96} for dwarfs and subgiants and \citet{alon99} for giants;
BC$_l$(V): bolometric corrections using \citet{lcb98};
$M_\ast/M_\odot$: stellar masses from curve of growth;
$M_i/M_\odot$: stellar masses from isochrones.
Numbers in parenthesis are errors in last decimals.}
{\scriptsize
\label{result2}
\setlength\tabcolsep{2pt}
\begin{tabular}{lccrrcccc}
\hline
star & BC$_a$(V) & BC$_l$(V) & M$_v$ & M$_{bol}$ & 
$L_\ast/L_\odot$ & $R_\ast/R_\odot$ & $M_\ast/M_\odot$ & 
$M_i/M_\odot$ \\
\hline
HD 749     & -0.43(6) & -0.47(7) &  1.9(3) &  1.4(3) &  21(6)  &  7(1)  & 0.4(1) & 1.2 \\
HR 107     & -0.08(1) & -0.04(1) &  3.3(1) &  3.2(1) & 4.2(5) &  1.6(1) & 1.0(3) & 1.2 \\
HD 5424    & -0.44(6) & -0.48(7) & -0.5(8) & -0.9(8) & 184(130) & 22(8) & 1.1(8) & 1.9 \\
HD 8270    & -0.13(1) & -0.12(1) &  4.4(2) &  4.2(2) &   1.6(3) &  1.2(1) & 0.8(3) & 0.9 \\
HD 12392   & -0.26(3) & -0.28(2) &  1.8(3) &  1.5(3) &  20(6) &  6.0(9) & 2.0(8) & 2.0 \\
HD 13551   & -0.14(1) & -0.12(1) &  4.0(2) &  3.8(2) &  2.3(5) &  1.5(2) & 0.4(1) & 0.9 \\
HD 22589   & -0.20(2) & -0.19(2) &  2.0(3) &  1.8(3) &  16(4) &  4.5(6) & 1.4(5) & 1.6 \\
HD 27271   & -0.32(4) & -0.35(5) &  1.3(2) &  1.0(2) &  32(7) &  8(1) & 0.5(2) & 1.9 \\
HD 48565   & -0.15(1) & -0.14(1) &  3.9(1) &  3.7(1) &   2.5(3) &  1.5(1) & 0.5(1) & 0.9 \\
HD 76225   & -0.08(1) & -0.09(1) &  2.6(3) &  2.5(3) &   8(2) &  2.5(4) & 1.1(5) & 1.4 \\
HD 87080   & -0.20(2) & -0.19(2) &  3.2(4) &  3.0(4) &   5(2) &  2.5(5) & 1.1(3) & 1.2 \\
HD 89948   & -0.11(1) & -0.10(1) &  4.3(1) &  4.2(1) &   1.6(2) &  1.18(9) & 0.8(2) & 1.0 \\
HD 92545   & -0.08(1) & -0.06(1) &  3.1(3) &  3.0(3) &   5(1) &  1.9(2) & 1.3(5) & 1.3 \\
HD 106191  & -0.11(1) & -0.11(1) &  4.2(3) &  4.0(3) &   1.9(6) &  1.3(2) & 1.0(4) & 1.0 \\
HD 107574  & -0.08(1) & -0.06(1) &  2.0(5) &  1.9(5) &  13(6) &  3.0(6) & 1.3(6) & 1.4 \\
HD 116869  & -0.37(4) & -0.38(5) &  0.2(5) & -0.1(5) &  88(40) & 14(4) & 0.9(5) & 1.2 \\
HD 123396  & -0.57(7) & -0.61(7) & -0.7(8) & -1.3(8) & 255(180) & 28(10) & 0.4(3) & 0.8 \\
HD 123585  & -0.10(4) & -0.06(1) &  3.7(2) &  3.6(2) &   2.8(6) &  1.4(2) & 1.1(4) & 1.1 \\
HD 147609  & -0.10(1) & -0.09(1) &  4.7(2) &  4.6(2) &   1.2(2) &  1.02(9) & 0.07(2) & 1.0 \\
HD 150862  & -0.07(1) & -0.06(1) &  4.7(3) &  4.7(3) &   1.1(3) &  0.9(1) & 1.1(4) & 1.1 \\
HD 188985  & -0.10(1) & -0.09(1) &  4.1(2) &  4.0(2) &   1.9(4) &  1.2(1) & 1.1(2) & 1.1 \\
HD 210709  & -0.53(6) & -0.45(7) &  1.1(4) &  0.5(4) &  48(20) & 11(2) & 0.8(4) & 1.1 \\
HD 210910  & -0.45(6) & -0.50(7) &  2.0(4) &  1.6(4) &  19(6) &  7(1) & 0.2(1) & 1.0 \\
HD 222349  & -0.12(1) & -0.11(1) &  3.1(2) &  2.9(2) &   5(1) &  2.0(2) & 1.2(4) & 1.2 \\
BD+18 5215 & -0.10(1) & -0.08(1) &  3.8(3) &  3.7(3) & 3(1) &  1.4(2) & 1.1(4) & 1.1 \\
HD 223938  & -0.31(4) & -0.28(2) &  1.9(4) &  1.6(4) &  18(6) &  6(1) & 0.6(2) & 1.4 \\
\noalign{\smallskip}
\hline
\end{tabular}
\\
}
\end{table}

\begin{table}
\caption{Atmospheric parameters for barium stars collected in the literature.
References:
E93: \citet{edv93}; G96: \citet{gcc96}; T99: \citet{ti99};
P05: \citet{claudio05}; P03: \citet{claudio03};
B92: \citet{bjra92}; N94: \citet{north94}; S86: \citet{sl86};
L91: \citet{lb91}; S93: \citet{scl93}. Numbers in parenthesis are errors
in last decimals.}
   $$ 
\label{lit}
\begin{tabular}{lccccc}
\hline
\noalign{\smallskip}
star & T$_{eff}(K)$ & log g & [M/H] & v$_t$(km/s) & ref \\
\noalign{\smallskip}
\hline
\noalign{\smallskip}
HR 107    & 6488 & 4.08 & -0.37 & -      & E93 \\
HR 107    & 6431 & 4.06 & -0.37 & 1.98   & G96 \\
HR 107    & 6462 & 4.17 & -0.21 & 1.00   & T99 \\
HD 8270   & 6100 & 4.2 & -0.53 & 1.4  & P05 \\
HD 13551  & 6400 & 4.4 & -0.28 & 1.6 & P05 \\
HD 22589  & 5600 & 3.8 & -0.16 & 1.4  & P05 \\
HD 27271  & 5350 & 2.30 & -0.50 & 3.0    & B92 \\
HD 48565  & 5910 & 3.50 & -0.90 & 1.2(2) & N94 \\
HD 48565  & 5929 & 3.72 & -0.54 & 1.00   & T99 \\
HD 76225  & 6010 & 3.50 & -0.50 & 1.7(2) & N94 \\
HD 87080  & 5600 & 4.00 & -0.51 & 1.2 & P03 \\
HD 89948  & 5929 & 4.10 & -0.31 & 1.00   & S86 \\
HD 89948  & 6000 & 4.00 & -0.13 & 1.80   & L91 \\
HD 89948  & 5950 & 4.10 & -0.27 & 0.80   & S93 \\
HD 92545  & 6240 & 3.90 & -0.33 & 1.7(1) & N94 \\
HD 106191 & 5840 & 4.05 & -0.40 & 1.0(1) & N94 \\
HD 107574 & 6340 & 3.63 & -0.80 & 1.9(3) & N94 \\
HD 123585 & 6047 & 3.50 & -0.50 & 1.8(3) & N94 \\
HD 123585 & 6000 & 3.50 & -0.50 & 2.00   & L91 \\
HD 147609 & 6270 & 3.50 & -0.50 & 1.2(2) & N94 \\
HD 147609 & 6300 & 3.61 & -0.36 & 1.20   & T99 \\
HD 150862 & 6135 & 4.05 & -0.30 & 1.2(1) & N94 \\
HD 150862 & 6200 & 4.00 & -0.22 & 2.20   & L91 \\
HD 188985 & 5960 & 3.78 & -0.30 & 1.4(1) & N94 \\
HD 222349 & 6000 & 3.76 & -0.90 & 1.8 & N94 \\
BD+18 5215& 6290 & 4.49 & -0.50 & 1.2(2) & N94 \\
\noalign{\smallskip}
\hline
\end{tabular}
   $$ 
\label{tab1}
\end{table}

%

\section{Abundances}

The LTE abundance analysis and the spectrum synthesis calculations
were performed using the codes by Spite (1967, and   subsequent
improvements in the  last thirty years), described in \citet{cay91} and
\citet{barb03}. 

Table \ref{abun} shows the resulting abundances ($\log\epsilon$(X) and [X/Fe])
for all atomic lines, whereas Tables \ref{medxfe1} and \ref{medxfe2} 
show the mean abundance of each element, obtained for the 26 sample barium stars.
Figures \ref{imalfefe1} to \ref{rsfe2} show [X/Fe] vs. [Fe/H].
For most elements, the solar abundances used 
were extracted from \citet{gs98}, otherwise references are indicated 
in Table \ref{medxfe2}.

For the stars HD 749, HD 13551, HD 27271, HD 123396, HD 147609, HD 210910 and HD 223938,
the difference between the metallicities derived from \ion{Fe}{i} and \ion{Fe}{ii} lines,
$\Delta$[Fe/H] = [\ion{Fe}{ii}/H] - [\ion{Fe}{i}/H] $\geq$ 0.2 dex (see Table \ref{result1}).
There are several possible explanations for it, such as the NLTE effect in 
lines of \ion{Fe}{i}, imprecision in stellar parameters 
(T$\sb {eff}$, log g, v$\sb t$), blends in \ion{Fe}{i} and \ion{Fe}{ii} lines. 
\citet{simm04} also observed this effect in their sample
of 159 giants and dwarfs. They established relations between $\Delta$[Fe/H] and T$\sb {eff}$ 
or log g and, despite having a dispersion, it is possible to note that the differences seem
to be larger at lower temperatures. \citet{kraft03} attributed this effect to 
an inadequacy of model atmospheres.
\citet{yong03} observed a similar behaviour in the relation between $\Delta$[Fe/H] and 
T$\sb {eff}$ for giants of the metal-poor globular cluster NGC 6752, also becoming
more pronounced for the cooler stars.

For the 7 stars mentioned before, the resulting [X/Fe] were very low, as shown
by the starred symbols of Figures \ref{imalfefe1} to \ref{rsfe2}, and for this
reason, the abundances were also determined by using metallicities from \ion{Fe}{i} 
lines, as shown in Figures \ref{imalfefe1} to \ref{rsfe2} and Tables \ref{abun},
\ref{medxfe1} and \ref{medxfe2}, and they were used in 
Figures 4, 5, 6, 9, 10, 15, 16, 17, 18. For the remaining stars, abundances in
these figures were obtained using metallicities derived from \ion{Fe}{ii} lines.

The abundance results of the sample stars are essentially homogeneous,
even considering the different luminosity classes, as shown in 
Figures \ref{imalfefe1} to \ref{rsfe2}. Regarding Al, Na, $\alpha$- and iron
peak elements, the behaviour of [X/Fe] vs. [Fe/H] is in agreement with
disk stars. For heavy elements, there is
a variation that could be explained by the amount of enriched material that each star
received from the more evolved companion. The sample is too small to reveal
differences between the 4 halo stars and 
the disk ones. The overabundance found for the s-elements in
the sample stars is expected for barium stars and this peculiarity is
independent of the luminosity class. Literature abundance data for the present sample 
barium stars are shown in Table \ref{ablit}.

\begin{figure*}[ht!]
\centerline{\includegraphics[totalheight=7.0cm]{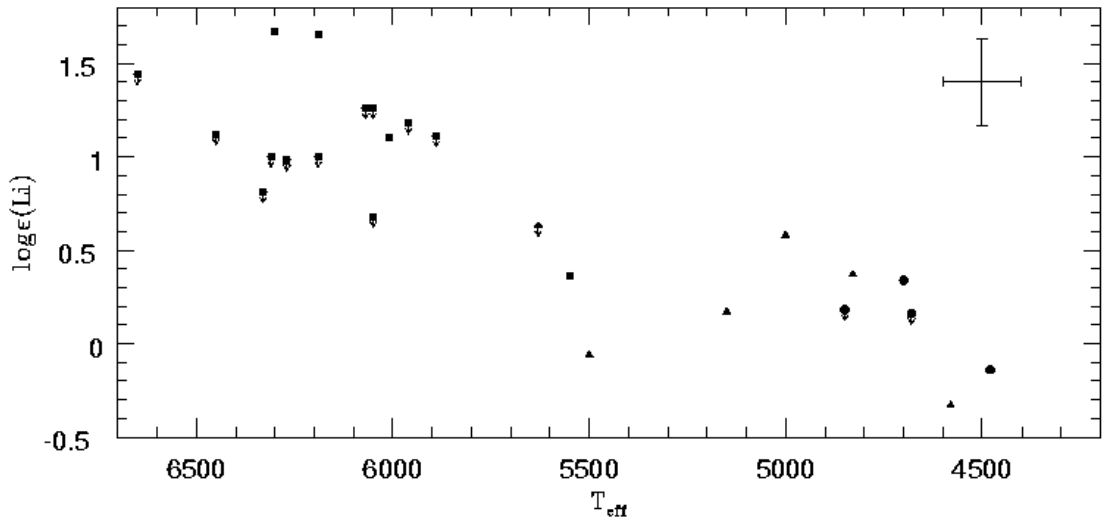}}
\caption{\label{teff_Li}Lithium abundance as a function of temperatures.
Symbols: squares: dwarf stars ($\log g \ge 3.7$); 
triangles: subgiants ($2.4 < \log g < 3.7$);
circles: giants ($\log g \le 2.4$). The arrows indicate an upper limit.}
\end{figure*}

\begin{figure*}[ht!]
\centerline{\includegraphics[totalheight=5.0cm]{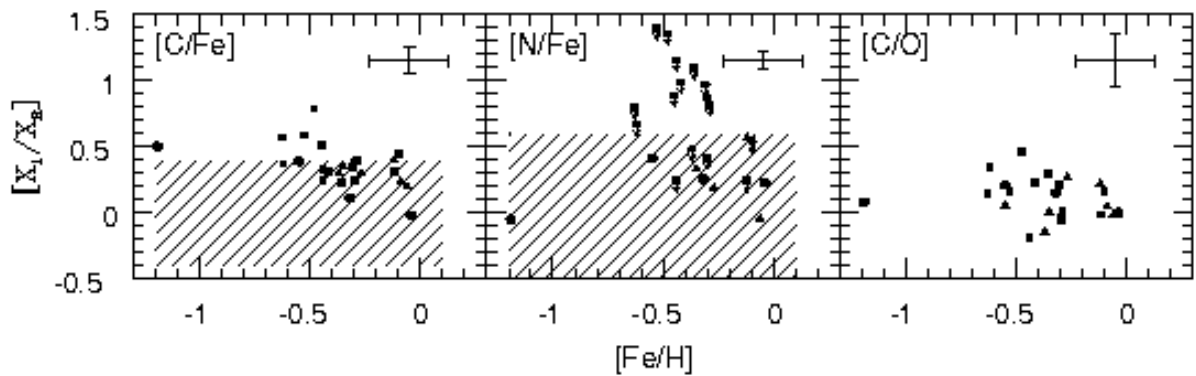}}
\caption{\label{cno}[C,N/Fe] and [C/O] as a function of [Fe/H]. Symbols
are the same as in Figure \ref{teff_Li}. The arrows indicate an upper limit.
In the dashed region are the disk and halo stars of \protect\citet{goswami00}.}
\end{figure*}

\begin{figure*}[ht!]
\centerline{\includegraphics[totalheight=5.0cm]{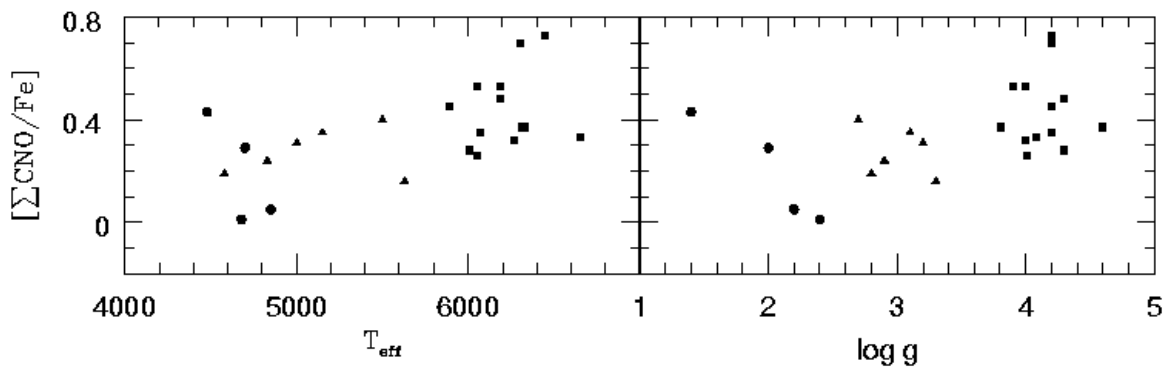}}
\caption{\label{cnoteflog}[$\sum${CNO}/Fe] as a function of T$\sb {eff}$ or $\log g$.
Symbols are the same as in Figure \ref{teff_Li}.}
\end{figure*}

\begin{figure}[ht!]
\centerline{\includegraphics[totalheight=9.0cm]{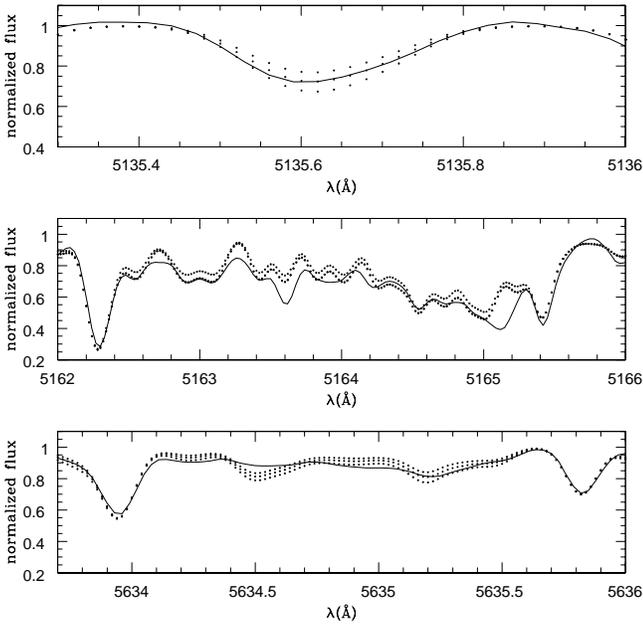}}
\caption{\label{12392c2}Example of fits of the C$\sb 2$ bands for the star
HD 12392. Symbols: solid line: observed spectrum; dotted lines: synthetic spectra
with several C abundances.
Upper panel: $\log\epsilon(C)$ = 8.84, 8.79 e 8.74;
Middle panel: $\log\epsilon(C)$ = 8.89, 8.84 e 8.79;
Lower panel: $\log\epsilon(C)$ = 8.84, 8.79 e 8.74.}
\end{figure}

\begin{figure}
\centerline{\includegraphics[totalheight=9.0cm]{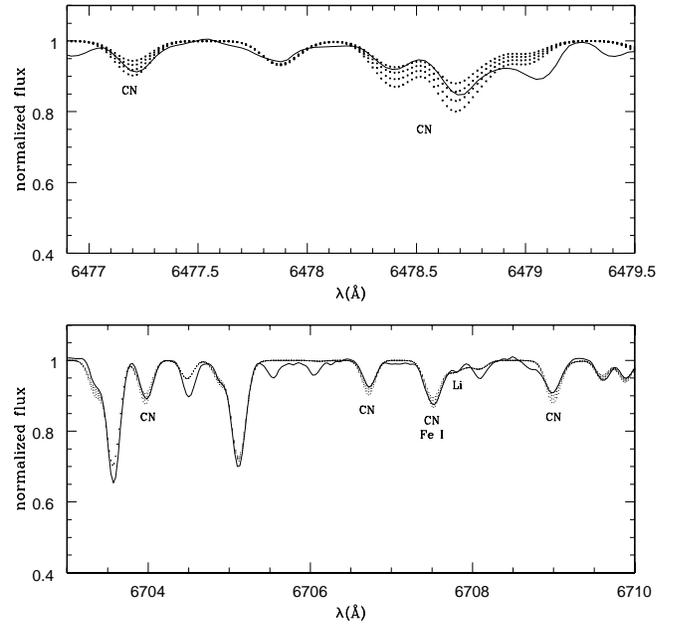}}
\caption{\label{12392cn}Example of fits of CN bands for the star HD 12392.
Symbols: solid line: observed spectrum; dotted lines: synthetic spectra with 
several N abundances.
Upper panel:  $\log\epsilon(N)$ = 8.60, 8.50, 8.40, 8.30;
Lower panel: $\log\epsilon(N)$ = 8.50, 8.45, 8.40, 8.35.}
\end{figure}

\begin{table*}
\caption{Derived abundances of Li, C and N. The stars codes in the header are:
e1 - HD 749; e2 - HR 107*; e3 - HD 5424; e4 - HD8270*; e5 - 12392;
e6 - 13551*; e7 - HD 22589; e8 - HD 27271; e9 - HD 48565*; e10 - HD 76225*; e11 - HD 87080*; e12 - HD 89948*;
e13 - HD 92545*; e14 - HD 106191*; e15 - HD 107574*; e16 - HD 116869; e17 - HD 123396; e18 - HD 123585*;
e19 - HD 147609*; e20 - HD 150862*; e21 - HD 188985; e22 - HD 210709; e23 - HD 210910*; e24 - HD 222349*;
e25 - BD+18 5215*; e26 - HD 223938. The G band covers the region 
$\lambda\lambda$4295 - 4315 $\rm \AA$. The stars with '*' show an upper limit for N.}
{\tiny
\label{abmol}
 $$
\setlength\tabcolsep{2pt}
\begin{tabular}{lcccccccccccccccccccccccccccc}
\hline
\noalign{\smallskip}
el & mol & $\lambda$ & e1 & e2 & e3 & e4 & e5 & e6 & e7 & e8 & e9 & e10 & e11 & e12 & e13 & e14 & 
e15 & e16 & e17 & e18 & e19 & e20 & e21 & e22 & e23 & e24 & e25 & e26 \\
\noalign{\smallskip}
\hline
\noalign{\smallskip}
Li &   ...    & 6707.776 &  -0.33 &$<$1.44 & 0.10 &$<$1.08 & 0.58 &$<$1.36 &$<$0.63 & 0.37 &$<$0.68 &$<$0.94 & 0.36 & 1.10 &$<$0.98 &$<$1.11 &  ...  &$<$0.18 & -0.14 &$<$1.12 &$<$1.18 &$<$1.00 &$<$1.00 &$<$0.16 & -0.06 & 1.47 & 1.67 & 0.17 \\
Li &   ...    & 6707.927 &  -0.33 &$<$1.44 & 0.50 &$<$1.38 & 0.58 &$<$1.12 &$<$0.63 & 0.37 &$<$0.68 &$<$0.63 & 0.36 & 1.10 &$<$0.98 &$<$1.11 &  ...  &$<$0.18& -0.14 &$<$1.12 &$<$1.18 &$<$1.00 &$<$1.00 &$<$0.16 & -0.06 & 1.77 & 1.67 & 0.17 \\
C  & C$\sb 2$ & 5135.600 &  8.62 &  8.39 & 8.35 &  8.30 &  8.79 &  8.18 &  8.55 &  8.59 & 8.27 & 8.58 & 8.38 &  8.47 &  8.63 &  8.60 &  8.37 & 8.31 &  7.83 &  8.83 &  8.60 &  8.86 & 8.61 & 8.46 & 8.42 &  8.28 &  8.53 &  8.53 \\
C  & C$\sb 2$ & 5165.254 &  8.67 &  8.39 & 8.35 &  8.40 &  8.84 &  8.28 &  8.55 &  8.69 & 8.27 & 8.58 & 8.40 &  8.47 &  8.68 &  8.60 &  8.37 & 8.31 &  7.83 &  8.83 &  8.60 &  8.86 & 8.61 & 8.46 & 8.52 &  8.48 &  8.53 &  8.53 \\
C  & C$\sb 2$ & 5635.500 &  8.67 &  8.39 & 8.40 &  8.50 &  8.79 &  8.48 &  8.55 &  8.69 & 8.27 & 8.58 & 8.42 &  ...  &  8.78 &  8.65 &  8.37 & 8.31 &  7.81 &  8.83 &  8.60 &  8.86 & 8.61 & 8.46 & 8.52 &  8.68 &  8.63 &  8.53 \\
C  & CH       & G band   &  8.67 &  8.39 & 8.35 &  8.40 &  8.79 &  8.28 &  8.55 &  8.69 & 8.27 & 8.47 & 8.42 &  8.47 &  8.78 &  8.65 &  8.32 & 8.31 &  7.83 &  8.83 &  8.50 &  8.86 & 8.61 & 8.46 & 8.32 &  8.28 &  8.63 &  8.53 \\
N  & CN       & 6477.200 &  7.89 &  8.66 & 7.65 &  8.80 &  8.50 &  8.75 &  8.13 &  8.35 & 7.98 & 8.74 & 7.71 &  ...  &  ...  &  ...  &  ...  & 7.93 &  ...  &  8.80 &  8.36 &  8.38 & 8.68 & 8.13 & 8.07 &  ...  &  ...  &  7.94 \\
N  & CN       & 6478.400 &  ...  &  8.66 & 7.65 &  ...  &  8.30 &  9.15 &  7.63 &  8.35 & 7.88 & 8.74 & 7.71 &  ...  &  ...  &  ...  &  ...  & 7.83 &  ...  &  8.80 &  8.36 &  8.38 & 8.68 & 8.13 & 8.07 &  ...  &  ...  &  ...  \\
N  & CN       & 6478.700 &  ...  &  8.66 & 7.95 &  ...  &  8.40 &  8.75 &  7.63 &  8.35 & 7.88 & 8.74 & 7.71 &  ...  &  ...  &  ...  &  ...  & 7.88 &  ...  &  8.80 &  8.36 &  8.38 & 8.68 & 8.13 & 8.07 &  ...  &  ...  &  ...  \\
N  & CN       & 6479.000 &  8.09 &  8.66 & 8.15 &  ...  &  ...  &  8.75 &  ...  &  ...  & 7.88 & 8.74 & 7.71 &  ...  &  ...  &  ...  &  ...  & 8.08 &  ...  &  ...  &  8.36 &  8.38 & 8.68 & 8.13 & 8.07 &  ...  &  ...  &  ...  \\
N  & CN       & 6703.968 &  7.79 &  8.66 & 7.71 &  8.60 &  8.45 &  8.55 &  7.93 &  8.35 & 7.88 & 8.64 & 7.71 &  ...  &  8.04 &  ...  &  9.31 & 7.83 &  6.76 &  ...  &  8.36 &  8.38 & 8.48 & 8.13 & 8.07 &  ...  &  ...  &  7.94 \\
N  & CN       & 6706.733 &  7.79 &  8.66 & 7.71 &  8.60 &  8.40 &  8.55 &  7.93 &  8.35 & 7.88 & 8.64 & 7.71 &  ...  &  8.04 &  ...  &  9.31 & 7.83 &  6.76 &  ...  &  8.36 &  8.38 & 8.48 & 8.13 & 8.07 &  ...  &  ...  &  7.94 \\
N  & CN       & 6708.993 &  7.79 &  8.66 & 7.71 &  8.60 &  8.40 &  8.75 &  ...  &  8.35 & 7.88 & 8.64 & 7.71 &  ...  &  8.04 &  ...  &  9.31 & 7.83 &  6.76 &  ...  &  8.36 &  8.38 & 8.48 & 8.13 & 8.07 &  ...  &  ...  &  7.94 \\
N  & CN       & 8030.410 &  7.74 &  ...  & 7.71 &  8.40 &  8.30 &  8.25 &  7.83 &  8.15 & 7.78 & 8.44 & 7.71 &  8.10 &  8.04 &  8.77 &  9.31 & 7.73 &  6.51 &  8.80 &  8.36 &  8.38 & 8.58 & 8.13 & 7.97 &  ...  &  8.79 &  7.84 \\
N  & CN       & 8030.720 &  7.79 &  ...  & 7.81 &  8.40 &  8.40 &  ...  &  7.83 &  8.25 & 7.78 & 8.14 & 7.81 &  8.10 &  8.04 &  8.47 &  9.21 & 7.93 &  6.75 &  8.80 &  8.36 &  8.38 & 8.18 & 8.13 & 7.97 &  ...  &  8.79 &  7.94 \\
N  & CN       & 8034.970 &  7.79 &  ...  & 7.75 &  7.90 &  8.30 &  7.55 &  7.78 &  8.35 & 7.78 & 7.94 & 7.71 &  7.85 &  8.04 &  8.07 &  ...  & 7.93 &  6.66 &  ...  &  ...  &  8.08 & 7.88 & 8.13 & 7.97 &  8.09 &  8.79 &  7.84 \\
N  & CN       & 8040.100 &  7.69 &  ...  & 7.55 &  8.25 &  8.25 &  7.95 &  7.73 &  8.20 & 8.28 & 8.44 & 7.71 &  8.05 &  8.04 &  8.32 &  9.31 & 7.68 &  6.51 &  8.80 &  8.36 &  ...  & 8.08 & 7.93 & 7.97 &  ...  &  8.79 &  7.84 \\
N  & CN       & 8040.220 &  7.69 &  ...  & 7.65 &  8.25 &  8.30 &  7.95 &  7.73 &  8.25 & 8.28 & 8.44 & 7.71 &  8.05 &  8.04 &  8.32 &  8.91 & 7.73 &  6.61 &  8.80 &  8.36 &  ...  & 8.08 & 8.03 & 7.97 &  ...  &  8.79 &  7.84 \\
\noalign{\smallskip}
\hline
\end{tabular}														      
$$
}
\end{table*}

\begin{table}
\caption{Average abundances from molecular lines and Li synthesis.
'$<$' indicates an upper limit.}
{\scriptsize
\label{mamol}
 $$
\setlength\tabcolsep{2pt}
\begin{tabular}{lccrrrcr}
\hline
\noalign{\smallskip}
star & $\log\epsilon$(C) & [C/Fe] & $\log\epsilon$(N) & [N/Fe] & 
[C/O] & [$\sum$CNO/Fe] & $\log\epsilon$(Li) \\
\noalign{\smallskip}
\hline
\noalign{\smallskip}
HD 749    & 8.66 & 0.20 &    7.82 &   -0.04 & -0.01$\pm$0.20 &  0.19 &   -0.33 \\
HR 107    & 8.39 & 0.23 & $<$8.66 & $<$1.10 &  0.30$\pm$0.14 &  0.33 & $<$1.44 \\
HD 5424   & 8.36 & 0.39 &    7.78 &    0.41 &  0.21$\pm$0.20 &  0.29 &    0.34 \\
HD 8270   & 8.41 & 0.31 & $<$8.49 & $<$0.99 &  0.23$\pm$0.14 &  0.35 & $<$1.26 \\
HD 12392  & 8.80 & 0.40 &    8.37 &    0.57 &  0.22$\pm$0.20 &  0.31 &    0.58 \\
HD 13551  & 8.32 & 0.24 & $<$8.64 & $<$1.16 & -0.19$\pm$0.14 &  0.53 & $<$1.26 \\
HD 22589  & 8.55 & 0.30 &    7.84 &    0.19 &  0.27$\pm$0.14 &  0.16 & $<$0.63 \\
HD 27271  & 8.67 & 0.24 &    8.31 &    0.48 &  0.05$\pm$0.20 &  0.24 &    0.37 \\
HD 48565  & 8.27 & 0.37 & $<$7.97 & $<$0.67 &  0.34$\pm$0.14 &  0.26 & $<$0.68 \\
HD 76225  & 8.56 & 0.35 & $<$8.58 & $<$0.97 &  0.22$\pm$0.14 &  0.37 & $<$0.81 \\
HD 87080  & 8.41 & 0.33 & $<$7.72 & $<$0.24 &	 ...	     &  ...  &    0.36 \\
HD 89948  & 8.47 & 0.25 & $<$8.04 & $<$0.42 & -0.03$\pm$0.14 &  0.28 &    1.10 \\
HD 92545  & 8.72 & 0.32 & $<$8.04 & $<$0.24 & -0.01$\pm$0.14 &  0.32 & $<$0.98 \\
HD 106191 & 8.63 & 0.40 & $<$8.45 & $<$0.82 &  0.02$\pm$0.14 &  0.45 & $<$1.11 \\
HD 107574 & 8.36 & 0.39 & $<$9.26 & $<$1.89 &  0.06$\pm$0.14 &  0.94 &   ...   \\
HD 116869 & 8.31 & 0.11 &    7.86 &    0.26 &  0.15$\pm$0.20 &  0.05 & $<$0.18 \\
HD 123396 & 7.83 & 0.50 &    6.68 &   -0.05 &  0.08$\pm$0.20 &  0.43 &   -0.14 \\
HD 123585 & 8.83 & 0.79 & $<$8.80 & $<$1.36 &  0.46$\pm$0.14 &  0.73 & $<$1.12 \\
HD 147609 & 8.58 & 0.51 & $<$8.36 & $<$0.89 &	    ...      &  ...  & $<$1.18 \\
HD 150862 & 8.86 & 0.44 & $<$8.36 & $<$0.54 &  0.16$\pm$0.14 &  0.37 & $<$1.00 \\
HD 188985 & 8.61 & 0.39 &    8.49 &    0.87 & -0.05$\pm$0.14 &  0.48 & $<$1.00 \\
HD 210709 & 8.46 &-0.02 &    8.11 &    0.23 &  0.00$\pm$0.20 &  0.01 & $<$0.16 \\
HD 210910 & 8.45 & 0.30 & $<$8.03 & $<$0.48 & -0.14$\pm$0.20 &  0.40 &   -0.06 \\
HD 222349 & 8.46 & 0.57 & $<$8.09 & $<$0.80 &  0.14$\pm$0.14 &  0.53 &    1.65 \\
BD+18 5215& 8.58 & 0.59 & $<$8.79 & $<$1.40 &  0.16$\pm$0.14 &  0.70 &    1.67 \\
HD 223938 & 8.53 & 0.36 &    7.90 &    0.33 &  0.01$\pm$0.20 &  0.35 &    0.17 \\
\noalign{\smallskip}
\hline
\end{tabular}														      
$$
}
\end{table}

\subsection{Lithium}

Lithium is among the yields of primordial nucleosynthesis. It can also be produced 
in stars with M $\leq$ 8 M$\sb \odot$, through spallation by cosmic rays \citep{ww95} 
and the $\nu$-process suggested for the first time by \citet{domoga77} and later by 
\citet{woosley90} and \citet{times95}. Red Giant Branch or Asymptotic 
Giant Branch stars possibly produce Li \citep{sack92,sack99} and some of them become 
Lithium Rich Giants \citep{brown89,cast98}. However, Li is mainly destroyed during
the life of a star, and for this reason, it can be used as an appraiser of its age.
Cooler stars have a deeper convective zone, consequently, Li is brought to inner regions 
where it is completely destroyed.

For low metallicity dwarfs there is an upper limit of the temperature where the 
convective zone acts such that surface Li can be preserved, forming the 
Spite's plateau \citep{spite82}.

\citet{cast00} showed that the globular cluster NGC 6397 presents a Li dilution curve 
where for T$\sb {eff}$ $\approx$ 6000 K, $\log\epsilon$(Li) $\approx$ 2.2 
and T$\sb {eff}$ $\approx$ 4200 K, $\log\epsilon$(Li) $\approx$ -0.8, indicating
an additional Li destruction. Other causes for Li depletion 
have been suggested such as rotationally induced mixing and mass loss.

Most AGB stars rich in O and C are poor in Li \citep{boesg70,denn91,kipper90}. 
Therefore, a barium star considered as a result of the mass transfer from the 
companion AGB
should be also Li-poor. The Li I lines at $\lambda$6708 $\rm \AA$ in the present
sample are very weak. For 14 of them it was possible to derive an upper limit
for Li abundances. Table \ref{mamol} shows the Li abundances and Figure \ref{teff_Li} 
shows their behaviour with temperature. The abundances usually obtained for dwarfs are higher
than for giants, even taking into account that in several cases it was possible to 
compute
only an upper limit. For the dwarf star BD+18 5215, T$_{eff}$ = 6300 K and 
$\log\epsilon$(Li) = 1.67 and for the coolest star in the sample HD 123396, 
T$_{eff}$ = 4360 K and $\log\epsilon$(Li) = -0.14.

\subsection{Carbon}

C abundances were derived using molecular synthesis of the
C$_2$ Swan (0,0) $\lambda$5165.2 $\rm \AA$, C$_2$ Swan (0,1) $\lambda$5635.5 $\rm \AA$
band heads and the G band (CH A$\sp 3$$\Delta$ - X$\sp 3$$\pi$) 
at $\lambda$4290-4315 $\rm \AA$. Examples of spectrum synthesis are shown in  
Figure \ref{12392c2}.

Figure \ref{cno} shows that the sample stars are C-rich as compared with
normal stars of the disk and halo, this being characteristic of barium stars 
\citep{bk51,warn65}. Despite the dispersion, the results show an increasing trend in 
[C/Fe] toward lower metallicities, except for the halo giant HD 123396.

\citet{bjra92} derived abundances of C, N and O for a sample of barium stars,
including the star HD 27271 in common with the present sample. Through the synthesis
of the C$\sb 2$ line at $\lambda$5136 $\rm \AA$, they found [C/Fe] = 0.15, 
compatible with the present work, [C/Fe] = 0.23, that is an average of several lines.
They obtained the range -0.25 $\leq$ [C/Fe] $\leq$ 0.3 taking into account all stars
in their sample and all indicators, and only 3 stars could be deficient in C. 
For the present sample the range -0.02 $\leq$ [C/Fe] $\leq$ 0.79 was found, with
[C/Fe] $<$ 0 for one star. \citet{claudio05} determined C abundances for the
stars HD 8270, HD 13551 and HD 22589 from C I lines at $\lambda$5380 $\rm \AA$, 
$\lambda$7113 $\rm \AA$, $\lambda$7115 $\rm \AA$, 
$\lambda$7117 $\rm \AA$, $\lambda$7120 $\rm \AA$, and found, respectively,
[C/Fe] = 0.71, 0.46, 0.64 (see Table \ref{ablit}), which values are higher than
the present analysis for the same stars (0.31, 0.24, 0.30, Table \ref{mamol}).
For the star HD 87080, \citet{claudio03} found [C I/Fe] = 0.61, also higher than
in this work (0.33), but it is well known that the triplet at 
$\lambda$7115 - $\lambda$7120 $\rm \AA$ used by them is subject to strong NLTE effects
\citep{przybilla01}. 
Comparing the results of [C/Fe] by \citet{north94} with
the stars in common with the present sample shown in Tables \ref{ablit} and
\ref{mamol}, a good agreement is seen, though their results
for the stars HD 48565, HD 76225 and HD 107574 are higher. Their results
are in the range -0.15 $\leq$ [C/Fe] $\leq$ 0.89, confirming the good
agreement.

Figure \ref{cno} shows a decreasing trend of [C/Fe] toward higher metallicities
for the sample barium stars. \citet{reddy03} found similar behaviour for their 
sample of F and G disk dwarfs, whereas \citet{goswami00} found a constant
behaviour centered in [C/Fe] $\sim$ 0 for their sample of halo and disk stars, 
in the ranges -0.4 $<$ [C/Fe] $<$ +0.4 and -1 $<$ [Fe/H] $<$ 0.
It is also possible to note that even
the metal-poor barium stars are richer in C than normal stars of similar
metallicities. The star HD 123396 does not follow the decreasing trend of [C/Fe] 
of Figure \ref{cno}, and, if it is a halo giant, then it is C-rich
relative to most stars of the halo and instead compatible with C-rich halo stars
\citep{rossi05}. Figure \ref {cno} also shows that the less evolved stars tend to 
have higher C abundances than more evolved ones, since they did not reach yet 
the first dredge-up (see Sect. 4.3).

[C/O] vs. [Fe/H] (Figure \ref{cno}) shows a large dispersion. The range shown 
is -0.19 $\leq$ [C/O] $\leq$ 0.46, and for 6 stars 
[C/O] $<$ 0 (see Table \ref{mamol}). \citet{bjra92} found the range 
0.19 $\leq$ [C/O] $\leq$ 0.47. \citet{north94} found higher values for [C/O],
inside the range 0.4 $\leq$ [C/O] $\leq$ 1.5.

\subsection{Nitrogen}

\citet{reddy03} derived N abundances from two N I lines for F and G dwarfs, and
observed a large dispersion -0.05 $\leq$ [N/Fe] $\leq$ 0.6 at
-0.4 $\leq$ [Fe/H] $\leq$ 0.15. The sample of \citet{clegg81} resulted in 
[N/Fe] $\approx$ 0 in the same range of metallicities.

In the present work, nitrogen abundances were determined by synthesis of
CN lines in the regions $\lambda$$\lambda$6476 - 6480 $\rm \AA$,
$\lambda$$\lambda$6703 - 6709 $\rm \AA$ and $\lambda$$\lambda$8030 - 8041 $\rm \AA$,
with C abundance previously established from the average of the synthesis of 
C$\sb 2$ and CH (Sect. 4.2). Figure \ref{12392cn} illustrates the fits to CN bands.
The CN bands are very weak in stars
with higher temperatures, and for them only an upper limit was given,
as indicated in Figure \ref{cno} and Table \ref{mamol}.
The results are shown in Tables \ref{abmol} and \ref{mamol}. 
[N/Fe] seems to increase toward lower metallicities, except for the star HD 123396, as shown 
in Figure \ref{cno}. The dispersion is probably due to different degrees of mixing.

\citet{bjra92} noted that the barium giants of their sample are rich in both N and C.
Figure \ref{cno} and Table \ref{mamol} show that the same applies to the present
sample. For the star in common HD 27271, a nitrogen abundance
0.23 dex lower than \citet{bjra92} was found. Figure \ref{cno} suggests 
that the less evolved barium stars show larger N overabundances.

The CNO excess provides clues on the origin of the material rich in heavy elements that
polluted the envelope of the barium stars, by comparing them with normal stars. During 
the main sequence, the H burning through CNO cycle conserves the number of nuclei of C, N 
and O. During the first dredge-up the atoms of $\sp {12}$C 
capture protons forming $\sp {13}$C or $\sp {14}$N, and, as a result, the $\sp {12}$C 
is depleted by $\approx$ 0.2 dex and the N increases by $\approx$ 0.3 dex. Some normal 
giants can reach [N/Fe] = 0.55 and in metal-poor stars the mixing is more efficient. 
The barium giants have already passed by the first 
dredge-up, and therefore they should carry the characteristics of this event in addition
to their peculiarities. Once the O abundance is not modified and N increases for every
star after the first dredge-up, the C is considered the main responsible for the CNO
excess in barium stars relative to normal stars, suggesting that the origin of the 
material that has polluted the envelope is the He burning shell. The fact that the less 
evolved barium stars are rich in N suggests that N is also responsible for the CNO 
excess in these
stars. Table \ref{mamol} shows [$\sum$CNO/Fe] $>$ 0 for all barium stars, and 
Figure \ref{cnoteflog} shows a clear increasing trend of [$\sum$CNO/Fe] with T$\sb {eff}$
and log g. The range found by \citet{bjra92} for their giants correspond to 
0.06 $\leq$ [$\sum$CNO/Fe] $\leq$ 0.24 and for the present work,  
0.01 $\leq$ [$\sum$CNO/Fe] $\leq$ 0.73, with the less evolved stars showing the higher 
values.

\begin{figure*}[ht!]
\centerline{\includegraphics[totalheight=6.0cm]{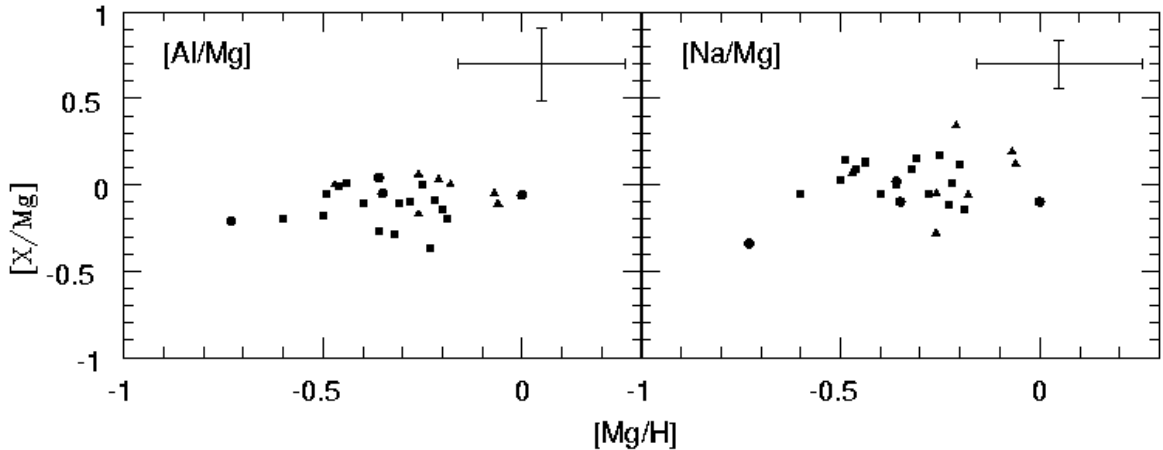}}
\caption{\label{alnamg}[Al/Mg] and [Na/Mg] vs. [Mg/H].
Symbols are the same as in Figure \ref{teff_Li}.}
\end{figure*}

\begin{figure*}[ht!]
\centerline{\includegraphics[totalheight=6.0cm]{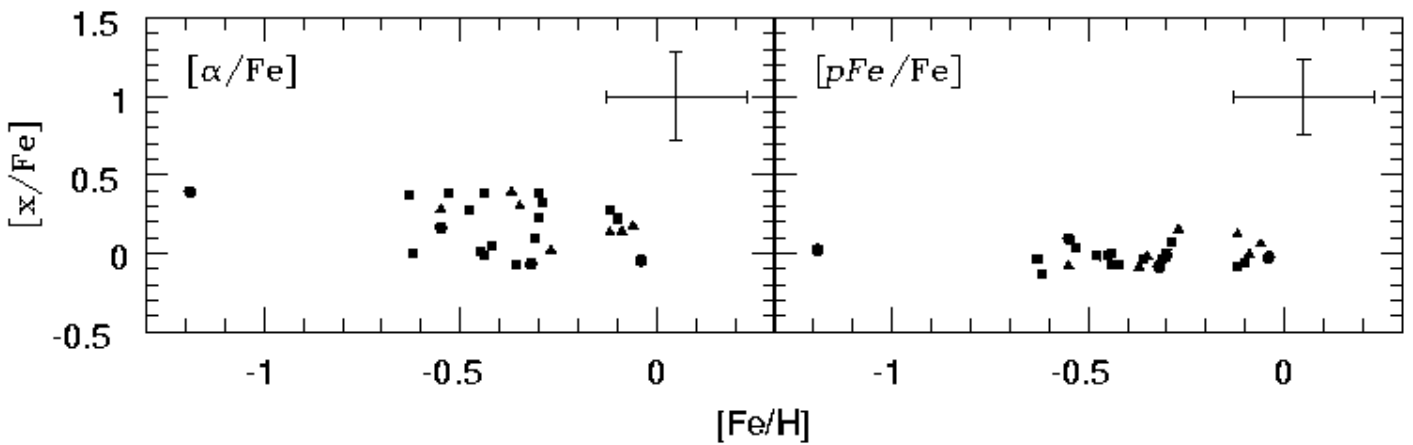}}
\caption{\label{relpal}[$\alpha$,$pFe$/Fe] vs. [Fe/H]. 
Symbols are the same as in Figure \ref{teff_Li}.}
\end{figure*}

\begin{table*}
\caption{Abundance ratios of Al and Na relative to Mg, and the $\alpha$ and $pFe$ relative to Fe.}
{\footnotesize
\label{alpfe}
 $$
\setlength\tabcolsep{3pt}
\begin{tabular}{lrrrrrrrrrr}
\hline
\noalign{\smallskip}
star & [Al/Mg] & $\sigma_{[Al/Mg]}$ & [Na/Mg] & $\sigma_{[Na/Mg]}$ & [Mg/H] & 
$\sigma_{[Mg/H]}$ & [$\alpha$/Fe] & $\sigma_{[\alpha/Fe]}$ & [$pFe$/Fe] & 
$\sigma_{[pFe/Fe]}$ \\
\noalign{\smallskip}
\hline
\noalign{\smallskip}
HD 749     &  0.00 & 0.20 & -0.06 &  0.14 & -0.18 &  0.21 &  0.17 & 0.28 &  0.06 & 0.24 \\
HR 107     & -0.10 & 0.10 & -0.05 &  0.07 & -0.28 &  0.07 & -0.08 & 0.11 & -0.04 & 0.06 \\
HD 5424    & -0.05 & 0.20 & -0.10 &  0.14 & -0.35 &  0.21 &  0.16 & 0.28 &  0.09 & 0.24 \\
HD 8270    & -0.01 & 0.10 &  0.09 &  0.07 & -0.46 &  0.07 &  0.04 & 0.11 & -0.07 & 0.06 \\
HD 12392   & -0.05 & 0.20 &  0.19 &  0.14 & -0.07 &  0.21 &  0.13 & 0.28 &  0.12 & 0.24 \\
HD 13551   & -0.20 & 0.10 & -0.14 &  0.07 & -0.19 &  0.07 &  0.38 & 0.11 &  0.00 & 0.06 \\
HD 22589   & -0.11 & 0.10 &  0.12 &  0.07 & -0.06 &  0.07 &  0.02 & 0.11 &  0.15 & 0.06 \\
HD 27271   &  0.03 & 0.20 &  0.34 &  0.14 & -0.21 &  0.21 &  0.14 & 0.28 & -0.01 & 0.24 \\
HD 48565   & -0.20 & 0.10 & -0.05 &  0.07 & -0.60 &  0.07 &  0.00 & 0.11 & -0.14 & 0.06 \\
HD 76225   & -0.29 & 0.10 &  0.09 &  0.07 & -0.32 &  0.07 &  0.09 & 0.11 & -0.04 & 0.06 \\
HD 87080   & -0.11 & 0.10 & -0.05 &  0.07 & -0.40 &  0.07 & -0.01 & 0.11 & -0.07 & 0.06 \\
HD 89948   & -0.37 & 0.10 & -0.12 &  0.07 & -0.23 &  0.07 &  0.23 & 0.11 & -0.03 & 0.06 \\
HD 92545   & -0.14 & 0.10 &  0.12 &  0.07 & -0.20 &  0.07 &  0.27 & 0.11 & -0.09 & 0.06 \\
HD 106191  & -0.09 & 0.10 &  0.01 &  0.07 & -0.22 &  0.07 &  0.32 & 0.11 &  0.07 & 0.06 \\
HD 107574  &  0.00 & 0.10 &  0.07 &  0.07 & -0.47 &  0.07 &  0.28 & 0.11 & -0.08 & 0.06 \\
HD 116869  &  0.04 & 0.20 &  0.02 &  0.14 & -0.36 &  0.21 & -0.07 & 0.28 & -0.09 & 0.24 \\
HD 123396  & -0.21 & 0.20 & -0.34 &  0.14 & -0.73 &  0.21 &  0.39 & 0.28 &  0.02 & 0.24 \\
HD 123585  &  0.01 & 0.10 &  0.13 &  0.07 & -0.44 &  0.07 &  0.27 & 0.11 & -0.02 & 0.06 \\
HD 147609  & -0.05 & 0.10 &  0.14 &  0.07 & -0.49 &  0.07 &  0.01 & 0.11 & -0.02 & 0.06 \\
HD 150862  &  0.00 & 0.10 &  0.17 &  0.07 & -0.25 &  0.07 &  0.22 & 0.11 & -0.06 & 0.06 \\
HD 188985  & -0.11 & 0.10 &  0.15 &  0.07 & -0.31 &  0.07 &  0.38 & 0.11 &  0.00 & 0.06 \\
HD 210709  & -0.06 & 0.20 & -0.10 &  0.14 &  0.00 &  0.21 & -0.05 & 0.28 & -0.03 & 0.24 \\
HD 210910  &  0.06 & 0.20 & -0.28 &  0.14 & -0.26 &  0.21 &  0.39 & 0.28 & -0.09 & 0.24 \\
HD 222349  & -0.18 & 0.10 &  0.03 &  0.07 & -0.50 &  0.07 &  0.37 & 0.11 & -0.04 & 0.06 \\
BD+18 5215 & -0.27 & 0.10 &  0.00 &  0.07 & -0.36 &  0.07 &  0.38 & 0.11 &  0.03 & 0.06 \\
HD 223938  & -0.17 & 0.20 & -0.05 &  0.14 & -0.26 &  0.21 &  0.30 & 0.28 & -0.02 & 0.24 \\
\noalign{\smallskip}
\hline
\end{tabular}														      
$$
}
\end{table*}

\subsection{Sodium and Aluminum}

\begin{figure*}
\centering
\includegraphics[width=17cm]{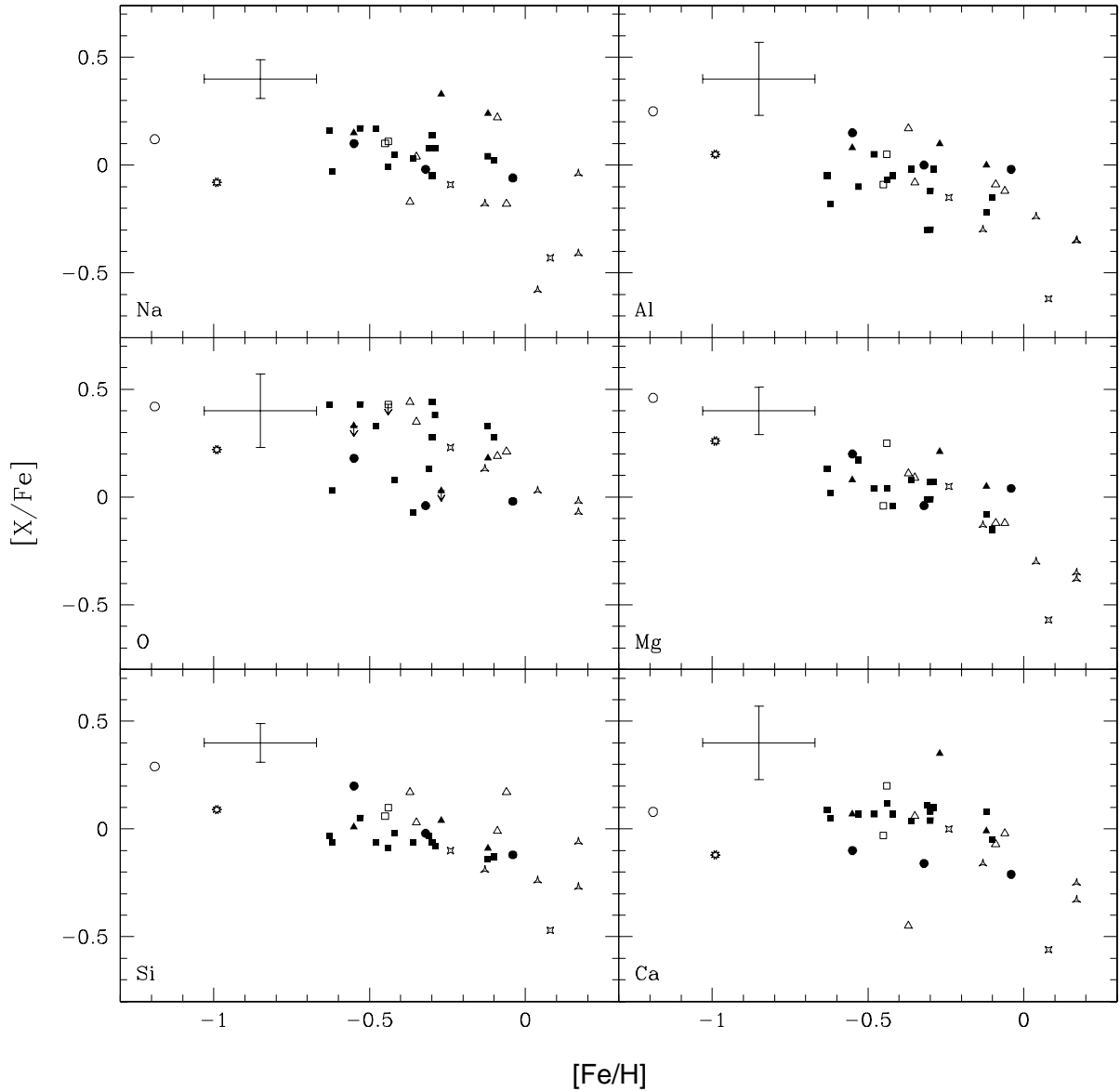}
\caption {[X/Fe] vs. [Fe/H]. The symbols carry a double information: luminosity class and 
$\Delta$[Fe/H]=[FeI/H]-[FeII/H]. 
4-pointed symbols (squares and stars): dwarf stars with $\log g \ge 3.7$;
3-pointed symbols (triangles and stars): subgiants stars with $2.4 < \log g < 3.7$;
round symbols (circles and star): giants stars with $\log g \le 2.4$.
Filled symbols indicate $\Delta$[Fe/H] $<$ 0.2 dex, being the adopted metallicities derived from \ion{Fe}{ii} lines.
Open symbols indicate the seven stars for which $\Delta$[Fe/H] $>$ 0.2 dex. These stars are represented twice
being circles, triangles and squares corresponding to metallicities derived from \ion{Fe}{i} lines and
starred symbols, the same stars with metallicities derived from \ion{Fe}{ii} lines. The arrows in the oxygen
panel indicate an upper limit for HD 13551, HD 22589 and HD 107574.}
\label{imalfefe1}
\end{figure*}

It is probable that the production of Na, Mg and Al occurs through C and N 
burning in massive stars, and for this reason, probably the SN II are the main
sources of $\alpha$-, Na and Al in the disk. Sharing the same production site, the
pattern of abundances are expected to show similarities, and Figure \ref{imalfefe1} 
shows that it is true for the program stars. For the same reason, it is 
interesting to study the relations involving [Al/Mg] and [Na/Mg] in addition to
[Al/Fe] and [Na/Fe].


The \citet{baum97} analysis taking into account NLTE effects suggests that for
[Fe/H] $\approx$ -0.5 there is an overabundance of [Al/Fe] $\approx$ 0.15 dex. 
According to Figure \ref{imalfefe1} and Table \ref{alpfe}, some of the sample
stars are found in this region, one of them being the halo star HD 5424
([Fe/H] = -0.55). The others are below this value, except HD 210910 
([Fe/H] = -0.37), which is a subgiant with broad lines, and HD 123396 
([Fe/H] = -1.19), which is a halo giant. Most data are found in the range
-0.1 $\leq$ [Al/Fe] $\leq$ 0.1. Relative to Mg, Figure \ref{alnamg} shows that
in the range -0.75 $\leq$ [Mg/H] $\leq$ 0, 4 stars show overabundance and 19 a
deficiency of Al relative to Mg, with values in the range 
-0.4 $\leq$ [Al/Mg] $\leq$ +0.1. In the Al abundance calculation the lines
$\lambda$6696 $\rm \AA$ and $\lambda$6698.7 $\rm \AA$ were used.

\citet{baum98} analysed a sample of stars taking into account NLTE effects and verified
a decreasing trend of Na abundance toward decreasing metallicities. The 
present sample stars are found to be in the range -0.18 $\leq$ [Na/Fe] $\leq$ 0.31, 
showing no clear trend, but most data have
-0.1 $\leq$ [Na/Fe] $\leq$ 0.2, as shown in 
Figure \ref{imalfefe1}. Relative to Mg, Figure \ref{alnamg} shows that the results
are in the range -0.4 $\leq$ [Na/Mg] $\leq$ +0.35 with a larger dispersion than
[Al/Mg]. A Na deficiency relative to Mg was found in 11 stars, and for the other stars 
[Na/Mg] $\geq$ 0. Six lines of Na I with two doublets were used. For 7 stars the difference 
between the lines is within $\pm$ 0.25 and $\pm$ 0.30. For one star, the difference is 0.40 dex. 
For the other sample stars there is good agreement between the lines.

The Al and Na excesses relative to Ba show an increase toward higher
metallicities (see Figure \ref{relacsbafe} and Table \ref{elalbaeu}), while a
decreasing trend is seen relative to [Ba/H] (Figure \ref{relacsba}). In the
range 0 $\leq$ [Ba/H] $\leq$ 1.5, 
-0.2 $\leq$ [Na/Ba] $\leq$ -1.6 and -0.3 $\leq$ [Al/Ba] $\leq$ -1.8 were found. Otherwise,
the excesses of Al and Na relative to Eu show no trend as a function of [Fe/H],
with -0.7 $\leq$ [Na/Eu] $\leq$ 0.2 and -0.8 $\leq$ [Al/Eu] $\leq$ 0 
(Figure \ref{relacseufe}). In the range -0.7 $\leq$ [Eu/H] $\leq$ 0.4 there is a
small decreasing trend of [Al,Na/Eu] toward higher [Eu/H] (see Figure \ref{relacseu}
and Table \ref{elalbaeu}).

\begin{table}[ht!]
\caption{Lines, equivalent widths and abundances. Symbols: 
'+': abundances derived by using metallicities from \ion{Fe}{i} lines;
'*': multiple line; '**': Ni line blended with O line. 
$<$ indicates an upper limit for oxygen.
The gf-values sources are:
1 - \citet{allende01}, 2 - \citet{L78}, 3 - NIST, 4 - \citet{fuhrmann95}, 
5 - \citet{proc00}, 6 - \citet{M94}, 7 - \citet{barb99}, 8 - \citet{Bi75}, 
9 - \citet{BG80}, 10 - \citet{g94}, 11 - \citet{H82}, 12 - \citet{HL83}, 
13 - \citet{B81}, 14 - \citet{T90}, 15 - \citet{T89}, 16 - \citet{S00}, 
17 - \citet{M98}, 18 - \citet{rut78}, 19 - \citet{law01a}, 20 - \citet{PQ00}, 
21 - \citet{G91}, 22 - \citet{L76}, 23 - \citet{Hartog03}, 24 - \citet{MW77}, 
25 - \citet{S96}, 26 - \citet{B89}, 27 - \citet{law01b}, 28 - \citet{B88}, 
29 - \citet{C62}, 30 - mean between \citet{K92} and \citet{BL93}, 31 - \citet{Biemont00}
32 - \citet{B83}.
Full table is only available in electronic form.}
{\tiny
\label{abun}
 $$
\setlength\tabcolsep{3pt}
\begin{tabular}{rccccccccccccc}
\hline\hline
 &&&& &&  \multicolumn{4}{c}{HD 749} && \multicolumn{3}{c}{HR 107} \\
\cline{7-10} \cline{12-14} \\
el & $\lambda$ & $\chi\sb {ex}$ & log gf & ref && EW & $\log\epsilon$ & [X/Fe] &
[X/Fe]+ && EW & $\log\epsilon$ & [X/Fe] \\
\hline 
 O I  & 6300.31  &  0.00  &  -9.717 &  1    &&  27 &  8.94 &  0.03 &  0.26  &&   6 &  8.31 & -0.07 ... \\
 O I  & 6363.776 &  0.02  & -10.250 &  2    &&  17 &  8.84 & -0.07 &  0.16  && ... & ...   & ...       \\
Na I  & 4982.808 &  2.10  &  -1.913 &  3    && 102 &  6.03 & -0.47 & -0.24  &&  39 &  5.97 &  0.00 ... \\
Na I  & 4982.813 &  2.10  &  -0.961 &  3    &&   * &  6.03 & -0.47 & -0.24  &&   * &  5.97 &  0.00 ... \\
Na I  & 5682.65  &  2.10  &  -0.700 &  3    && 104 &  6.08 & -0.42 & -0.19  &&  55 &  5.94 & -0.03 ... \\
Na I  & 5688.193 &  2.104 &  -1.390 &  3    && 135 &  6.18 & -0.32 & -0.09  &&  86 &  6.04 &  0.07 ... \\
. & . & . & . & . && . & . & . & . && . & . & . \\
. & . & . & . & . && . & . & . & . && . & . & . \\
\noalign{\smallskip}                         
\hline
\end{tabular} 
$$                                           
}
\end{table}

\begin{table*}
\caption{Mean $\log$$\epsilon$(X) and [X/Fe]. The symbol ``*''
indicates that the metallicity was derived from \ion{Fe}{i} lines.}
{\scriptsize
\label{medxfe1}
   $$ 
\setlength\tabcolsep{3pt}
\begin{tabular}{rrrrrrrrrrrrrrrrrrrrrrrr}
\hline
\noalign{\smallskip}
&& \multicolumn{3}{c}{HD 749} && \multicolumn{2}{c}{HR 107} && \multicolumn{2}{c}{HD 5424} && 
\multicolumn{2}{c}{HD 8270} && \multicolumn{2}{c}{HD 12392} && \multicolumn{3}{c}{HD 13551} && \multicolumn{2}{c}{HD 22589} \\ 
\cline{3-5} \cline{7-8} \cline{10-11} \cline{13-14} \cline{16-17} \cline{19-21} \cline{23-24} \\
el && 
$\log\epsilon$(X) & [X/Fe] & [X/Fe]* && $\log\epsilon$(X) & [X/Fe] && $\log\epsilon$(X) & [X/Fe] && 
$\log\epsilon$(X) & [X/Fe] && $\log\epsilon$(X) & [X/Fe] && $\log\epsilon$(X) & [X/Fe] & [X/Fe]* && 
$\log\epsilon$(X) & [X/Fe] \\
\noalign{\smallskip}
\hline
\noalign{\smallskip}
 O    && 8.89 & -0.02 &  0.21 && 8.31 & -0.07 && 8.37 &  0.18 && 8.40 &  0.08 && 8.80 &  0.18 && $<$8.73 &  $<$0.23 &  $<$0.43 && $<$8.50 &  $<$0.03 \\
Na    && 6.09 & -0.41 & -0.18 && 6.00 &  0.03 && 5.88 &  0.10 && 5.96 &  0.05 && 6.45 &  0.24 && 6.00 & -0.09 &  0.11 && 6.39 &  0.33 \\
Mg    && 7.40 & -0.35 & -0.12 && 7.30 &  0.08 && 7.23 &  0.20 && 7.12 & -0.04 && 7.51 &  0.05 && 7.39 &  0.05 &  0.25 && 7.52 &  0.21 \\
Al    && 6.29 & -0.35 & -0.12 && 6.09 & -0.02 && 6.07 &  0.15 && 6.00 & -0.05 && 6.35 &  0.00 && 6.08 & -0.15 &  0.05 && 6.30 &  0.10 \\
Si    && 7.66 & -0.06 &  0.17 && 7.13 & -0.06 && 7.20 &  0.20 && 7.11 & -0.02 && 7.34 & -0.09 && 7.21 & -0.10 &  0.10 && 7.32 &  0.04 \\
Ca    && 6.28 & -0.25 & -0.02 && 6.04 &  0.04 && 5.71 & -0.10 && 6.01 &  0.07 && 6.23 & -0.01 && 6.12 &  0.00 &  0.20 && 6.44 &  0.35 \\
Sc    && 3.42 &  0.08 &  0.31 && 3.00 &  0.19 && 3.11 &  0.49 && 3.01 &  0.26 && 3.48 &  0.43 && 2.98 &  0.05 &  0.25 && 3.06 &  0.16 \\
Ti    && 4.75 & -0.44 & -0.21 && 4.69 &  0.03 && 4.41 & -0.06 && 4.47 & -0.13 && 5.03 &  0.13 && 4.53 & -0.25 & -0.05 && 4.77 &  0.02 \\
 V    && 3.93 & -0.24 & -0.01 && 3.70 &  0.06 && 3.28 & -0.17 && 3.36 & -0.22 && 3.94 &  0.06 && 3.58 & -0.18 &  0.02 && 3.65 & -0.08 \\
Cr    && 5.51 & -0.33 & -0.10 && 5.28 & -0.03 && 5.14 &  0.02 && 5.13 & -0.12 && 5.64 &  0.09 && 5.17 & -0.26 & -0.06 && 5.63 &  0.23 \\
Co    && 5.37 &  0.28 &  0.51 && 4.74 &  0.18 && 4.66 &  0.29 && 4.43 & -0.07 && 5.27 &  0.47 && 4.48 & -0.20 &  0.00 && 4.73 &  0.08 \\
Ni    && 6.25 & -0.17 &  0.06 && 5.83 & -0.06 && 5.79 &  0.09 && 5.77 & -0.06 && 6.23 &  0.10 && 5.82 & -0.19 &  0.01 && 6.11 &  0.13 \\
Cu    && 4.20 & -0.18 &  0.05 && 3.68 & -0.17 && 3.49 & -0.17 && 3.58 & -0.21 && 4.18 &  0.09 && 3.60 & -0.37 & -0.17 && 3.94 &  0.00 \\
Zn    && 4.53 & -0.24 & -0.01 && 4.14 & -0.10 && 4.15 &  0.10 && 4.14 & -0.04 && 4.50 &  0.02 && 4.26 & -0.10 &  0.10 && 4.44 &  0.11 \\
\ion{Sr}{i} && 3.49 &  0.35 &  0.58 && 2.91 &  0.30 && 3.37 &  0.95 && 3.38 &  0.83 && 3.90 &  1.05 && 3.29 &  0.56 &  0.76 && 3.53 &  0.83 \\
\ion{Sr}{ii} && 3.59 &  0.45 &  0.68 && 3.36 &  0.75 && 3.02 &  0.60 && 3.45 &  0.90 && 3.63 &  0.78 && 3.53 &  0.80 &  1.00 && 3.58 &  0.88 \\
 Y    && 3.36 &  0.95 &  1.18 && 2.48 &  0.60 && 2.72 &  1.03 && 2.77 &  0.95 && 3.33 &  1.21 && 2.88 &  0.88 &  1.08 && 2.80 &  0.83 \\
\ion{Zr}{i} && 2.86 &  0.09 &  0.32 && 2.84 &  0.60 && 2.54 &  0.49 && 3.02 &  0.84 && 3.20 &  0.72 && 3.05 &  0.69 &  0.89 && 2.82 &  0.49 \\
\ion{Zr}{ii} && 3.99 &  1.22 &  1.45 && 2.81 &  0.57 && 3.40 &  1.35 && 3.18 &  1.00 && 3.84 &  1.36 && 3.18 &  0.82 &  1.02 && 3.40 &  1.07 \\
Mo    && 1.99 & -0.10 &  0.13 && 2.16 &  0.60 && 1.57 &  0.20 && 2.10 &  0.60 && 2.30 &  0.50 && 2.28 &  0.60 &  0.80 && 1.85 &  0.20 \\
Ba    && 3.25 &  0.95 &  1.18 && 2.72 &  0.95 && 3.06 &  1.48 && 2.82 &  1.11 && 3.52 &  1.51 && 2.85 &  0.96 &  1.16 && 2.74 &  0.88 \\
La    && 2.33 &  1.03 &  1.26 && 1.44 &  0.67 && 2.13 &  1.55 && 1.71 &  1.00 && 2.62 &  1.61 && 1.68 &  0.79 &  0.99 && 1.56 &  0.70 \\
Ce    && 3.14 &  1.27 &  1.50 && 1.75 &  0.41 && 3.01 &  1.86 && 2.11 &  0.83 && 3.25 &  1.67 && 2.17 &  0.71 &  0.91 && 1.88 &  0.45 \\
Pr    && 1.49 &  0.66 &  0.89 && 0.82 &  0.52 && 1.49 &  1.38 && 0.72 &  0.48 && 1.93 &  1.39 && 0.79 &  0.37 &  0.57 && 0.66 &  0.27 \\
Nd    && 2.72 &  1.10 &  1.33 && 1.41 &  0.32 && 2.62 &  1.72 && 1.76 &  0.73 && 2.82 &  1.49 && 1.74 &  0.53 &  0.73 && 1.50 &  0.32 \\
Sm    && 1.88 &  0.70 &  0.93 && 0.95 &  0.30 && 1.73 &  1.27 && 0.96 &  0.37 && 2.36 &  1.47 && 1.02 &  0.25 &  0.45 && 0.82 &  0.08 \\
Eu    && 0.79 &  0.10 &  0.33 && 0.20 &  0.04 && 0.43 &  0.46 && 0.42 &  0.32 && 0.88 &  0.48 && 0.29 &  0.01 &  0.21 && 0.46 &  0.21 \\
Gd    && 1.38 &  0.09 &  0.32 && 1.31 &  0.55 && 1.02 &  0.45 && 0.95 &  0.25 && 1.57 &  0.57 && 1.23 &  0.35 &  0.55 && 0.88 &  0.03 \\
Dy    && 2.06 &  0.69 &  0.92 && ...  & ...   && 2.30 &  1.65 && 0.82 &  0.04 && 1.83 &  0.75 && 0.85 & -0.11 &  0.09 && 0.97 &  0.04 \\
Pb    && 2.27 &  0.15 &  0.38 && 2.49 &  0.90 && 2.50 &  1.10 && 2.03 &  0.50 && 2.98 &  1.15 && 2.01 &  0.30 &  0.50 && 1.53 & -0.15 \\
\noalign{\smallskip}
\hline\hline
\noalign{\smallskip}
&& \multicolumn{3}{c}{HD 27271} && \multicolumn{2}{c}{HD 48565} && \multicolumn{2}{c}{HD 76225} && \multicolumn{2}{c}{HD 87080} && 
\multicolumn{2}{c}{HD 89948} && \multicolumn{2}{c}{HD 92545} && \multicolumn{2}{c}{HD 106191} &\\ 
\cline{3-5} \cline{7-8} \cline{10-11} \cline{13-14} \cline{16-17} \cline{19-20} \cline{22-23} 
el && $\log\epsilon$(X) & [X/Fe] & [X/Fe]* && $\log\epsilon$(X) & [X/Fe] && $\log\epsilon$(X) & [X/Fe] && $\log\epsilon$(X) & [X/Fe] && 
$\log\epsilon$(X) & [X/Fe] && $\log\epsilon$(X) & [X/Fe] && $\log\epsilon$(X) & [X/Fe] &\\
\noalign{\smallskip}
\hline
\noalign{\smallskip}
 O    && 8.84 & -0.07 &  0.19 && 8.15 &  0.03 && 8.56 &  0.13 && ...  &  ...  && 8.72 &  0.28 && 8.95 &  0.33 && 8.83 &  0.38 &\\
Na    && 6.46 & -0.04 &  0.22 && 5.68 & -0.03 && 6.10 &  0.08 && 5.88 & -0.01 && 5.98 & -0.05 && 6.25 &  0.04 && 6.12 &  0.08 &\\
Mg    && 7.37 & -0.38 & -0.12 && 6.98 &  0.02 && 7.26 & -0.01 && 7.18 &  0.04 && 7.35 &  0.07 && 7.38 & -0.08 && 7.36 &  0.07 &\\
Al    && 6.29 & -0.35 & -0.09 && 5.67 & -0.18 && 5.86 & -0.30 && 5.96 & -0.07 && 5.87 & -0.30 && 6.13 & -0.22 && 6.16 & -0.02 &\\
Si    && 7.45 & -0.27 & -0.01 && 6.87 & -0.06 && 7.21 & -0.03 && 7.02 & -0.09 && 7.19 & -0.06 && 7.29 & -0.14 && 7.18 & -0.08 &\\
Ca    && 6.20 & -0.33 & -0.07 && 5.79 &  0.05 && 6.16 &  0.11 && 6.04 &  0.12 && 6.10 &  0.04 && 6.32 &  0.08 && 6.17 &  0.10 &\\
Sc    && 3.42 &  0.08 &  0.34 && 2.82 &  0.27 && 3.09 &  0.23 && 3.13 &  0.40 && 2.95 &  0.08 && 3.23 &  0.18 && 3.03 &  0.15 &\\
Ti    && 4.69 & -0.50 & -0.24 && 4.35 & -0.05 && 4.59 & -0.12 && 4.53 & -0.05 && 4.56 & -0.16 && 4.78 & -0.12 && 4.70 & -0.03 &\\
 V    && 3.63 & -0.54 & -0.28 && 3.19 & -0.19 && 3.46 & -0.23 && 3.42 & -0.14 && 3.60 & -0.10 && 3.77 & -0.11 && 3.77 &  0.06 &\\
Cr    && 5.35 & -0.49 & -0.23 && 4.88 & -0.17 && 5.24 & -0.12 && 5.13 & -0.10 && 5.37 &  0.00 && 5.42 & -0.13 && 5.45 &  0.07 &\\
Co    && 5.00 & -0.09 &  0.17 && 4.22 & -0.08 && 4.64 &  0.03 && 4.48 &  0.00 && 4.60 & -0.02 && 4.82 &  0.02 && 4.78 &  0.15 &\\
Ni    && 6.19 & -0.23 &  0.03 && 5.50 & -0.13 && 5.91 & -0.03 && 5.74 & -0.07 && 5.91 & -0.04 && 6.04 & -0.09 && 6.03 &  0.07 &\\
Cu    && 4.01 & -0.37 & -0.11 && 3.27 & -0.32 && 3.78 & -0.12 && 3.52 & -0.25 && 3.74 & -0.17 && 3.82 & -0.27 && 3.88 & -0.04 &\\
Zn    && 4.47 & -0.30 & -0.04 && 4.01 &  0.03 && 4.27 & -0.02 && 4.26 &  0.10 && 4.32 &  0.02 && 4.33 & -0.15 && 4.39 &  0.08 &\\
\ion{Sr}{i}  && 3.37 &  0.23 &  0.49 && 3.32 &  0.97 && 3.74 &  1.08 && 3.51 &  0.98 && 3.65 &  0.98 && 3.52 &  0.67 && 3.63 &  0.95 &\\
\ion{Sr}{ii} && 3.52 &  0.38 &  0.64 && 3.39 &  1.04 && 3.92 &  1.26 && 3.53 &  1.00 && 3.75 &  1.08 && 3.58 &  0.73 && 3.33 &  0.65 &\\
 Y    && 3.04 &  0.63 &  0.89 && 2.63 &  1.01 && 3.10 &  1.17 && 2.91 &  1.11 && 2.96 &  1.02 && 2.76 &  0.64 && 2.86 &  0.91 &\\
\ion{Zr}{i}  && 2.65 & -0.12 &  0.14 && 2.71 &  0.73 && 3.47 &  1.18 && 2.96 &  0.80 && 3.15 &  0.85 && 3.03 &  0.55 && ...  &  ...  &\\
\ion{Zr}{ii} && 3.55 &  0.78 &  1.04 && 3.17 &  1.19 && 3.55 &  1.26 && 3.51 &  1.35 && 3.33 &  1.03 && 3.23 &  0.75 && 3.38 &  1.07 &\\
Mo    && 1.79 & -0.30 & -0.04 && 1.80 &  0.50 && 2.21 &  0.60 && 1.88 &  0.40 && 2.12 &  0.50 && 2.20 &  0.40 && 2.13 &  0.50 &\\
Ba    && 2.92 &  0.62 &  0.88 && 2.80 &  1.29 && 3.17 &  1.35 && 3.17 &  1.48 && 2.82 &  0.99 && 3.05 &  1.04 && 2.72 &  0.88 &\\
La    && 1.76 &  0.46 &  0.72 && 1.90 &  1.39 && 2.04 &  1.22 && 2.43 &  1.74 && 1.76 &  0.93 && 1.73 &  0.72 && 1.50 &  0.66 &\\
Ce    && 2.27 &  0.40 &  0.66 && 2.68 &  1.60 && 2.45 &  1.06 && 2.99 &  1.73 && 2.11 &  0.71 && 2.18 &  0.60 && 2.02 &  0.61 &\\
Pr    && 1.06 &  0.23 &  0.49 && 1.13 &  1.09 && 1.10 &  0.75 && 1.46 &  1.24 && 0.96 &  0.60 && 0.98 &  0.44 && 1.21 &  0.84 &\\
Nd    && 1.89 &  0.27 &  0.53 && 2.14 &  1.31 && 1.91 &  0.77 && 2.57 &  1.56 && 1.80 &  0.65 && 1.75 &  0.42 && 1.64 &  0.48 &\\
Sm    && 1.33 &  0.15 &  0.41 && 1.34 &  0.95 && 1.23 &  0.53 && 1.69 &  1.12 && 1.14 &  0.43 && 1.13 &  0.24 && 1.18 &  0.46 &\\
Eu    && 0.74 &  0.05 &  0.31 && 0.25 &  0.35 && 0.46 &  0.25 && 0.74 &  0.66 && 0.38 &  0.16 && 0.72 &  0.32 && 0.43 &  0.20 &\\
Gd    && 1.28 & -0.01 &  0.25 && 1.19 &  0.69 && 1.25 &  0.44 && 1.58 &  0.90 && 1.07 &  0.25 && 1.14 &  0.14 && 1.23 &  0.40 &\\
Dy    && 1.31 & -0.06 &  0.20 && 1.27 &  0.69 && 1.18 &  0.29 && 1.90 &  1.14 && 0.59 & -0.31 && 1.17 &  0.09 && 1.20 &  0.29 &\\
Pb    && 2.17 &  0.05 &  0.31 && 2.68 &  1.35 && 2.49 &  0.85 && 2.56 &  1.05 && 2.00 &  0.35 && 2.53 &  0.70 && 2.31 &  0.65 &\\
\noalign{\smallskip}
\hline
\end{tabular}
   $$ 
}
The usual notations were adopted: $\log\epsilon$(A) = $\log(N_A/N_H)$+12 and
[A/B] = $\log(N_A/N_B)_\ast$-$\log(N_A/N_B)_\odot$,
where $N_A$ and $N_B$ are the numerical particle density 
of ``A'' and ``B'', respectively.
\end{table*}

\begin{table*}[ht!]
\caption{Same as Table \ref{medxfe1} for other 12 sample stars and $\log\epsilon$(X) for the Sun. 
For solar abundances the errors on last decimals are given in parenthesis.}
{\scriptsize
\label{medxfe2}
   $$ 
\setlength\tabcolsep{3pt}
\begin{tabular}{rrrrrrrrrrrrrrrrrrrlrrrr}
\hline
\noalign{\smallskip}
&& \multicolumn{2}{c}{HD 107574} && \multicolumn{2}{c}{HD 116869} && \multicolumn{3}{c}{HD 123396} && \multicolumn{2}{c}{HD 123585} && 
\multicolumn{3}{c}{HD 147609} && \multicolumn{2}{c}{HD 150862} && \multicolumn{2}{c}{HD 188985} \\ 
\cline{3-4} \cline{6-7} \cline{9-11} \cline{13-14} \cline{16-18} \cline{20-21} \cline{23-24} \\
el && $\log\epsilon$(X) & [X/Fe] && $\log\epsilon$(X) & [X/Fe] && $\log\epsilon$(X) & [X/Fe] & [X/Fe]* && 
$\log\epsilon$(X) & [X/Fe] && $\log\epsilon$(X) & [X/Fe] & [X/Fe]* && $\log\epsilon$(X) & [X/Fe] && 
$\log\epsilon$(X) & [X/Fe] \\
\noalign{\smallskip}
\hline
\noalign{\smallskip}
 O    && $<$8.52 &  $<$0.33 && 8.38 & -0.04 &&  7.97 &  0.22 &  0.42 && 8.59 &  0.33 && ...  &  ...  &  ...  && 8.92 &  0.28 && 8.88 &  0.44 \\
Na    && 5.93 &  0.15 && 5.99 & -0.02 &&  5.26 & -0.08 &  0.12 && 6.02 &  0.17 && 5.98 & -0.43 &  0.10 && 6.25 &  0.02 && 6.17 &  0.14 \\
Mg    && 7.11 &  0.08 && 7.22 & -0.04 &&  6.85 &  0.26 &  0.46 && 7.14 &  0.04 && 7.09 & -0.57 & -0.04 && 7.33 & -0.15 && 7.27 & -0.01 \\
Al    && 6.00 &  0.08 && 6.15 &  0.00 &&  5.53 &  0.05 &  0.25 && 6.04 &  0.05 && 5.93 & -0.62 & -0.09 && 6.22 & -0.15 && 6.05 & -0.12 \\
Si    && 7.01 &  0.01 && 7.21 & -0.02 &&  6.65 &  0.09 &  0.29 && 7.01 & -0.06 && 7.16 & -0.47 &  0.06 && 7.32 & -0.13 && 7.19 & -0.06 \\
Ca    && 5.88 &  0.07 && 5.88 & -0.16 &&  5.25 & -0.12 &  0.08 && 5.95 &  0.07 && 5.88 & -0.56 & -0.03 && 6.21 & -0.05 && 6.14 &  0.08 \\
Sc    && 2.70 &  0.08 && 2.98 &  0.13 &&  2.45 &  0.27 &  0.47 && 2.89 &  0.20 && 3.43 &  0.18 &  0.71 && 3.32 &  0.25 && 3.11 &  0.24 \\
Ti    && 4.43 & -0.04 && 4.54 & -0.16 &&  3.76 & -0.27 & -0.07 && 4.59 &  0.05 && 4.41 & -0.69 & -0.16 && 4.86 & -0.06 && 4.70 & -0.02 \\
 V    && 3.56 &  0.11 && 3.41 & -0.27 &&  2.48 & -0.53 & -0.33 && 3.63 &  0.11 && 3.35 & -0.73 & -0.20 && 3.78 & -0.12 && 3.75 &  0.05 \\
Cr    && 4.97 & -0.15 && 5.12 & -0.23 &&  4.25 & -0.43 & -0.23 && 5.12 & -0.07 && 5.12 & -0.63 & -0.10 && 5.50 & -0.07 && 5.41 &  0.04 \\
Co    && 4.44 &  0.07 && 4.73 &  0.13 &&  3.90 & -0.03 &  0.17 && 4.78 &  0.34 && 4.46 & -0.54 & -0.01 && 4.90 &  0.08 && 4.68 &  0.06 \\
Ni    && 5.63 & -0.07 && 5.86 & -0.07 &&  5.12 & -0.14 &  0.06 && 5.74 & -0.03 && 5.80 & -0.53 &  0.00 && 6.09 & -0.06 && 5.93 & -0.02 \\
Cu    && 3.44 & -0.22 && 3.65 & -0.24 &&  2.83 & -0.39 & -0.19 && 3.77 &  0.04 && 3.66 & -0.63 & -0.10 && 3.99 & -0.12 && 3.80 & -0.11 \\
Zn    && 4.01 & -0.04 && 4.25 & -0.03 &&  3.84 &  0.23 &  0.43 && 4.22 &  0.10 && 4.29 & -0.39 &  0.14 && 4.50 &  0.00 && 4.37 &  0.07 \\
\ion{Sr}{i}  && 3.02 &  0.60 && 3.08 &  0.43 &&  2.40 &  0.42 &  0.62 && 3.59 &  1.10 && 3.56 &  0.51 &  1.04 && 3.68 &  0.81 && 3.72 &  1.05 \\
\ion{Sr}{ii} && 3.52 &  1.10 && 2.90 &  0.25 &&  2.18 &  0.20 &  0.40 && 3.70 &  1.21 && 4.11 &  1.06 &  1.59 && 3.57 &  0.70 && 3.65 &  0.98 \\
 Y    && 2.65 &  0.96 && 2.51 &  0.59 &&  1.75 &  0.50 &  0.70 && 3.10 &  1.34 && 3.36 &  1.04 &  1.57 && 3.22 &  1.08 && 2.96 &  1.02 \\
\ion{Zr}{i} && ...  &  ...  && 2.31 &  0.03 &&  1.31 & -0.30 & -0.10 && 3.57 &  1.45 && 3.04 &  0.36 &  0.89 && 3.51 &  1.01 && 2.95 &  0.65 \\
\ion{Zr}{ii} && 3.00 &  0.95 && 2.96 &  0.68 &&  2.60 &  0.99 &  1.19 && 3.48 &  1.36 && 3.71 &  1.03 &  1.56 && 3.62 &  1.12 && 3.48 &  1.18 \\
Mo    && 2.27 &  0.90 && 1.40 & -0.20 &&  0.53 & -0.40 & -0.20 && 2.44 &  1.00 && 2.20 &  0.20 &  0.73 && 2.32 &  0.50 && 2.12 &  0.50 \\
Ba    && 3.29 &  1.71 && 2.83 &  1.02 &&  2.24 &  1.10 &  1.30 && 3.44 &  1.79 && 3.25 &  1.04 &  1.57 && 3.06 &  1.03 && 3.03 &  1.20 \\
La    && 1.74 &  1.16 && 1.77 &  0.96 &&  1.21 &  1.07 &  1.27 && 2.30 &  1.65 && 2.31 &  1.10 &  1.63 && 1.83 &  0.80 && 1.99 &  1.16 \\
Ce    && 2.17 &  1.02 && 2.28 &  0.90 &&  2.13 &  1.42 &  1.62 && 2.91 &  1.69 && 2.89 &  1.11 &  1.64 && 2.15 &  0.55 && 2.63 &  1.23 \\
Pr    && 0.81 &  0.70 && 0.98 &  0.64 &&  0.67 &  1.00 &  1.20 && 1.49 &  1.31 && 1.43 &  0.69 &  1.22 && 1.01 &  0.45 && 1.18 &  0.82 \\
Nd    && 1.76 &  0.86 && 1.99 &  0.86 &&  1.81 &  1.35 &  1.55 && 2.38 &  1.41 && 2.32 &  0.79 &  1.32 && 1.69 &  0.34 && 2.19 &  1.04 \\
Sm    && 1.10 &  0.64 && 1.25 &  0.56 &&  1.01 &  0.99 &  1.19 && 1.73 &  1.20 && 1.65 &  0.56 &  1.09 && 1.14 &  0.23 && 1.43 &  0.72 \\
Eu    && 0.44 &  0.47 && 0.36 &  0.16 && -0.17 &  0.30 &  0.50 && 0.87 &  0.83 && 0.81 &  0.21 &  0.74 && 0.62 &  0.20 && 0.51 &  0.29 \\
Gd    && 0.93 &  0.36 && 1.02 &  0.22 &&  0.74 &  0.61 &  0.81 && 1.19 &  0.55 && 1.47 &  0.27 &  0.80 && 1.25 &  0.23 && 1.36 &  0.54 \\
Dy    && 0.99 &  0.34 && 1.43 &  0.55 &&  1.20 &  0.99 &  1.19 && 1.36 &  0.64 && 1.45 &  0.17 &  0.70 && 1.24 &  0.14 && 1.09 &  0.19 \\
Pb    && 2.45 &  1.05 && 2.48 &  0.85 &&  1.96 &  1.00 &  1.20 && 3.02 &  1.55 && 2.28 &  0.25 &  0.78 && 2.55 &  0.70 && 2.60 &  0.95 \\
\noalign{\smallskip}
\hline\hline
\noalign{\smallskip}
&& \multicolumn{2}{c}{HD 210709} && \multicolumn{3}{c}{HD 210910} && \multicolumn{2}{c}{HD 222349} && 
\multicolumn{2}{c}{BD+18 5215} && \multicolumn{3}{c}{HD 223938} && \multicolumn{2}{c}{Sun} &&&\\ 
\cline{3-4} \cline{6-8} \cline{10-11} \cline{13-14} \cline{16-18} \cline{20-21} \\
el && $\log\epsilon$(X) & [X/Fe] && $\log\epsilon$(X) & [X/Fe] & [X/Fe]* && $\log\epsilon$(X) & [X/Fe] && 
$\log\epsilon$(X) & [X/Fe] && $\log\epsilon$(X) & [X/Fe] & [X/Fe]* && $\log\epsilon$(X) & ref &&&\\
\noalign{\smallskip}
\hline
\noalign{\smallskip}
 O    && 8.68 & -0.02 && 8.81 &  0.03 &  0.44 && 8.54 &  0.43 && 8.64 &  0.43 && 8.74 &  0.13 &  0.35 && 8.74(6) & 1 &&&\\
Na    && 6.23 & -0.06 && 5.79 & -0.58 & -0.17 && 5.86 &  0.16 && 5.97 &  0.17 && 6.02 & -0.18 &  0.04 && 6.33(3) & 2 &&&\\
Mg    && 7.58 &  0.04 && 7.32 & -0.30 &  0.11 && 7.08 &  0.13 && 7.22 &  0.17 && 7.32 & -0.13 &  0.09 && 7.58(5) & 2 &&&\\
Al    && 6.41 & -0.02 && 6.27 & -0.24 &  0.17 && 5.79 & -0.05 && 5.84 & -0.10 && 6.04 & -0.30 & -0.08 && 6.47(7) & 2 &&&\\
Si    && 7.39 & -0.12 && 7.35 & -0.24 &  0.17 && 6.89 & -0.03 && 7.07 &  0.05 && 7.23 & -0.19 &  0.03 && 7.55(5) & 2 &&&\\
Ca    && 6.11 & -0.21 && 5.54 & -0.86 & -0.45 && 5.82 &  0.09 && 5.90 &  0.07 && 6.07 & -0.16 &  0.06 && 6.36(2) & 2 &&&\\
Sc    && 3.31 &  0.18 && 3.08 & -0.13 &  0.28 && 2.76 &  0.22 && 2.82 &  0.18 && 3.14 &  0.10 &  0.32 && 3.17(10) & 2 &&&\\
Ti    && 4.73 & -0.25 && 4.14 & -0.92 & -0.51 && 4.34 & -0.05 && 4.45 & -0.04 && 4.51 & -0.38 & -0.16 && 5.02(6) & 2 &&&\\
 V    && 3.92 & -0.04 && 3.35 & -0.69 & -0.28 && 3.51 &  0.14 && 3.56 &  0.09 && 3.46 & -0.41 & -0.19 && 4.00(2) & 2 &&&\\
Cr    && 5.60 & -0.03 && 4.62 & -1.09 & -0.68 && 4.90 & -0.14 && 5.14 &  0.00 && 5.14 & -0.40 & -0.18 && 5.67(3) & 2 &&&\\
Co    && 5.21 &  0.33 && 4.69 & -0.27 &  0.14 && 4.26 & -0.03 && 4.74 &  0.35 && 4.70 & -0.09 &  0.13 && 4.92(4) & 2 &&&\\
Ni    && 6.15 & -0.06 && 5.86 & -0.43 & -0.02 && 5.60 & -0.02 && 5.74 &  0.02 && 5.91 & -0.21 &  0.01 && 6.25(4) & 2 &&&\\
Cu    && 4.00 & -0.17 && 3.68 & -0.57 & -0.16 && 3.39 & -0.19 && 3.59 & -0.09 && 3.71 & -0.37 & -0.15 && 4.21(4) & 2 &&&\\
Zn    && 4.40 & -0.16 && 4.41 & -0.23 &  0.18 && 4.01 &  0.04 && 4.12 &  0.05 && 4.32 & -0.15 &  0.07 && 4.60(8) & 2 &&&\\
\ion{Sr}{i}  && 3.26 &  0.33 && 3.06 &  0.05 &  0.46 && 3.24 &  0.90 && 3.55 &  1.11 && 3.21 &  0.37 &  0.59 && 2.97(7) & 2 &&&\\
\ion{Sr}{ii} && 3.02 &  0.09 && 3.46 &  0.45 &  0.86 && 3.39 &  1.05 && 3.59 &  1.15 && 3.23 &  0.39 &  0.61 && 2.97(7) & 2 &&&\\
 Y    && 2.73 &  0.53 && 2.41 &  0.13 &  0.54 && 2.64 &  1.03 && 2.72 &  1.01 && 2.63 &  0.52 &  0.74 && 2.24(3) & 2 &&&\\
\ion{Zr}{i} && 2.47 & -0.09 && 2.12 & -0.52 & -0.11 && 3.20 &  1.23 && 3.37 &  1.30 && 2.41 & -0.06 &  0.16 && 2.60(2) & 2 &&&\\
\ion{Zr}{ii} && 3.35 &  0.79 && 2.59 & -0.05 &  0.36 && 3.19 &  1.22 && 3.32 &  1.25 && 3.31 &  0.84 &  1.06 && 2.60(2) & 2 &&&\\
Mo    && 1.58 & -0.30 && 1.36 & -0.60 & -0.19 && 1.79 &  0.50 && 2.09 &  0.70 && 1.49 & -0.30 & -0.08 && 1.92(5) & 2 &&&\\
Ba    && 2.82 &  0.73 && 2.75 &  0.58 &  0.99 && 2.88 &  1.38 && 3.06 &  1.46 && 3.00 &  1.00 &  1.22 && 2.13(5) & 2 &&&\\
La    && 1.77 &  0.68 && 1.50 &  0.33 &  0.74 && 1.85 &  1.35 && 1.79 &  1.19 && 1.82 &  0.82 &  1.04 && 1.13(3) & 3 &&&\\
Ce    && 2.40 &  0.74 && 1.98 &  0.24 &  0.65 && 2.47 &  1.40 && 2.28 &  1.11 && 2.39 &  0.82 &  1.04 && 1.70(4) & 4 &&&\\
Pr    && 1.01 &  0.39 && 1.15 &  0.45 &  0.86 && 0.92 &  0.89 && 1.01 &  0.88 && 0.88 &  0.35 &  0.57 && 0.66(15) & 5 &&&\\
Nd    && 2.04 &  0.63 && 1.63 &  0.14 &  0.55 && 2.08 &  1.26 && 1.75 &  0.83 && 2.18 &  0.86 &  1.08 && 1.45(1) & 6 &&&\\
Sm    && 1.27 &  0.30 && 1.05 &  0.00 &  0.41 && 1.26 &  0.88 && 1.18 &  0.70 && 1.41 &  0.53 &  0.75 && 1.01(6) & 7 &&&\\
Eu    && 0.56 &  0.08 && 0.69 &  0.13 &  0.54 && 0.13 &  0.24 && 0.23 &  0.24 && 0.54 &  0.15 &  0.37 && 0.52(1) & 8 &&&\\
Gd    && 1.03 & -0.05 && 1.31 &  0.15 &  0.56 && 1.06 &  0.57 && 1.74 &  1.15 && 1.14 &  0.15 &  0.37 && 1.12(4) & 2 &&&\\
Dy    && 1.31 &  0.15 && 1.13 & -0.11 &  0.30 && 1.16 &  0.59 && 1.21 &  0.54 && 1.36 &  0.29 &  0.51 && 1.20(6) & 9 &&&\\
Pb    && 2.36 &  0.45 && 1.54 & -0.45 & -0.04 && 2.77 &  1.45 && 1.87 &  0.45 && 2.57 &  0.75 &  0.97 && 1.95(8) & 2 &&&\\
\noalign{\smallskip}
\hline
 \end{tabular}
   $$ 
}
References on solar abundances: 1 - \citet{asplund05}; 2 - \citet{gs98}; 3 - \citet{law01a}; 
4 - \citet{PQ00}; 5 - \citet{L76}; 6 - \citet{Hartog03}; 7 - \citet{B89}; 8 - \citet{law01b}; 9 - \citet{BL93}
\end{table*}

\begin{figure*}
\centering
\includegraphics[width=17cm]{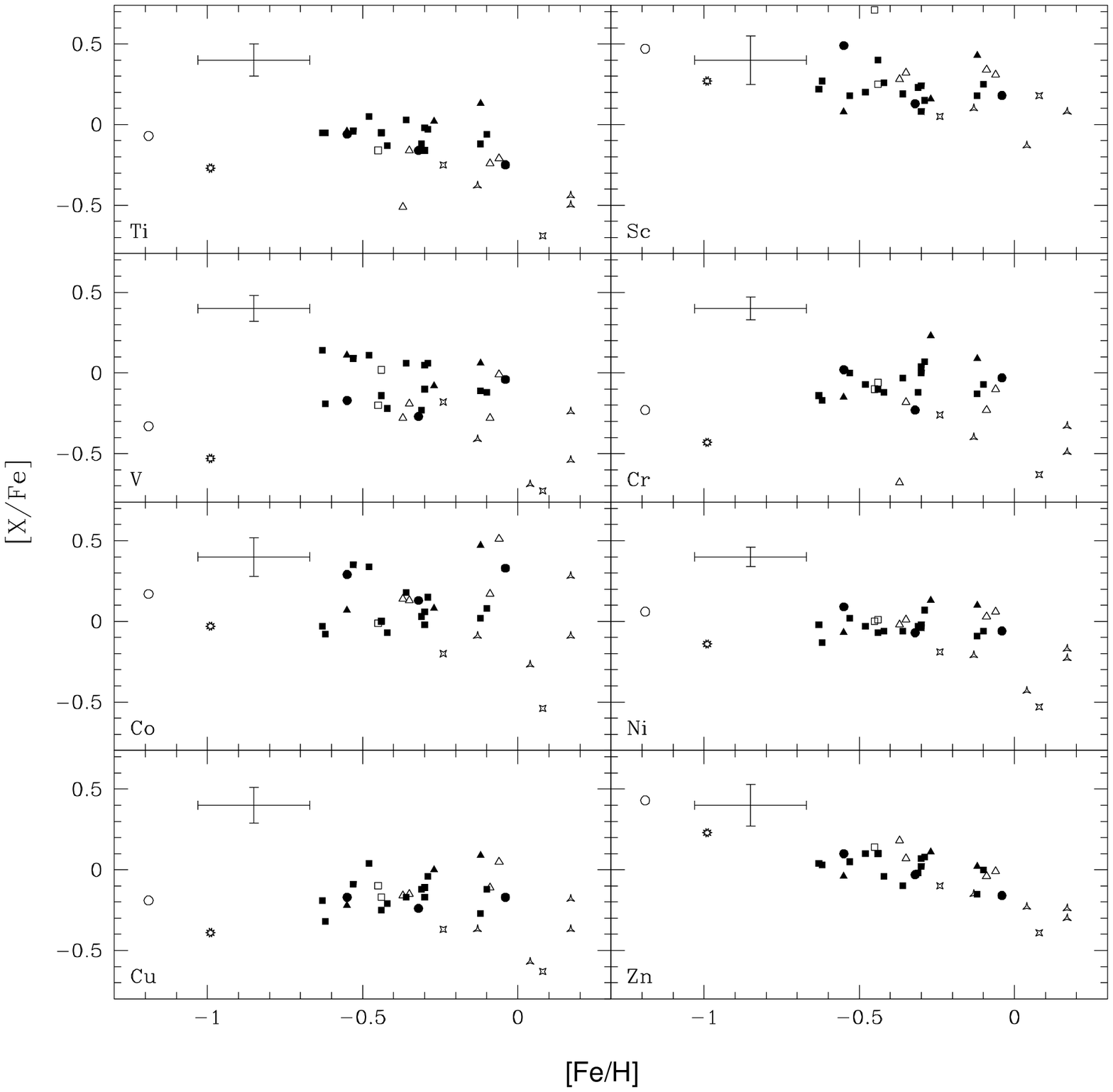}
\caption{[X/Fe] vs. [Fe/H]. Symbols are the same as in Figure \ref{imalfefe1}.}
\label{imalfefe2}
\end{figure*}

\begin{figure*}
\centering
\includegraphics[width=17cm]{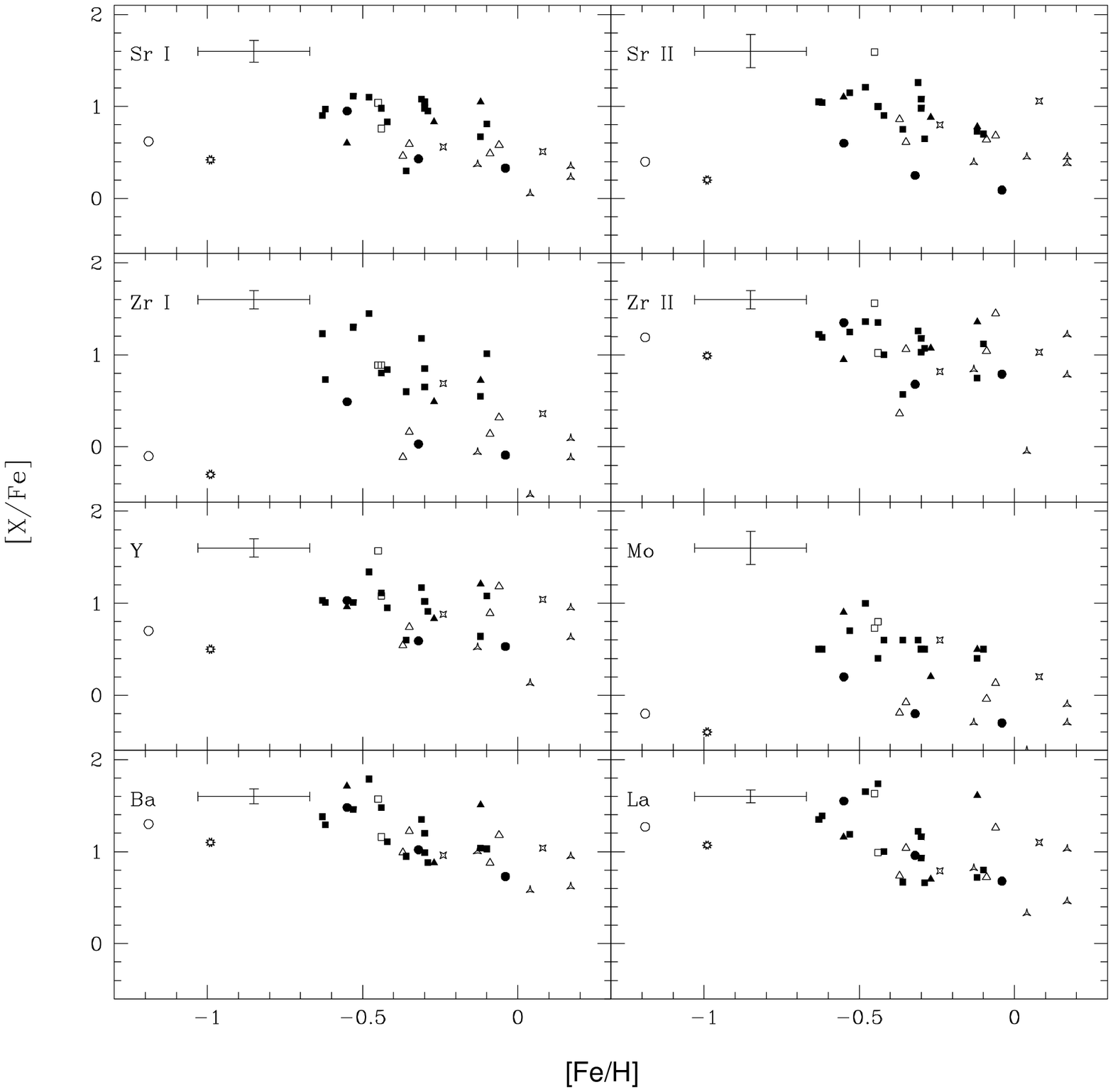}
\caption{[X/Fe] vs. [Fe/H] for heavy elements. 
Symbols are the same as in Figure \ref{imalfefe1}.}
\label{rsfe1}
\end{figure*}

\begin{figure*}
\centering
\includegraphics[width=17cm]{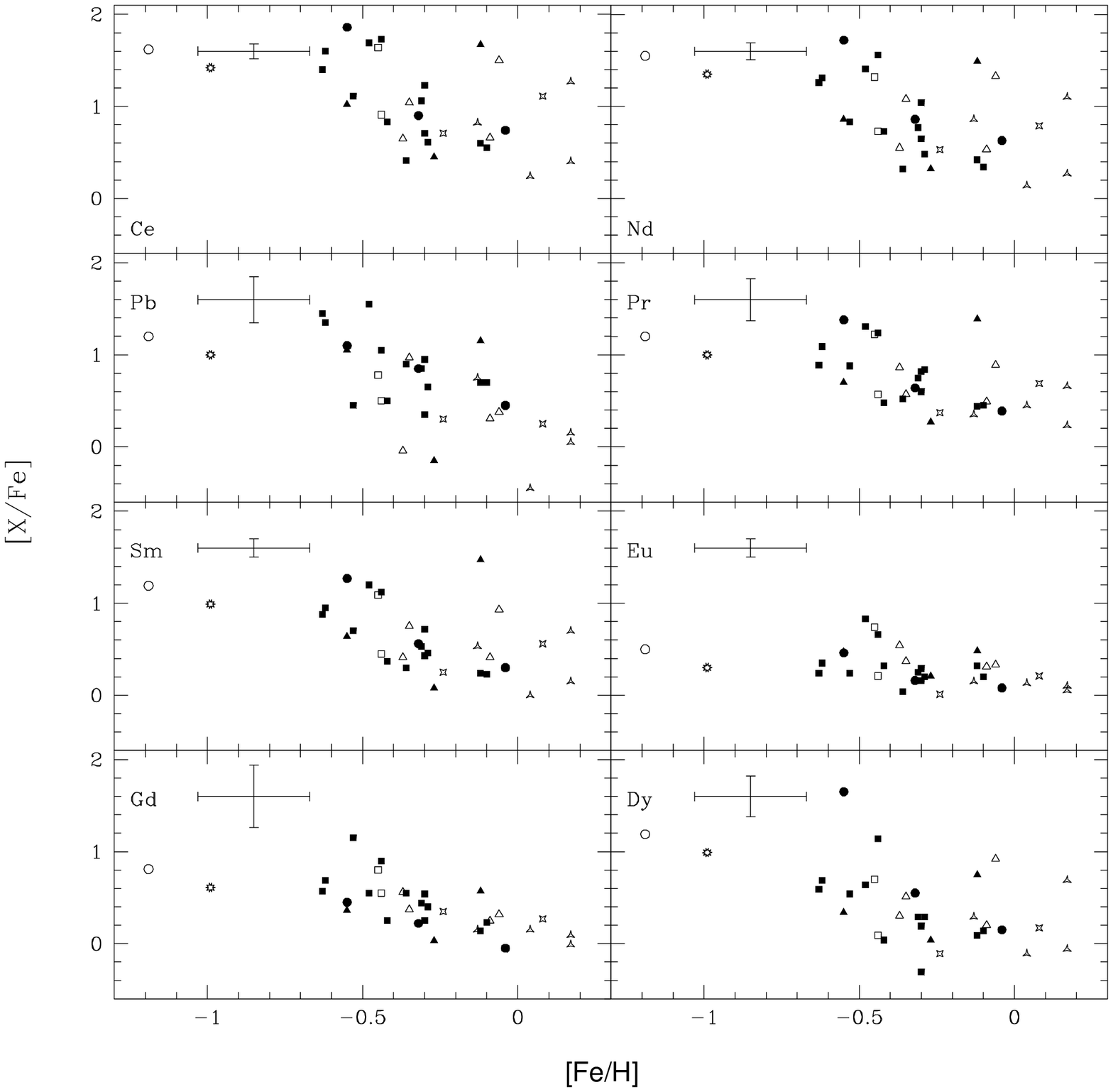}
\caption{[X/Fe] vs. [Fe/H] for heavy elements. 
Symbols are the same as in Figure \ref{imalfefe1}.}
\label{rsfe2}
\end{figure*}

\begin{figure*}[ht!]
\centerline{\includegraphics[totalheight=18.0cm]{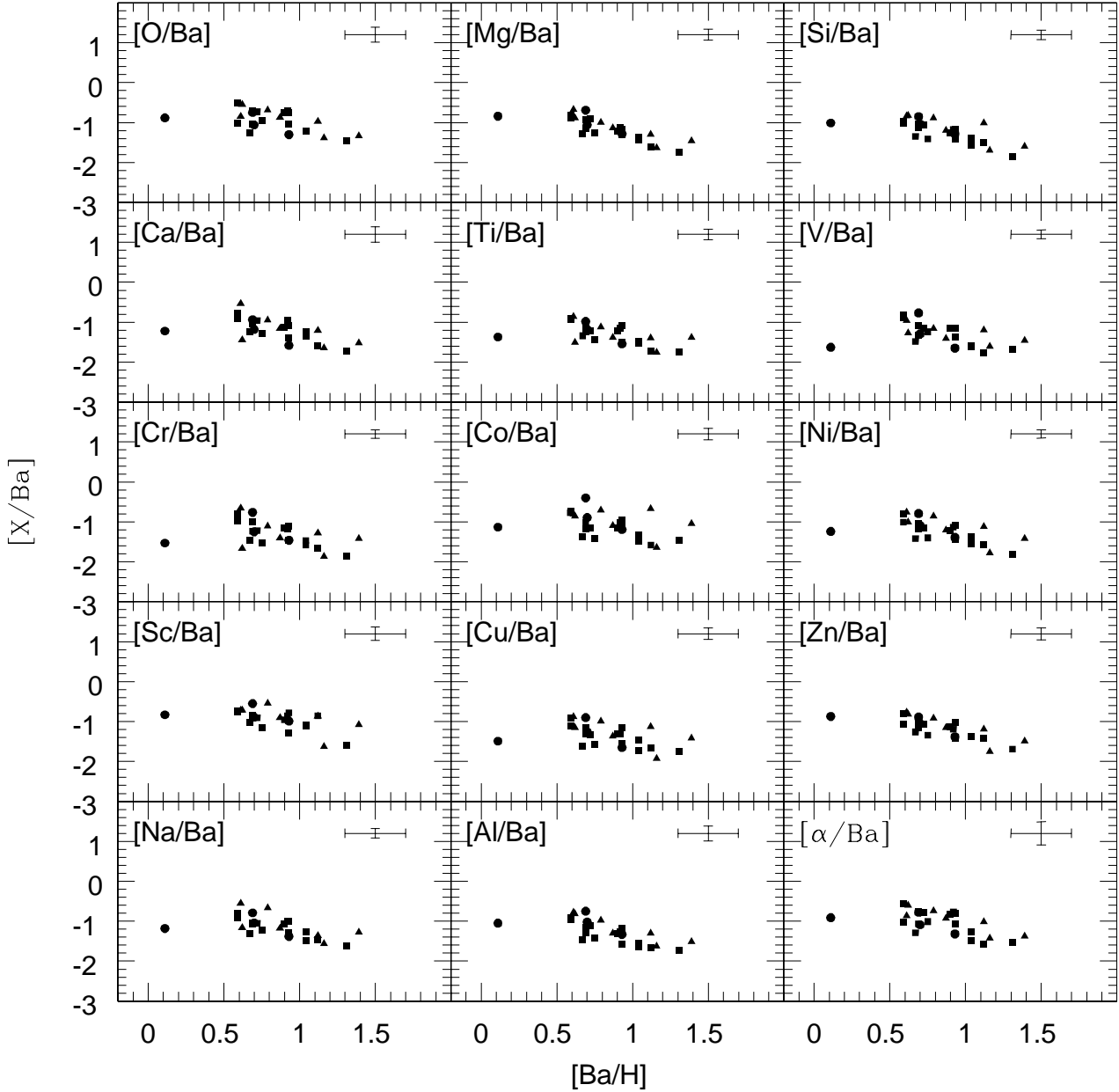}}
\caption{\label{relacsba} [X/Ba] vs. [Ba/H]. 
Symbols: squares: dwarf stars ($\log g \ge 3.7$); triangles: subgiants ($2.4 < \log g < 3.7$);
circles: giants ($\log g \le 2.4$).}
\end{figure*}

\begin{figure*}[ht!]
\centerline{\includegraphics[totalheight=18.0cm]{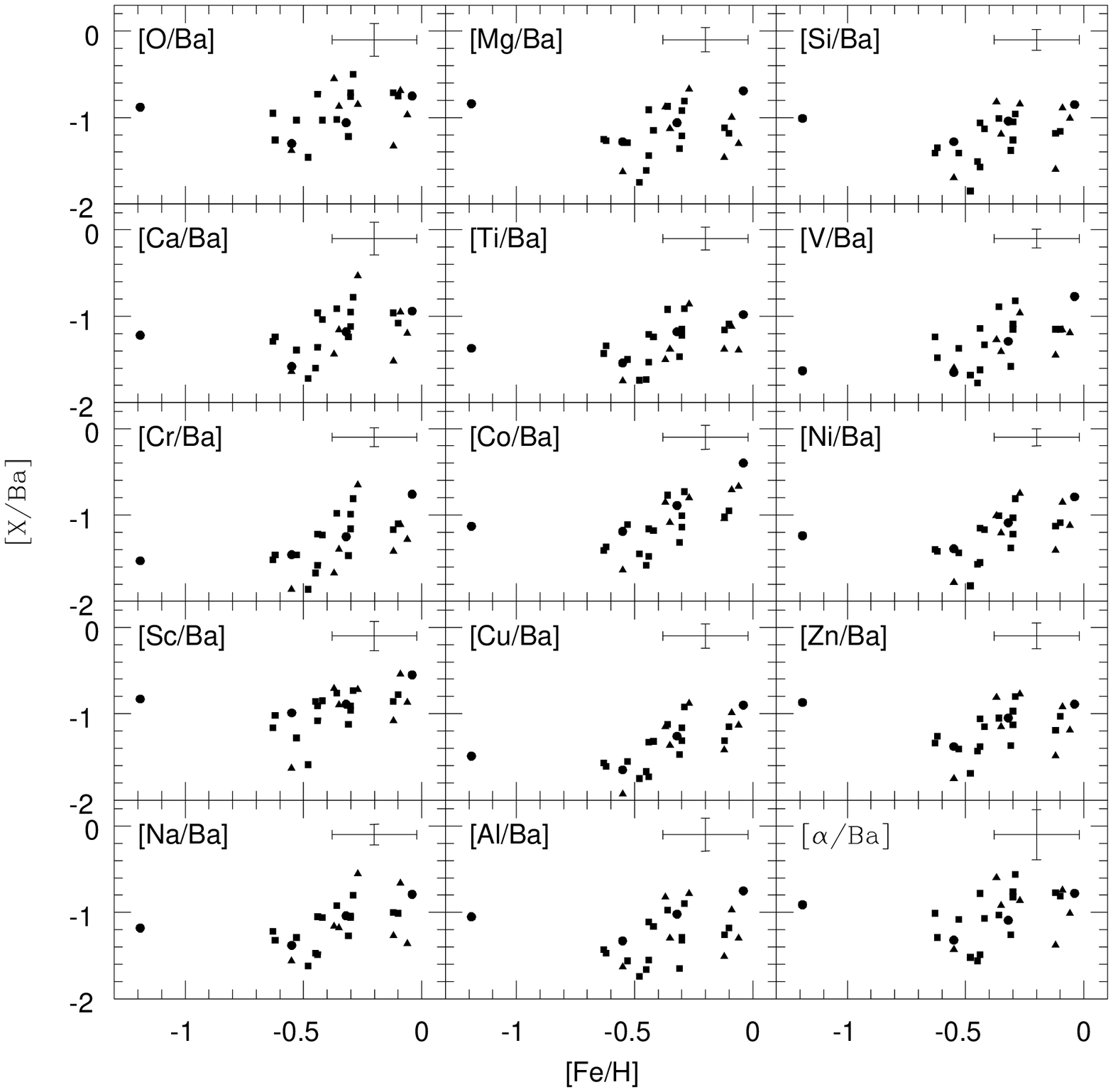}}
\caption{\label{relacsbafe} [X/Ba] vs. [Fe/H]. Symbols are the same as in Figure \ref{relacsba}.}
\end{figure*}

\begin{figure*}[ht!]
\centerline{\includegraphics[totalheight=18.0cm]{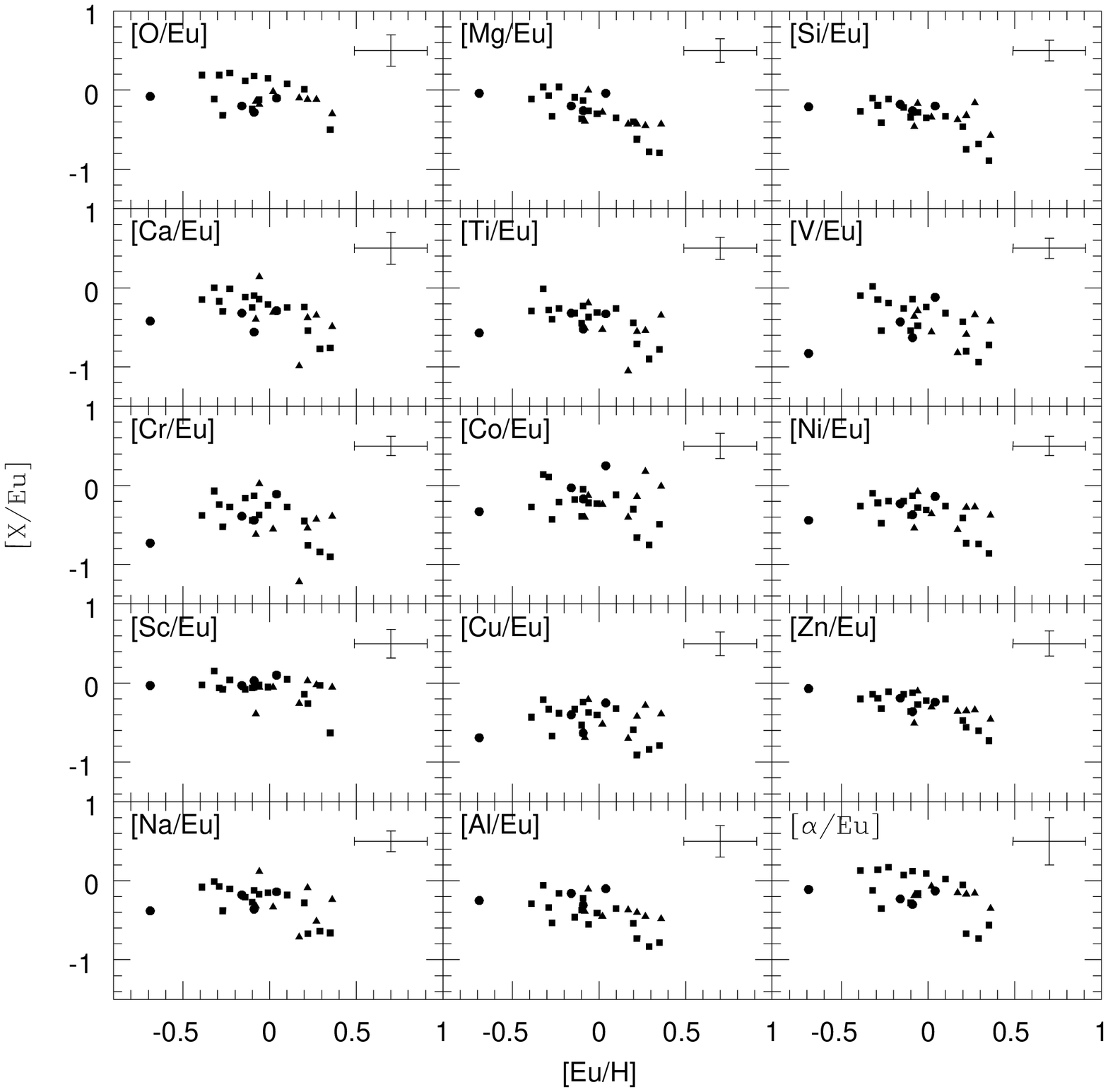}}
\caption{\label{relacseu} [X/Eu] vs. [Eu/H]. Symbols are the same as in Figure \ref{relacsba}.}
\end{figure*}

\begin{figure*}[ht!]
\centerline{\includegraphics[totalheight=18.0cm]{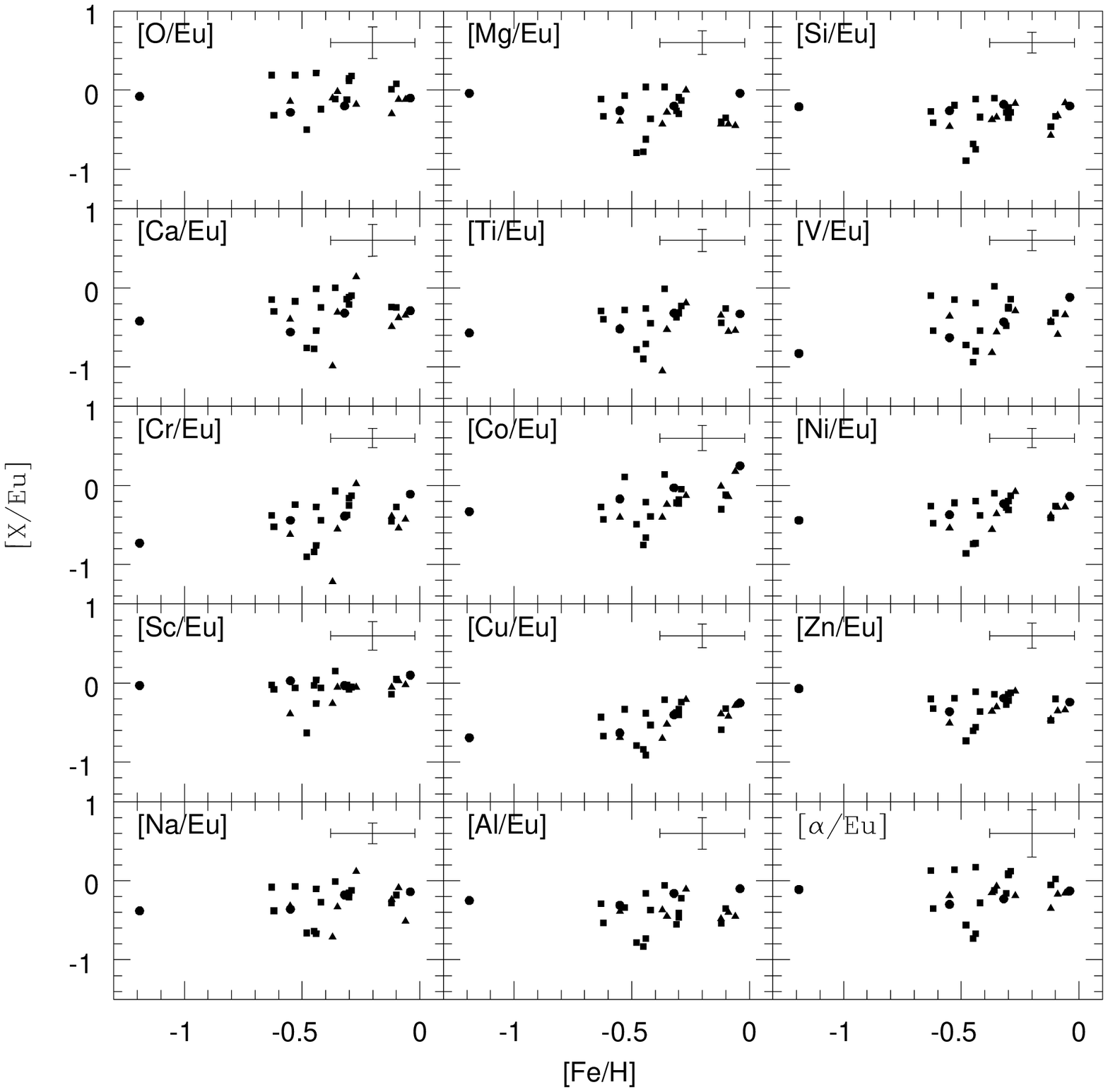}}
\caption{\label{relacseufe}[X/Eu] vs. [Fe/H]. Symbols are the same as in Figure \ref{relacsba}.}
\end{figure*}

\subsection{$\alpha$-elements}

The observations show that the $\alpha$-elements tend to be overabundant at low 
metallicities, with [X/Fe] reaching  $\approx$ 0.5 at -4 $<$ [Fe/H] $<$ -1 
\citep[e.g.][]{barbuy88,cay04}. 
At [Fe/H] $\approx$ -1 the overabundance starts to decrease toward higher 
metallicities, and [X/Fe] can be subsolar at [Fe/H] $\approx$ 0. This behaviour 
of the $\alpha$-elements relative to iron has been attributed to the time delay 
between core-collapse supernovae and SNIa.

No substantial difference between $\alpha$-elements in normal and barium stars
were seen up to now. Figure \ref{relpal} shows [$\alpha$/Fe] vs. [Fe/H] for the 
sample barium stars where $\alpha$'s included are O, Mg, Si, Ca and Ti.

{\it Oxygen}: It is the third most abundant element in the Universe after 
H and He. It is produced during He burning in the interior of massive stars,
and is released through SN II events. 

Abundances of oxygen for the present sample were determined through
spectrum synthesis of the forbidden lines of [O I] at 
$\lambda$6300.3 $\rm \AA$ and $\lambda$6363.8 $\rm \AA$. These lines are 
reliable since they are not subject to NLTE effects. 
The resolution of the program stars spectra 
is such that the \ion{Sc}{ii} line at $\lambda$6300.7 $\rm \AA$ is not 
blended with the [O I] line at $\lambda$6300.3 $\rm \AA$. 
Telluric lines are displaced thanks to their radial velocities and do not blend 
the oxygen line in the sample stars. For 2 stars, HD 147609 and HD 87080, the sky 
emission line is present preventing the determination of the O abundance.
For HD 13551, HD 22589 and HD 107574, an upper limit was derived.

The Ni I line at $\lambda$6300.335 $\rm \AA$ was taken into account in the
oxygen abundance calculations. The adopted abundance for Ni was 
the average of other ten lines and the 
atomic constants are shown in Table \ref{abun}.

The solar abundance adopted was $\log\epsilon$(O) = 8.74, which is
the value by \citet{asplund05} through a 3D atmosphere model, 
$\log\epsilon$(O) = 8.66, corrected by 0.08 for 1D, according to
\citet{allende01}.

The O abundance results obtained shown in
Figure \ref{imalfefe1}, are in good agreement with Figure 4 by \citet{francois04} 
in the same range of metallicities.

{\it Magnesium}: Similarly to oxygen, magnesium is also produced by
massive stars, but in this case carbon and neon
burning is responsible for its production.
Figure 4 of \citet{francois04} shows the evolution of [Mg/Fe] relative 
to [Fe/H], where [Mg/Fe] decreases toward increasing metallicities.

The 3 lines used in the determination of Mg abundance for the present
sample present a good agreement, as shown in Table \ref{abun}. 
Figure \ref{imalfefe1} shows that the range of [Mg/Fe] 
is similar to those of [Na/Fe] and [Al/Fe], differing from [Ti/Fe], that reaches
lower abundances, in agreement with \citet[][ and references therein]{francois04}.

{\it Silicon}: SNae of $\approx$ 20 M$\sb \odot$ could be the main sources
of Si \citep{ww95}.
\citet{proc00} found a high overabundance of Si and an increasing 
abundance trend toward lower metallicities for stars of the thick disk
in the range of metallicities -1.2 $<$ [Fe/H] $<$ -0.3, in agreement 
with \citet{M97}.

For the present sample 5 lines were used. All lines give
similar abundances, except the line $\lambda$5948.5 $\rm \AA$ that gives
values 0.5 dex higher for some stars, possibly due to an unknown 
blend. Most results give
[Si/Fe] $\approx$ 0, except for the star HD 123396, a halo giant, for which the
abundance is higher, as shown in Figure \ref{imalfefe1}. The results are
in agreement with the chemical evolution model by \citet{francois04} for
this range of metallicities. The odd-even effect observed by \citet{arnett71}, 
where Mg and Si are overabundant relative to Na and Al, is confirmed for some
sample stars. For several stars, Na and Al abundances are higher than
Si and Mg, with -0.1 $\leq$ [Mg/Fe] $\leq$ 0.2,
-0.15 $\leq$ [Si/Fe] $\leq$ 0.2, -0.1 $\leq$ [Na/Fe] $\leq$ 0.2 and 
-0.1 $\leq$ [Al/Fe] $\leq$ 0.1.

{\it Calcium}: According to \citet{M97}, the abundance of Ca is expected to 
behave similarly to Si, given that SN II of moderate mass is the main source of both elements 
\citep{ww95}, and they can be also released by SN Ia \citep{nomoto97b}.
The similarity of behaviour between Ca and Si was verified by \citet{proc00} as well as for
the barium stars of the present sample (Figure \ref{imalfefe1}), except for
the star HD 210910. Six Ca lines were used, being some of them very strong in some 
sample stars (Table \ref{abun}). For 16 stars the difference between abundances derived from
those lines is from 0.2 to 0.5 dex, and for the others, there is a better agreement.

{\it Titanium}: Ti is usually included in the $\alpha$-elements list because
its overabundance in metal-poor stars is similar to that of $\alpha$-elements \citep{gs91},
but its nucleosynthesis is unclear \citep{ww95}. For this reason, Ti abundance 
can be different from that of Ca and Si, as observed by
\citet{proc00}. According to \citet{francois04}, Ti and Mg abundances have a similar behaviour,
though the dispersion is larger for Mg in their Figure 4. 

In the present work, Ti abundances
are usually lower than those of Ca, Si and Mg, as shown in Figures \ref{imalfefe1} and 
\ref{imalfefe2}.
Twelve lines of Ti I were used. For 4 stars, the abundance 
from the $\lambda$5210.4 $\rm \AA$ line is different from the other lines. The reason for
that difference is unclear. If the problem were in the atomic constants, the difference would be
observed for all stars. In general, the abundance results from the 12 lines are in
agreement.

\begin{table*}
\caption{Abundance excesses of Na, Al, $\alpha$- and iron peak elements 
relative to Ba (upper table) and Eu (lower table).}
{\tiny
\label{elalbaeu}
 $$
\setlength\tabcolsep{2pt}
\begin{tabular}{lrrrrrrrrrrrrrrrr}
\hline\hline
\noalign{\smallskip}
star & [O/Ba] & [Mg/Ba] & [Si/Ba] & [Ca/Ba] & [Ti/Ba] & [V/Ba] & [Cr/Ba] & [Co/Ba] & [Ni/Ba] & [Sc/Ba] & [Cu/Ba] & [Zn/Ba] & [Na/Ba]	& [Al/Ba] & [$\alpha$/Ba] & [Ba/H] \\
\noalign{\smallskip}
\hline
\noalign{\smallskip}
HD 749     & -0.97$\pm$0.19 & -1.30$\pm$0.14 & -1.01$\pm$0.12 & -1.20$\pm$0.19 & -1.39$\pm$0.13 & -1.19$\pm$0.11 & -1.28$\pm$0.11 & -0.67$\pm$0.14 & -1.12$\pm$0.10 & -0.87$\pm$0.17 & -1.13$\pm$0.14 & -1.19$\pm$0.15 & -1.36$\pm$0.12 & -1.30$\pm$0.19 & -1.01$\pm$0.29 & 1.12$\pm$0.20 \\
HR 107     & -1.02$\pm$0.13 & -0.87$\pm$0.08 & -1.01$\pm$0.08 & -0.91$\pm$0.07 & -0.92$\pm$0.09 & -0.89$\pm$0.08 & -0.98$\pm$0.07 & -0.77$\pm$0.08 & -1.01$\pm$0.07 & -0.76$\pm$0.13 & -1.12$\pm$0.11 & -1.05$\pm$0.11 & -0.92$\pm$0.07 & -0.97$\pm$0.10 & -1.03$\pm$0.13 & 0.59$\pm$0.07 \\
HD 5424    & -1.30$\pm$0.19 & -1.28$\pm$0.14 & -1.28$\pm$0.12 & -1.58$\pm$0.19 & -1.54$\pm$0.13 & -1.65$\pm$0.11 & -1.46$\pm$0.11 & -1.19$\pm$0.14 & -1.39$\pm$0.10 & -0.99$\pm$0.17 & -1.65$\pm$0.14 & -1.38$\pm$0.15 & -1.38$\pm$0.12 & -1.33$\pm$0.19 & -1.32$\pm$0.29 & 0.93$\pm$0.20 \\
HD 8270    & -1.03$\pm$0.13 & -1.15$\pm$0.08 & -1.13$\pm$0.08 & -1.04$\pm$0.07 & -1.24$\pm$0.09 & -1.33$\pm$0.08 & -1.23$\pm$0.07 & -1.18$\pm$0.08 & -1.17$\pm$0.07 & -0.85$\pm$0.13 & -1.32$\pm$0.11 & -1.15$\pm$0.11 & -1.06$\pm$0.07 & -1.16$\pm$0.10 & -1.07$\pm$0.13 & 0.69$\pm$0.07 \\
HD 12392   & -1.33$\pm$0.19 & -1.46$\pm$0.14 & -1.60$\pm$0.12 & -1.52$\pm$0.19 & -1.38$\pm$0.13 & -1.45$\pm$0.11 & -1.42$\pm$0.11 & -1.04$\pm$0.14 & -1.41$\pm$0.10 & -1.08$\pm$0.17 & -1.42$\pm$0.14 & -1.49$\pm$0.15 & -1.27$\pm$0.12 & -1.51$\pm$0.19 & -1.38$\pm$0.29 & 1.39$\pm$0.20 \\
HD 13551   & -0.73$\pm$0.13 & -0.91$\pm$0.08 & -1.06$\pm$0.08 & -0.96$\pm$0.07 & -1.21$\pm$0.09 & -1.14$\pm$0.08 & -1.22$\pm$0.07 & -1.16$\pm$0.08 & -1.15$\pm$0.07 & -0.91$\pm$0.13 & -1.33$\pm$0.11 & -1.06$\pm$0.11 & -1.05$\pm$0.07 & -1.11$\pm$0.10 & -0.78$\pm$0.13 & 0.72$\pm$0.07 \\
HD 22589   & -0.85$\pm$0.13 & -0.67$\pm$0.08 & -0.84$\pm$0.08 & -0.53$\pm$0.07 & -0.86$\pm$0.09 & -0.96$\pm$0.08 & -0.65$\pm$0.07 & -0.80$\pm$0.08 & -0.75$\pm$0.07 & -0.72$\pm$0.13 & -0.88$\pm$0.11 & -0.77$\pm$0.11 & -0.55$\pm$0.07 & -0.78$\pm$0.10 & -0.86$\pm$0.13 & 0.61$\pm$0.07 \\
HD 27271   & -0.69$\pm$0.19 & -1.00$\pm$0.14 & -0.89$\pm$0.12 & -0.95$\pm$0.19 & -1.12$\pm$0.13 & -1.16$\pm$0.11 & -1.11$\pm$0.11 & -0.71$\pm$0.14 & -0.85$\pm$0.10 & -0.54$\pm$0.17 & -0.99$\pm$0.14 & -0.92$\pm$0.15 & -0.66$\pm$0.12 & -0.97$\pm$0.19 & -0.74$\pm$0.29 & 0.79$\pm$0.20 \\
HD 48565   & -1.26$\pm$0.13 & -1.27$\pm$0.08 & -1.35$\pm$0.08 & -1.24$\pm$0.07 & -1.34$\pm$0.09 & -1.48$\pm$0.08 & -1.46$\pm$0.07 & -1.37$\pm$0.08 & -1.42$\pm$0.07 & -1.02$\pm$0.13 & -1.61$\pm$0.11 & -1.26$\pm$0.11 & -1.32$\pm$0.07 & -1.47$\pm$0.10 & -1.29$\pm$0.13 & 0.67$\pm$0.07 \\
HD 76225   & -1.22$\pm$0.13 & -1.36$\pm$0.08 & -1.38$\pm$0.08 & -1.24$\pm$0.07 & -1.47$\pm$0.09 & -1.58$\pm$0.08 & -1.47$\pm$0.07 & -1.32$\pm$0.08 & -1.38$\pm$0.07 & -1.12$\pm$0.13 & -1.47$\pm$0.11 & -1.37$\pm$0.11 & -1.27$\pm$0.07 & -1.65$\pm$0.10 & -1.26$\pm$0.13 & 1.04$\pm$0.07 \\
HD 87080   &    ...         & -1.44$\pm$0.08 & -1.57$\pm$0.08 & -1.36$\pm$0.07 & -1.53$\pm$0.09 & -1.62$\pm$0.08 & -1.58$\pm$0.07 & -1.48$\pm$0.08 & -1.55$\pm$0.07 & -1.08$\pm$0.13 & -1.73$\pm$0.11 & -1.38$\pm$0.11 & -1.49$\pm$0.07 & -1.55$\pm$0.10 & -1.49$\pm$0.13 & 1.04$\pm$0.07 \\
HD 89948   & -0.71$\pm$0.13 & -0.92$\pm$0.08 & -1.05$\pm$0.08 & -0.95$\pm$0.07 & -1.15$\pm$0.09 & -1.09$\pm$0.08 & -0.99$\pm$0.07 & -1.01$\pm$0.08 & -1.03$\pm$0.07 & -0.91$\pm$0.13 & -1.16$\pm$0.11 & -0.97$\pm$0.11 & -1.04$\pm$0.07 & -1.29$\pm$0.10 & -0.76$\pm$0.13 & 0.69$\pm$0.07 \\
HD 92545   & -0.71$\pm$0.13 & -1.12$\pm$0.08 & -1.18$\pm$0.08 & -0.96$\pm$0.07 & -1.16$\pm$0.09 & -1.15$\pm$0.08 & -1.17$\pm$0.07 & -1.02$\pm$0.08 & -1.13$\pm$0.07 & -0.86$\pm$0.13 & -1.31$\pm$0.11 & -1.19$\pm$0.11 & -1.00$\pm$0.07 & -1.26$\pm$0.10 & -0.77$\pm$0.13 & 0.92$\pm$0.07 \\
HD 106191  & -0.50$\pm$0.13 & -0.81$\pm$0.08 & -0.96$\pm$0.08 & -0.78$\pm$0.07 & -0.91$\pm$0.09 & -0.82$\pm$0.08 & -0.81$\pm$0.07 & -0.73$\pm$0.08 & -0.81$\pm$0.07 & -0.73$\pm$0.13 & -0.92$\pm$0.11 & -0.80$\pm$0.11 & -0.80$\pm$0.07 & -0.90$\pm$0.10 & -0.56$\pm$0.13 & 0.59$\pm$0.07 \\
HD 107574  & -1.38$\pm$0.13 & -1.63$\pm$0.08 & -1.70$\pm$0.08 & -1.64$\pm$0.07 & -1.75$\pm$0.09 & -1.60$\pm$0.08 & -1.86$\pm$0.07 & -1.64$\pm$0.08 & -1.78$\pm$0.07 & -1.63$\pm$0.13 & -1.93$\pm$0.11 & -1.75$\pm$0.11 & -1.56$\pm$0.07 & -1.63$\pm$0.10 & -1.43$\pm$0.13 & 1.16$\pm$0.07 \\
HD 116869  & -1.06$\pm$0.19 & -1.06$\pm$0.14 & -1.04$\pm$0.12 & -1.18$\pm$0.19 & -1.18$\pm$0.13 & -1.29$\pm$0.11 & -1.25$\pm$0.11 & -0.89$\pm$0.14 & -1.09$\pm$0.10 & -0.89$\pm$0.17 & -1.26$\pm$0.14 & -1.05$\pm$0.15 & -1.04$\pm$0.12 & -1.02$\pm$0.19 & -1.09$\pm$0.29 & 0.70$\pm$0.20 \\
HD 123396  & -0.88$\pm$0.19 & -0.84$\pm$0.14 & -1.01$\pm$0.12 & -1.22$\pm$0.19 & -1.37$\pm$0.13 & -1.63$\pm$0.11 & -1.53$\pm$0.11 & -1.13$\pm$0.14 & -1.24$\pm$0.10 & -0.83$\pm$0.17 & -1.49$\pm$0.14 & -0.87$\pm$0.15 & -1.18$\pm$0.12 & -1.05$\pm$0.19 & -0.91$\pm$0.29 & 0.11$\pm$0.20 \\
HD 123585  & -1.46$\pm$0.13 & -1.75$\pm$0.08 & -1.85$\pm$0.08 & -1.72$\pm$0.07 & -1.74$\pm$0.09 & -1.68$\pm$0.08 & -1.86$\pm$0.07 & -1.45$\pm$0.08 & -1.82$\pm$0.07 & -1.59$\pm$0.13 & -1.75$\pm$0.11 & -1.69$\pm$0.11 & -1.62$\pm$0.07 & -1.74$\pm$0.10 & -1.52$\pm$0.13 & 1.31$\pm$0.07 \\
HD 147609  &    ...         & -1.61$\pm$0.08 & -1.51$\pm$0.08 & -1.60$\pm$0.07 & -1.73$\pm$0.09 & -1.77$\pm$0.08 & -1.67$\pm$0.07 & -1.58$\pm$0.08 & -1.57$\pm$0.07 & -0.86$\pm$0.13 & -1.67$\pm$0.11 & -1.43$\pm$0.11 & -1.47$\pm$0.07 & -1.66$\pm$0.10 & -1.56$\pm$0.13 & 1.12$\pm$0.07 \\
HD 150862  & -0.75$\pm$0.13 & -1.18$\pm$0.08 & -1.16$\pm$0.08 & -1.08$\pm$0.07 & -1.09$\pm$0.09 & -1.15$\pm$0.08 & -1.10$\pm$0.07 & -0.95$\pm$0.08 & -1.09$\pm$0.07 & -0.78$\pm$0.13 & -1.15$\pm$0.11 & -1.03$\pm$0.11 & -1.01$\pm$0.07 & -1.18$\pm$0.10 & -0.81$\pm$0.13 & 0.93$\pm$0.07 \\
HD 188985  & -0.76$\pm$0.13 & -1.21$\pm$0.08 & -1.26$\pm$0.08 & -1.12$\pm$0.07 & -1.22$\pm$0.09 & -1.15$\pm$0.08 & -1.16$\pm$0.07 & -1.14$\pm$0.08 & -1.22$\pm$0.07 & -0.96$\pm$0.13 & -1.31$\pm$0.11 & -1.13$\pm$0.11 & -1.06$\pm$0.07 & -1.32$\pm$0.10 & -0.82$\pm$0.13 & 0.90$\pm$0.07 \\
HD 210709  & -0.75$\pm$0.19 & -0.69$\pm$0.14 & -0.85$\pm$0.12 & -0.94$\pm$0.19 & -0.98$\pm$0.13 & -0.77$\pm$0.11 & -0.76$\pm$0.11 & -0.40$\pm$0.14 & -0.79$\pm$0.10 & -0.55$\pm$0.17 & -0.90$\pm$0.14 & -0.89$\pm$0.15 & -0.79$\pm$0.12 & -0.75$\pm$0.19 & -0.78$\pm$0.29 & 0.69$\pm$0.20 \\
HD 210910  & -0.55$\pm$0.19 & -0.88$\pm$0.14 & -0.82$\pm$0.12 & -1.44$\pm$0.19 & -1.50$\pm$0.13 & -1.27$\pm$0.11 & -1.67$\pm$0.11 & -0.85$\pm$0.14 & -1.01$\pm$0.10 & -0.71$\pm$0.17 & -1.15$\pm$0.14 & -0.81$\pm$0.15 & -1.16$\pm$0.12 & -0.82$\pm$0.19 & -0.60$\pm$0.29 & 0.62$\pm$0.20 \\
HD 222349  & -0.95$\pm$0.13 & -1.25$\pm$0.08 & -1.41$\pm$0.08 & -1.29$\pm$0.07 & -1.43$\pm$0.09 & -1.24$\pm$0.08 & -1.52$\pm$0.07 & -1.41$\pm$0.08 & -1.40$\pm$0.07 & -1.16$\pm$0.13 & -1.57$\pm$0.11 & -1.34$\pm$0.11 & -1.22$\pm$0.07 & -1.43$\pm$0.10 & -1.01$\pm$0.13 & 0.75$\pm$0.07 \\
BD+18 5215 & -1.03$\pm$0.13 & -1.29$\pm$0.08 & -1.41$\pm$0.08 & -1.39$\pm$0.07 & -1.50$\pm$0.09 & -1.37$\pm$0.08 & -1.46$\pm$0.07 & -1.11$\pm$0.08 & -1.44$\pm$0.07 & -1.28$\pm$0.13 & -1.55$\pm$0.11 & -1.41$\pm$0.11 & -1.29$\pm$0.07 & -1.56$\pm$0.10 & -1.08$\pm$0.13 & 0.93$\pm$0.07 \\
HD 223938  & -0.87$\pm$0.19 & -1.13$\pm$0.14 & -1.19$\pm$0.12 & -1.16$\pm$0.19 & -1.38$\pm$0.13 & -1.41$\pm$0.11 & -1.40$\pm$0.11 & -1.09$\pm$0.14 & -1.21$\pm$0.10 & -0.90$\pm$0.17 & -1.37$\pm$0.14 & -1.15$\pm$0.15 & -1.18$\pm$0.12 & -1.30$\pm$0.19 & -0.92$\pm$0.29 & 0.87$\pm$0.20 \\
\hline\hline
\noalign{\smallskip}
star & [O/Eu] & [Mg/Eu] & [Si/Eu] & [Ca/Eu] & [Ti/Eu] & [V/Eu] & [Cr/Eu] & [Co/Eu] & [Ni/Eu] & [Sc/Eu] & [Cu/Eu] & [Zn/Eu] & [Na/Eu]	& [Al/Eu] & [$\alpha$/Eu] & [Eu/H] \\
\noalign{\smallskip}
\hline
\noalign{\smallskip}
HD 749     & -0.12$\pm$0.20 & -0.45$\pm$0.15 & -0.16$\pm$0.13 & -0.35$\pm$0.20 & -0.54$\pm$0.14 & -0.34$\pm$0.13 & -0.43$\pm$0.12 &  0.18$\pm$0.16 & -0.27$\pm$0.12 & -0.02$\pm$0.18 & -0.28$\pm$0.15 & -0.34$\pm$0.16 & -0.51$\pm$0.13 & -0.45$\pm$0.20 & -0.16$\pm$0.30 &  0.27$\pm$0.21 \\
HR 107     & -0.11$\pm$0.14 &  0.04$\pm$0.11 & -0.10$\pm$0.10 &  0.00$\pm$0.09 & -0.01$\pm$0.11 &  0.02$\pm$0.10 & -0.07$\pm$0.10 &  0.14$\pm$0.10 & -0.10$\pm$0.10 &  0.15$\pm$0.14 & -0.21$\pm$0.13 & -0.14$\pm$0.13 & -0.01$\pm$0.10 & -0.06$\pm$0.12 & -0.12$\pm$0.14 & -0.32$\pm$0.10 \\
HD 5424    & -0.28$\pm$0.20 & -0.26$\pm$0.15 & -0.26$\pm$0.13 & -0.56$\pm$0.20 & -0.52$\pm$0.14 & -0.63$\pm$0.13 & -0.44$\pm$0.12 & -0.17$\pm$0.16 & -0.37$\pm$0.12 &  0.03$\pm$0.18 & -0.63$\pm$0.15 & -0.36$\pm$0.16 & -0.36$\pm$0.13 & -0.31$\pm$0.20 & -0.30$\pm$0.30 & -0.09$\pm$0.21 \\
HD 8270    & -0.24$\pm$0.14 & -0.36$\pm$0.11 & -0.34$\pm$0.10 & -0.25$\pm$0.09 & -0.45$\pm$0.11 & -0.54$\pm$0.10 & -0.44$\pm$0.10 & -0.39$\pm$0.10 & -0.38$\pm$0.10 & -0.06$\pm$0.14 & -0.53$\pm$0.13 & -0.36$\pm$0.13 & -0.27$\pm$0.10 & -0.37$\pm$0.12 & -0.28$\pm$0.14 & -0.10$\pm$0.10 \\
HD 12392   & -0.30$\pm$0.20 & -0.43$\pm$0.15 & -0.57$\pm$0.13 & -0.49$\pm$0.20 & -0.35$\pm$0.14 & -0.42$\pm$0.13 & -0.39$\pm$0.12 & -0.01$\pm$0.16 & -0.38$\pm$0.12 & -0.05$\pm$0.18 & -0.39$\pm$0.15 & -0.46$\pm$0.16 & -0.24$\pm$0.13 & -0.48$\pm$0.20 & -0.35$\pm$0.30 &  0.36$\pm$0.21 \\
HD 13551   &  0.22$\pm$0.14 &  0.04$\pm$0.11 & -0.11$\pm$0.10 & -0.01$\pm$0.09 & -0.26$\pm$0.11 & -0.19$\pm$0.10 & -0.27$\pm$0.10 & -0.21$\pm$0.10 & -0.20$\pm$0.10 &  0.04$\pm$0.14 & -0.38$\pm$0.13 & -0.11$\pm$0.13 & -0.10$\pm$0.10 & -0.16$\pm$0.12 &  0.17$\pm$0.14 & -0.23$\pm$0.10 \\
HD 22589   & -0.18$\pm$0.14 &  0.00$\pm$0.11 & -0.17$\pm$0.10 &  0.14$\pm$0.09 & -0.19$\pm$0.11 & -0.29$\pm$0.10 &  0.02$\pm$0.10 & -0.13$\pm$0.10 & -0.08$\pm$0.10 & -0.05$\pm$0.14 & -0.21$\pm$0.13 & -0.10$\pm$0.13 &  0.12$\pm$0.10 & -0.11$\pm$0.12 & -0.19$\pm$0.14 & -0.06$\pm$0.10 \\
HD 27271   & -0.12$\pm$0.20 & -0.43$\pm$0.15 & -0.32$\pm$0.13 & -0.38$\pm$0.20 & -0.55$\pm$0.14 & -0.59$\pm$0.13 & -0.54$\pm$0.12 & -0.14$\pm$0.16 & -0.28$\pm$0.12 &  0.03$\pm$0.18 & -0.42$\pm$0.15 & -0.35$\pm$0.16 & -0.09$\pm$0.13 & -0.40$\pm$0.20 & -0.17$\pm$0.30 &  0.22$\pm$0.21 \\
HD 48565   & -0.32$\pm$0.14 & -0.33$\pm$0.11 & -0.41$\pm$0.10 & -0.30$\pm$0.09 & -0.40$\pm$0.11 & -0.54$\pm$0.10 & -0.52$\pm$0.10 & -0.43$\pm$0.10 & -0.48$\pm$0.10 & -0.08$\pm$0.14 & -0.67$\pm$0.13 & -0.32$\pm$0.13 & -0.38$\pm$0.10 & -0.53$\pm$0.12 & -0.35$\pm$0.14 & -0.27$\pm$0.10 \\
HD 76225   & -0.12$\pm$0.14 & -0.26$\pm$0.11 & -0.28$\pm$0.10 & -0.14$\pm$0.09 & -0.37$\pm$0.11 & -0.48$\pm$0.10 & -0.37$\pm$0.10 & -0.22$\pm$0.10 & -0.28$\pm$0.10 & -0.02$\pm$0.14 & -0.37$\pm$0.13 & -0.27$\pm$0.13 & -0.17$\pm$0.10 & -0.55$\pm$0.12 & -0.16$\pm$0.14 & -0.06$\pm$0.10 \\
HD 87080   &     ...        & -0.62$\pm$0.11 & -0.75$\pm$0.10 & -0.54$\pm$0.09 & -0.71$\pm$0.11 & -0.80$\pm$0.10 & -0.76$\pm$0.10 & -0.66$\pm$0.10 & -0.73$\pm$0.10 & -0.26$\pm$0.14 & -0.91$\pm$0.13 & -0.56$\pm$0.13 & -0.67$\pm$0.10 & -0.73$\pm$0.12 & -0.67$\pm$0.14 &  0.22$\pm$0.10 \\
HD 89948   &  0.12$\pm$0.14 & -0.09$\pm$0.11 & -0.22$\pm$0.10 & -0.12$\pm$0.09 & -0.32$\pm$0.11 & -0.26$\pm$0.10 & -0.16$\pm$0.10 & -0.18$\pm$0.10 & -0.20$\pm$0.10 & -0.08$\pm$0.14 & -0.33$\pm$0.13 & -0.14$\pm$0.13 & -0.21$\pm$0.10 & -0.46$\pm$0.12 &  0.07$\pm$0.14 & -0.14$\pm$0.10 \\
HD 92545   &  0.01$\pm$0.14 & -0.40$\pm$0.11 & -0.46$\pm$0.10 & -0.24$\pm$0.09 & -0.44$\pm$0.11 & -0.43$\pm$0.10 & -0.45$\pm$0.10 & -0.30$\pm$0.10 & -0.41$\pm$0.10 & -0.14$\pm$0.14 & -0.59$\pm$0.13 & -0.47$\pm$0.13 & -0.28$\pm$0.10 & -0.54$\pm$0.12 & -0.05$\pm$0.14 &  0.20$\pm$0.10 \\
HD 106191  &  0.18$\pm$0.14 & -0.13$\pm$0.11 & -0.28$\pm$0.10 & -0.10$\pm$0.09 & -0.23$\pm$0.11 & -0.14$\pm$0.10 & -0.13$\pm$0.10 & -0.05$\pm$0.10 & -0.13$\pm$0.10 & -0.05$\pm$0.14 & -0.24$\pm$0.13 & -0.12$\pm$0.13 & -0.12$\pm$0.10 & -0.22$\pm$0.12 &  0.12$\pm$0.14 & -0.09$\pm$0.10 \\
HD 107574  & -0.14$\pm$0.14 & -0.39$\pm$0.11 & -0.46$\pm$0.10 & -0.40$\pm$0.09 & -0.51$\pm$0.11 & -0.36$\pm$0.10 & -0.62$\pm$0.10 & -0.40$\pm$0.10 & -0.54$\pm$0.10 & -0.39$\pm$0.14 & -0.69$\pm$0.13 & -0.51$\pm$0.13 & -0.32$\pm$0.10 & -0.39$\pm$0.12 & -0.19$\pm$0.14 & -0.08$\pm$0.10 \\
HD 116869  & -0.20$\pm$0.20 & -0.20$\pm$0.15 & -0.18$\pm$0.13 & -0.32$\pm$0.20 & -0.32$\pm$0.14 & -0.43$\pm$0.13 & -0.39$\pm$0.12 & -0.03$\pm$0.16 & -0.23$\pm$0.12 & -0.03$\pm$0.18 & -0.40$\pm$0.15 & -0.19$\pm$0.16 & -0.18$\pm$0.13 & -0.16$\pm$0.20 & -0.23$\pm$0.30 & -0.16$\pm$0.21 \\
HD 123396  & -0.08$\pm$0.20 & -0.04$\pm$0.15 & -0.21$\pm$0.13 & -0.42$\pm$0.20 & -0.57$\pm$0.14 & -0.83$\pm$0.13 & -0.73$\pm$0.12 & -0.33$\pm$0.16 & -0.44$\pm$0.12 & -0.03$\pm$0.18 & -0.69$\pm$0.15 & -0.07$\pm$0.16 & -0.38$\pm$0.13 & -0.25$\pm$0.20 & -0.11$\pm$0.30 & -0.69$\pm$0.21 \\
HD 123585  & -0.50$\pm$0.14 & -0.79$\pm$0.11 & -0.89$\pm$0.10 & -0.76$\pm$0.09 & -0.78$\pm$0.11 & -0.72$\pm$0.10 & -0.90$\pm$0.10 & -0.49$\pm$0.10 & -0.86$\pm$0.10 & -0.63$\pm$0.14 & -0.79$\pm$0.13 & -0.73$\pm$0.13 & -0.66$\pm$0.10 & -0.78$\pm$0.12 & -0.56$\pm$0.14 &  0.35$\pm$0.10 \\
HD 147609  &    ...  	    & -0.78$\pm$0.11 & -0.68$\pm$0.10 & -0.77$\pm$0.09 & -0.90$\pm$0.11 & -0.94$\pm$0.10 & -0.84$\pm$0.10 & -0.75$\pm$0.10 & -0.74$\pm$0.10 & -0.03$\pm$0.14 & -0.84$\pm$0.13 & -0.60$\pm$0.13 & -0.64$\pm$0.10 & -0.83$\pm$0.12 & -0.73$\pm$0.14 &  0.29$\pm$0.10 \\
HD 150862  &  0.08$\pm$0.14 & -0.35$\pm$0.11 & -0.33$\pm$0.10 & -0.25$\pm$0.09 & -0.26$\pm$0.11 & -0.32$\pm$0.10 & -0.27$\pm$0.10 & -0.12$\pm$0.10 & -0.26$\pm$0.10 &  0.05$\pm$0.14 & -0.32$\pm$0.13 & -0.20$\pm$0.13 & -0.18$\pm$0.10 & -0.35$\pm$0.12 &  0.02$\pm$0.14 &  0.10$\pm$0.10 \\
HD 188985  &  0.15$\pm$0.14 & -0.30$\pm$0.11 & -0.35$\pm$0.10 & -0.21$\pm$0.09 & -0.31$\pm$0.11 & -0.24$\pm$0.10 & -0.25$\pm$0.10 & -0.23$\pm$0.10 & -0.31$\pm$0.10 & -0.05$\pm$0.14 & -0.40$\pm$0.13 & -0.22$\pm$0.13 & -0.15$\pm$0.10 & -0.41$\pm$0.12 &  0.09$\pm$0.14 & -0.01$\pm$0.10 \\
HD 210709  & -0.10$\pm$0.20 & -0.04$\pm$0.15 & -0.20$\pm$0.13 & -0.29$\pm$0.20 & -0.33$\pm$0.14 & -0.12$\pm$0.13 & -0.11$\pm$0.12 &  0.25$\pm$0.16 & -0.14$\pm$0.12 &  0.10$\pm$0.18 & -0.25$\pm$0.15 & -0.24$\pm$0.16 & -0.14$\pm$0.13 & -0.10$\pm$0.20 & -0.13$\pm$0.30 &  0.04$\pm$0.21 \\
HD 210910  & -0.10$\pm$0.20 & -0.43$\pm$0.15 & -0.37$\pm$0.13 & -0.99$\pm$0.20 & -1.05$\pm$0.14 & -0.82$\pm$0.13 & -1.22$\pm$0.12 & -0.40$\pm$0.16 & -0.56$\pm$0.12 & -0.26$\pm$0.18 & -0.70$\pm$0.15 & -0.36$\pm$0.16 & -0.71$\pm$0.13 & -0.37$\pm$0.20 & -0.15$\pm$0.30 &  0.17$\pm$0.21 \\
HD 222349  &  0.19$\pm$0.14 & -0.11$\pm$0.11 & -0.27$\pm$0.10 & -0.15$\pm$0.09 & -0.29$\pm$0.11 & -0.10$\pm$0.10 & -0.38$\pm$0.10 & -0.27$\pm$0.10 & -0.26$\pm$0.10 & -0.02$\pm$0.14 & -0.43$\pm$0.13 & -0.20$\pm$0.13 & -0.08$\pm$0.10 & -0.29$\pm$0.12 &  0.13$\pm$0.14 & -0.39$\pm$0.10 \\
BD+18 5215 &  0.19$\pm$0.14 & -0.07$\pm$0.11 & -0.19$\pm$0.10 & -0.17$\pm$0.09 & -0.28$\pm$0.11 & -0.15$\pm$0.10 & -0.24$\pm$0.10 &  0.11$\pm$0.10 & -0.22$\pm$0.10 & -0.06$\pm$0.14 & -0.33$\pm$0.13 & -0.19$\pm$0.13 & -0.07$\pm$0.10 & -0.34$\pm$0.12 &  0.14$\pm$0.14 & -0.29$\pm$0.10 \\
HD 223938  & -0.02$\pm$0.20 & -0.28$\pm$0.15 & -0.34$\pm$0.13 & -0.31$\pm$0.20 & -0.53$\pm$0.14 & -0.56$\pm$0.13 & -0.55$\pm$0.12 & -0.24$\pm$0.16 & -0.36$\pm$0.12 & -0.05$\pm$0.18 & -0.52$\pm$0.15 & -0.30$\pm$0.16 & -0.33$\pm$0.13 & -0.45$\pm$0.20 & -0.07$\pm$0.30 &  0.02$\pm$0.21 \\
\noalign{\smallskip}
\hline
\end{tabular}														      
$$
}
\end{table*}

\begin{table*}[h!]
\caption{Abundance uncertainties.}
{\scriptsize
\label{errab}
   $$ 
\setlength\tabcolsep{1pt}
\begin{tabular}{cccccccccccccccccccc}
\hline\hline
\noalign{\smallskip}
&&& \multicolumn{8}{c}{HD 150862} && \multicolumn{8}{c}{HD 5424} \\
\cline{4-11} \cline{13-20}
X & $\sigma_{log\epsilon(X)_\odot}$ &&
n & $\log{A_p}$ & $\log{A_T}$& $\log{A_m}$ & $\log{A_{lg}}$ & $\log{A_v}$ & $\sigma_{\log\epsilon(X)}$ & $\sigma_{[X/Fe]}$ &&
n & $\log{A_p}$ & $\log{A_T}$& $\log{A_m}$ & $\log{A_{lg}}$ & $\log{A_v}$ & $\sigma_{\log\epsilon(X)}$ & $\sigma_{[X/Fe]}$ \\
\hline
\noalign{\smallskip}
Li & 0.10 &&  1 & 1.10 & 1.20 & 1.06 & 1.10 & 1.10 & 0.13 &  ... &&  2 & 1.05 & 0.90 & 0.86 & 1.00 & 1.05 & 0.23 &  ... \\
 C & 0.06 &&  1 & 8.96 & 9.01 & 8.92 & 8.96 & 8.96 & 0.08 & 0.09 &&  4 & 8.90 & 8.92 & 8.72 & 8.93 & 8.90 & 0.20 & 0.10 \\
 N & 0.06 &&  1 & 8.98 & 9.08 & 8.94 & 8.98 & 8.98 & 0.13 & 0.13 && 12 & 8.26 & 8.21 & 8.11 & 8.28 & 8.26 & 0.18 & 0.07 \\
 O & 0.06 &&  2 & 8.97 & 8.87 & 8.93 & 8.87 & 8.97 & 0.10 & 0.11 &&  2 & 8.92 & 8.97 & 8.78 & 9.07 & 8.92 & 0.24 & 0.17 \\
Na & 0.03 &&  6 & 6.41 & 6.46 & 6.37 & 6.41 & 6.41 & 0.05 & 0.04 &&  6 & 6.53 & 6.48 & 6.29 & 6.53 & 6.48 & 0.20 & 0.09 \\
Mg & 0.05 &&  3 & 7.55 & 7.60 & 7.51 & 7.55 & 7.55 & 0.05 & 0.06 &&  3 & 7.80 & 7.75 & 7.61 & 7.80 & 7.78 & 0.20 & 0.11 \\
Al & 0.07 &&  2 & 6.32 & 6.37 & 6.28 & 6.32 & 6.32 & 0.06 & 0.08 &&  2 & 6.57 & 6.47 & 6.38 & 6.67 & 6.62 & 0.24 & 0.17 \\
Si & 0.05 &&  5 & 7.36 & 7.38 & 7.32 & 7.36 & 7.36 & 0.04 & 0.05 &&  5 & 7.64 & 7.59 & 7.45 & 7.69 & 7.59 & 0.20 & 0.09 \\
Ca & 0.02 &&  6 & 6.26 & 6.31 & 6.22 & 6.26 & 6.26 & 0.05 & 0.03 &&  6 & 6.11 & 6.21 & 5.97 & 6.26 & 6.36 & 0.25 & 0.17 \\
Sc & 0.10 &&  4 & 3.42 & 3.32 & 3.38 & 3.42 & 3.42 & 0.06 & 0.11 &&  4 & 3.57 & 3.62 & 3.41 & 3.72 & 3.62 & 0.22 & 0.15 \\
Ti & 0.06 && 12 & 5.02 & 4.92 & 4.98 & 5.02 & 5.02 & 0.05 & 0.07 && 12 & 4.87 & 4.62 & 4.65 & 4.82 & 4.77 & 0.20 & 0.10 \\
 V & 0.02 &&  7 & 3.81 & 3.91 & 3.77 & 3.86 & 3.81 & 0.06 & 0.05 && 13 & 3.91 & 3.61 & 3.72 & 3.86 & 3.91 & 0.19 & 0.08 \\
Cr & 0.03 &&  6 & 5.62 & 5.67 & 5.58 & 5.62 & 5.62 & 0.05 & 0.04 &&  6 & 5.72 & 5.67 & 5.58 & 5.77 & 5.77 & 0.19 & 0.07 \\
Co & 0.04 &&  4 & 4.99 & 5.02 & 4.95 & 5.02 & 4.99 & 0.05 & 0.05 &&  5 & 5.27 & 5.32 & 5.13 & 5.42 & 5.37 & 0.21 & 0.12 \\
Ni & 0.04 && 10 & 6.30 & 6.29 & 6.26 & 6.25 & 6.30 & 0.04 & 0.04 && 10 & 6.50 & 6.45 & 6.36 & 6.55 & 6.55 & 0.19 & 0.06 \\
Cu & 0.04 &&  2 & 4.01 & 4.11 & 3.97 & 4.01 & 4.01 & 0.09 & 0.09 &&  3 & 4.06 & 4.01 & 3.87 & 4.11 & 4.01 & 0.21 & 0.11 \\
Zn & 0.08 &&  4 & 4.60 & 4.65 & 4.56 & 4.60 & 4.60 & 0.05 & 0.09 &&  4 & 4.60 & 4.70 & 4.46 & 4.70 & 4.65 & 0.21 & 0.13 \\
Ss & 0.07 &&  2 & 3.87 & 3.97 & 3.83 & 3.87 & 3.87 & 0.09 & 0.11 &&  3 & 3.92 & 3.82 & 3.73 & 3.92 & 3.93 & 0.21 & 0.12 \\
Sr & 0.07 &&  1 & 3.67 & 3.79 & 3.63 & 3.67 & 3.67 & 0.15 & 0.16 &&  2 & 3.57 & 3.47 & 3.38 & 3.67 & 3.47 & 0.24 & 0.18 \\
 Y & 0.03 && 12 & 3.30 & 3.35 & 3.26 & 3.30 & 3.30 & 0.04 & 0.04 && 12 & 3.54 & 3.49 & 3.35 & 3.64 & 3.53 & 0.19 & 0.06 \\
Zz & 0.02 &&  4 & 3.50 & 3.55 & 3.46 & 3.45 & 3.50 & 0.06 & 0.04 &&  5 & 3.30 & 3.30 & 3.11 & 3.20 & 3.15 & 0.20 & 0.10 \\
Zr & 0.02 &&  5 & 3.75 & 3.80 & 3.71 & 3.75 & 3.75 & 0.05 & 0.04 &&  5 & 3.90 & 3.90 & 3.71 & 4.00 & 3.80 & 0.20 & 0.10 \\
Mo & 0.05 &&  1 & 2.42 & 2.52 & 2.38 & 2.42 & 2.42 & 0.13 & 0.13 &&  1 & 2.12 & 2.02 & 1.95 & 2.17 & 2.14 & 0.25 & 0.18 \\
Ba & 0.05 &&  5 & 3.18 & 3.23 & 3.14 & 3.23 & 3.18 & 0.05 & 0.06 &&  5 & 3.73 & 3.68 & 3.63 & 3.78 & 3.78 & 0.19 & 0.08 \\
La & 0.03 &&  8 & 1.96 & 2.06 & 1.92 & 1.96 & 1.96 & 0.06 & 0.05 &&  8 & 2.67 & 2.62 & 2.53 & 2.77 & 2.67 & 0.19 & 0.07 \\
Ce & 0.04 && 10 & 2.28 & 2.33 & 2.24 & 2.33 & 2.28 & 0.05 & 0.05 && 11 & 3.08 & 3.13 & 2.95 & 3.23 & 3.09 & 0.19 & 0.08 \\
Pr & 0.15 &&  2 & 1.06 & 1.11 & 1.02 & 1.06 & 1.06 & 0.06 & 0.16 &&  3 & 2.16 & 2.21 & 1.97 & 2.36 & 2.15 & 0.25 & 0.23 \\
Nd & 0.01 &&  8 & 2.04 & 2.14 & 2.00 & 2.09 & 2.04 & 0.06 & 0.05 &&  9 & 3.14 & 3.04 & 2.95 & 3.04 & 2.94 & 0.20 & 0.09 \\
Sm & 0.06 &&  4 & 1.26 & 1.36 & 1.22 & 1.31 & 1.26 & 0.08 & 0.09 &&  5 & 2.26 & 2.16 & 2.15 & 2.36 & 2.21 & 0.20 & 0.10 \\
Eu & 0.01 &&  3 & 0.82 & 0.92 & 0.78 & 0.92 & 0.82 & 0.10 & 0.09 &&  4 & 1.19 & 1.21 & 1.00 & 1.29 & 1.14 & 0.21 & 0.10 \\
Gd & 0.04 &&  2 & 1.17 & 1.07 & 1.13 & 1.07 & 1.17 & 0.10 & 0.10 &&  3 & 1.67 & 1.87 & 1.48 & 1.97 & 1.87 & 0.38 & 0.34 \\
Dy & 0.06 &&  2 & 1.34 & 1.29 & 1.30 & 1.34 & 1.34 & 0.06 & 0.07 &&  2 & 2.94 & 2.69 & 2.75 & 2.79 & 2.79 & 0.28 & 0.22 \\
Pb & 0.08 &&  1 & 2.65 & 2.80 & 2.61 & 2.65 & 2.65 & 0.19 & 0.20 &&  1 & 3.05 & 2.85 & 2.86 & 3.00 & 3.00 & 0.29 & 0.25 \\
\noalign{\smallskip}
\hline
\end{tabular}
   $$ 
}
\end{table*}

\begin{table*}
\caption{Abundances found in the literature for barium stars. The stars codes in the header are the same as in Table \ref{abmol}.
m=average of \ion{Zr}{i} and \ion{Zr}{ii}. References: E93 - \citet{edv93}; T89 - \citet{tomk89}; P05 - \citet{claudio05}; 
B92 - \citet{bjra92}; L03 - \citet{liang03}; N94 - \citet{north94}; P03 - \citet{claudio03}; S86 - \citet{sl86}; 
L91 - \citet{lu91}; S93 - \citet{scl93}.}
{\tiny
\label{ablit}
   $$ 
\setlength\tabcolsep{3pt}
\begin{tabular}{rrrrrrrrrrrrrrrrrrrrrrrrrrrrrrrrrrrrrr}
\noalign{\smallskip}
\hline
\noalign{\smallskip}
 HD/HR & \multicolumn{2}{c}{e2} && e4 && e6 && e7 && \multicolumn{2}{c}{e8} && e9 && e10 && e11 && \multicolumn{3}{c}{e12} && 
 e13 && e14 && e15 && \multicolumn{2}{c}{e18} && e19 && \multicolumn{2}{c}{e20} && e21\\
\cline{2-3} \cline{5-5} \cline{7-7} \cline{9-9} \cline{11-12} \cline{14-14} \cline{16-16} \cline{18-18} \cline{20-22} \cline{24-24} \cline{26-26} 
\cline{28-28} \cline{30-31} \cline{33-33} \cline{35-36} \cline{38-38} \\
      & E93  & T89 && P05 && P05 && P05 &&  B92 & L03   && N94   && N94 && P03 && S86  & L91   & S93   &&  N94  &&  N94  &&   N94 &&   N94 & L91   
      &&  N94 &&   N94 & L91   &&   N94 \\
\noalign{\smallskip}
\hline
\noalign{\smallskip}

Li    & ...  & ... &&  ...  &&  ...  &&  ...  && ...  &  ...  &&  ...  &&  ...  &&  ...  &&$<$1.3&  ...  &  ...  &&  ...  &&  ...  &&  ...  &&  ...  &  ...  && ...  &&  ...  &  ...  &&  ...  \\
 C    & ...  & 0.1 &&  0.71 &&  0.46 &&  0.64 && 0.15 &  ...  &&  0.70 &&  0.68 &&  0.61 && ...  &  0.47 &  0.61 &&  0.26 &&  0.40 &&  0.77 &&  0.75 &  0.87 && 0.53 &&  0.42 &  ...  &&  0.36 \\
 N    & ...  & 0.0 &&  ...  &&  ...  &&  ...  && 0.70 &  ...  &&  ...  &&  ...  &&  ...  && ...  &  ...  &  ...  &&  ...  &&  ...  &&  ...  &&  ...  &  ...  && ...  &&  ...  &  ...  &&  ...  \\
 O    & ...  & 0.1 &&  ...  &&  ...  &&  ...  && 0.05 &  0.39 &&  0.76 &&  0.76 &&  ...  && ...  &  ...  &  0.27 &&  0.37 &&  0.40 &&  1.15 &&  0.57 &  ...  && 0.73 &&  0.58 &  ...  &&  0.34 \\
Na    & 0.13 & ... &&  0.08 &&  0.21 &&  0.02 && ...  &  0.16 &&  ...  &&  ...  &&  0.00 && ...  &  0.23 &  0.07 &&  ...  &&  ...  &&  ...  &&  ...  &  0.21 && ...  &&  ...  &  0.17 &&  ...  \\
Mg    & 0.20 & ... &&  0.07 && -0.09 && -0.03 && ...  &  0.10 &&  ...  &&  ...  &&  0.10 && ...  & -0.24 &  ...  &&  ...  &&  ...  &&  ...  &&  ...  &  0.25 && ...  &&  ...  &  0.49 &&  ...  \\
Al    & 0.06 & ... &&  ...  &&  ...  && -0.13 && ...  &  0.03 &&  0.29 &&  ...  &&  0.14 && 0.02 & -0.23 &  ...  &&  0.05 && -0.02 &&  ...  &&  ...  &  0.52 && 0.36 && -0.04 &  ...  && -0.13 \\
Si    & 0.12 & ... &&  0.32 &&  0.39 &&  0.24 && ...  &  0.21 &&  0.30 &&  ...  &&  0.21 && ...  &  0.15 &  0.16 &&  ...  &&  ...  &&  0.48 &&  ...  &  0.09 && 0.33 &&  0.28 &  ...  &&  0.15 \\
Ca    & 0.05 & ... &&  0.12 &&  0.03 &&  0.04 && ...  & -0.01 &&  0.41 &&  0.20 &&  0.07 && 0.11 &  0.20 &  0.15 &&  0.38 &&  0.52 &&  0.37 &&  0.03 &  0.04 && 0.54 &&  0.47 &  0.25 &&  0.23 \\
Sc    & ...  & ... && -0.15 &&  0.12 && -0.06 && ...  &  0.00 &&  ...  &&  ...  && -0.13 && ...  &  0.13 &  ...  &&  ...  &&  ...  &&  ...  &&  ...  & -0.08 && ...  &&  ...  & -0.30 &&  ...  \\
Ti    & 0.04 & ... &&  ...  &&  ...  &&  ...  && ...  & -0.17 &&  0.11 && -0.10 &&  0.25 && ...  &  0.29 &  ...  && -0.03 && -0.13 &&  0.12 && -0.14 &  0.37 && ...  && -0.05 & -0.12 && -0.28 \\
 V    & ...  & ... &&  ...  &&  ...  &&  ...  && ...  & -0.33 &&  ...  &&  ...  && -0.08 && ...  &  0.19 & -0.07 && ...   && ...   &&  ...  &&  ...  &  ...  && ...  &&  ...  &  0.13 &&  ...  \\
Cr    & ...  & ... &&  0.07 &&  ...  &&  0.12 && ...  &  0.26 &&  ...  && -0.11 &&  0.21 && ...  &  0.16 &  0.17 &&  0.01 && -0.04 && -0.07 && -0.32 &  0.29 && ...  &&  0.30 &  0.11 && -0.22 \\
Co    & ...  & ... &&  ...  &&  ...  &&  ...  && ...  & ...   &&  ...  &&  0.07 &&  0.25 && ...  &  0.39 &  0.09 &&  0.13 &&  0.29 &&  0.06 && -0.03 &  0.69 && ...  &&  0.57 & -0.15 &&  0.03 \\
Ni    & 0.12 & ... && -0.05 &&  0.07 && -0.02 && ...  & -0.01 && -0.18 && -0.04 &&  0.04 && ...  & -0.03 &  0.15 &&  0.02 &&  0.09 &&  0.06 && -0.21 & -0.01 && 0.10 &&  0.11 &  0.06 && -0.09 \\
Cu    & ...  & ... && -0.07 &&  0.03 &&  0.06 && ...  & ...   &&  ...  &&  ...  &&  0.01 && ...  &  ...  &  0.08 &&  ...  &&  ...  &&  ...  &&  ...  &  ...  && ...  &&  ...  &  ...  &&  ...  \\
Zn    & ...  & ... &&  0.04 &&  0.04 &&  0.06 && ...  & ...   &&  ...  &&  ...  &&  0.26 && ...  & -0.23 &  ...  &&  ...  &&  ...  &&  ...  &&  ...  &  ...  && ...  &&  ...  & -0.25 &&  ...  \\
\ion{Sr}{i} & ...  & ... &&  ...  &&  ...  &&  ...  && ...  & ...   &&  ...  &&  ...  &&  ...  && ...  &  ...  &  ...  &&  0.66 &&  0.96 &&  1.09 &&  1.14 &  ...  && ...  &&  1.23 &  ...  &&  0.89 \\
\ion{Sr}{ii} & ...  & ... &&  ...  &&  ...  &&  ...  && ...  & ...   &&  ...  &&  ...  &&  ...  && ...  &  ...  &  ...  &&  ...  &&  ...  &&  ...  &&  ...  &  ...  && ...  &&  ...  &  ...  &&  ...  \\
 Y    & 0.50 & 0.7 &&  0.75 &&  0.09 &&  0.72 && ...  &  0.47 &&  0.70 &&  1.21 &&  1.01 && ...  &  0.86 &  1.11 &&  0.75 &&  1.09 &&  1.29 &&  1.09 &  1.04 && 1.32 &&  1.26 &  0.65 &&  0.93 \\
\ion{Zr}{i}  & ...  & ... &&  ...  &&  ...  &&  ...  && ...  &  0.41 &&  ...  &&  ...  &&  1.22 && ...  &  0.62 &  ...  &&  ...  &&  ...  &&  ...  &&  ...  &  ...  && ...  &&  ...  &  ...  &&  ...  \\
\ion{Zr}{ii} & 0.65 & 0.4 &&  0.71 &&  ...  &&  ...  && ...  &  ...  &&  0.88 &&  0.90 &&  ...  && ...  &  ...  &0.89$\sp m$&&  0.52 &&  0.68 &&  0.95 &&  0.89 &  ...  && 1.13 &&  0.90 &  0.39 &&  0.83 \\
Mo    & ...  & ... &&  ...  &&  ...  &&  ...  && ...  &  ...  &&  ...  &&  ...  &&  ...  && ...  &  ...  &  ...  &&  ...  &&  ...  &&  ...  &&  ...  &  1.12 && ...  &&  ...  &  ...  &&  ...  \\
Ba    & 0.54 & 0.4 &&  1.17 &&  1.38 &&  0.75 && ...  &  0.67 &&  ...  &&  1.44 &&  1.51 && ...  &  0.87 &  0.83 &&  0.85 &&  0.77 &&  1.54 &&  1.40 &  1.32 && ...  &&  0.97 &  ...  &&  1.25 \\
La    & ...  & ... &&  0.75 &&  1.23 &&  0.56 && ...  &  0.81 &&  ...  &&  ...  &&  1.75 && ...  &  0.78 &  ...  &&  ...  &&  ...  &&  ...  &&  ...  &  0.96 && ...  &&  ...  &  0.38 &&  ...  \\
Ce    & ...  & 0.3 &&  0.68 &&  0.92 &&  0.29 && ...  &  ...  &&  ...  &&  1.02 &&  1.32 && ...  &  0.57 &  ...  &&  0.77 &&  0.54 &&  1.43 &&  1.35 &  1.42 && ...  &&  0.69 &  0.35 &&  0.90 \\
Pr    & ...  & ... &&  ...  &&  ...  &&  ...  && ...  &  ...  &&  ...  &&  ...  &&  ...  && ...  &  1.01 &  ...  &&  ...  &&  ...  &&  ...  &&  ...  &  ...  && ...  &&  ...  &  ...  &&  ...  \\
Nd    & ...  & ... &&  0.80 &&  0.53 &&  0.07 && ...  &  ...  &&  1.22 &&  0.87 &&  0.97 && ...  &  0.60 &  0.60 &&  0.63 &&  0.17 &&  1.15 &&  1.18 &  0.98 && 1.19 &&  0.55 &  0.32 &&  0.87 \\
Sm    & ...  & ... &&  ...  &&  ...  &&  ...  && ...  &  ...  &&  ...  &&  ...  &&  ...  && ...  &  ...  &  ...  &&  ...  &&  ...  &&  ...  &&  ...  &  1.00 && ...  &&  ...  &  0.18 &&  ...  \\
Eu    & ...  & 0.2 &&  0.19 &&  ...  &&  0.26 && ...  &  0.36 &&  ...  &&  ...  &&  0.61 && ...  &  ...  &  ...  &&  ...  &&  ...  &&  ...  &&  ...  &       && ...  &&  ...  &       &&  ...  \\
\noalign{\smallskip}
\hline
 \end{tabular}
   $$ 
}
\end{table*}

\subsection{Iron peak elements}

At the last moments of the life of a massive star, the iron and the iron peak elements are 
formed in large amounts \citep{ww95}. The process that precedes the explosion of SN Ia also
produces these elements, but in lower amounts relative to massive stars. However, the SN Ia 
ejecta contain larger amounts of those elements than SN II, because part of the yield is 
restrained in the neutron star newly formed.

Figure 12 of \citet{M97} shows abundances from previous work, and it is
possible to observe the trends of [X/Fe] relative to metallicity for iron peak elements.
In the present work, this trend is unclear. However, in average, [$pFe$/Fe] $\approx$ 0 at
-1.2 $<$ [Fe/H] $<$ 0, as shown in Figure \ref{relpal}, being $pFe$ the average of V, Cr, 
Co and Ni abundances. 

The stellar yield of iron peak elements is uncertain given that it depends on several
process not well established such as the amount of mass released during supernovae events,
the mass retained in the proto neutron star, the energy of the explosion and
the neutron flux.

{\it Scandium}: \citet{zhao90} have found Sc overabundances of $\approx$ 0.25 dex 
in metal-poor stars.
\citet{gs91} suggested that such result was a consequence of gf-values adopted.
\citet{proc00} found Sc overabundance $\approx$ 0.20 dex at [Fe/H] $\sim$ -0.5, 
with a decreasing trend for lower metallicities, using log gf from \citet{mart88} 
and \citet{law89} for disk stars in the range -1.2 $\leq$ [Fe/H] $\leq$ -0.3.
\citet{nissen00} found a trend of [Sc/Fe] similar that of $\alpha$-elements 
for 100 dwarf F and G stars with -1.4 $<$ [Fe/H] $<$ 0, by using hyperfine 
structure from \citet{steffen85}. 
In the chemical evolution model of Sc by \citet{francois04}, [Sc/Fe] $\approx$ 0.2 for
very metal-poor stars, with a trend toward zero for stars richer than [Fe/H] $\sim$ -2.

In the present work, 4 lines of Sc were used, with good agreement among their abundance 
results. The gf-values adopted were those from NIST. All results for
[Sc/Fe] shown in Figure \ref{imalfefe2} are above solar, reaching 0.7 for the
star HD 147609 ([\ion{Fe}{i}/H]=-0.45).

{\it Vanadium}: Few studies on vanadium were found. \citet{gs91} analysed a sample
of 20 stars in an extensive range of metallicities and found [V/Fe] $\sim$ 0 for all
metallicities, in agreement with the work by \citet{pagel68}. However, \citet{proc00} 
obtained 0.1 $<$ [V/Fe] $<$ 0.4.

The V I lines are very sensitive to temperature. They are usually weak for all stars
of the present sample, but they are weaker for hotter stars. As an example, only 2 lines
are available for the stars HD 123585 and BD+18 5215. Despite the strength of the lines,
the abundances derived show a good agreement. A larger difference, of 0.5 dex was
obtained for 2 lines of the star HD 210910 (see Table \ref{abun}). All abundances
are in the range -0.40 $<$ [V/Fe] $<$ 0.2, as shown in Figure \ref{imalfefe2}.

{\it Chromium}: \citet{cay04} found an increasing trend of [Cr/Fe] in the range
-4 $\leq$ [Cr/Fe] $\leq$ -2 for very metal-poor stars. Figure 6 from \citet{francois04} 
shows that for higher metallicities, the data remain around [Cr/Fe] $\sim$ 0. In the 
present work, 6 lines of Cr were used, resulting in the range -0.2 $\leq$ [Cr/Fe] $\leq$ 0.2, 
consistent with \citet{francois04}. For the star HD 210910 the result is lower than for
other stars.
 
{\it Cobalt}: In the same range of metallicity, \citet{proc00} found an overabundance of Co
relative to Fe reaching $\sim$ 0.2 dex, whereas \citet{gs91} found a deficiency of
0.1 dex. The lines of Co are sensitive to temperature. For the hotter stars of the
sample, the equivalent widths are very small and in some cases, they had to be
discarded. Differences in the abundances from different lines are shown in
Table \ref{abun}, and Figure \ref{imalfefe2} shows that -0.15 $<$ [Co/Fe] $<$ 0.4.

{\it Nickel}: \citet{gs91}, \citet{edv93}, \citet{peterson90},
\citet{mac95} and \citet{ryan96} find -0.1 $\leq$ [Ni/Fe] $\leq$ 0.1 in the range 
-4 $\leq$ [Fe/H] $\leq$ 0. In the present work, 10 lines of Ni were used with small
differences between them. Figure \ref{imalfefe2} 
shows that -0.13 $\leq$ [Ni/Fe] $\leq$ 0.12, in agreement with \citet{francois04}.

{\it Copper}: The analysis of Cu as a function of metallicity by \citet{sc88} and
\citet{sneden91} showed a linear decrease of [Cu/Fe] toward decreasing metallicities,
reaching [Cu/Fe] $\sim$ -1 at [Fe/H] $\sim$ -3. This trend was confirmed by 
\citet{mish02}, who extended the sample for halo field giants, and by \citet{simm03}, 
whose sample included 117 giant stars in 10 globular clusters in the range of
metallicities -2.4 $\leq$ [Fe/H] $\leq$ -0.8.

The nucleosynthetic sites of Cu are not well established yet. \citet{sneden91} 
suggested that the Cu nucleosynthesis occours mainly through the weak component of
the s-process and a small contribution of the explosive burning in SN II. The
additional source of Cu could be SN Ia or SN II \citep{mate93,times95,baraf93}.

In the present work, 3 lines of Cu were used taking into account the hyperfine structure.
For some stars, abundance results derived from the line $\lambda$5218.2 $\rm \AA$ 
are higher than those from the other 2 lines, $\lambda$5105.5 $\rm \AA$ and 
$\lambda$5782.1 $\rm \AA$. It could be due to the gf-value, taken from a
different source. However, for only one star the abundance
difference between $\lambda$5218.2 $\rm \AA$ and the other lines reaches 0.4 dex and for
3 stars, this difference is 0.25 to 0.3 dex. For the other stars, the agreement among
the abundance results derived from those lines is very satisfactory. Except for a few
stars, [Cu/Fe] is below solar for all stars (see Figure \ref{imalfefe2}), in agreement 
with previous work \citep{M97,francois04}.

{\it Zinc}: \citet{sc88} and \citet{sneden91} obtained a constant behaviour for
Zn with [Zn/Fe] = 0, and this result was confirmed by \citet{mish02} 
for [Fe/H] $>$ -2.0. \citet{cay04} observed an increasing trend
of [Zn/Fe] toward lower metallicities. The results of
\citet{cay04} are shown in Figure 5 of \citet{francois04}, including also observations
from previous work, showing that for [Fe/H] $>$ -2.5, [Zn/Fe] is approximately constant,
although with a dispersion in the range -0.25 $\leq$ [Zn/Fe] $\leq$ 0.3.

Similarly to Cu, Zn can be produced by a sum of nucleosynthetic processes \citep{mish02},
the weak component of the s-process \citep{sneden91}, SN Ia \citep{mate93} and 
SN II \citep{times95}.

In the present work, 4 atomic lines were used in order to compute Zn abundance. Generally,
the results derived from different lines are in good agreement. For 6 stars a
larger difference was observed from 0.25 to 0.6 dex, usually between the 
lines $\lambda$4680.1 $\rm \AA$ 
and $\lambda$6362.3 $\rm \AA$. This difference should not be due to
atomic constants, once for most stars a good agreement was found. 
[Zn/Fe] vs. [Fe/H] shown in Figure \ref{imalfefe2} 
is in good agreement with \citet{francois04}.

\subsection{s-elements}

In the s-elements list we included those elements that have more than 50\% of
s-process contribution for their abundances, according to \citet{arlandini99}. 
Molibdenium was also included in this list, given that the s-process 
contribution for its abundance, although lower than 50\%, 
is much larger than r- and p-processes. Details 
of the s-, r- and p-processes contributions for heavy elements abundances of 
sample stars will be found in the forthcoming paper by \citet[][paper II]{papII}.

{\it Strontium}: Few studies on strontium are found. \citet{mg01} 
found -0.2 $\leq$ [Sr/Fe] $\leq$ 0.1 for disk stars in the range
of metallicities -1.5 $\leq$ [Fe/H] $\leq$ 0. For one star of the sample with
[Fe/H] $\approx$ -1 they found [Sr/Fe] $\approx$ 0.2. In the same range of
metallicities, \citet{g94} found -0.2 $\leq$ [Sr/Fe] $\leq$ 0.2.
\citet{jehin99} obtained -0.4 $<$ [Sr/Fe] $<$ 0 in the metallicity range
-1.3 $\leq$ [Fe/H] $\leq$ -0.8.

The difficulty in computing Sr abundance is the strength of the line $\lambda$4077.7 $\rm \AA$,
which characterises barium stars. However, for most stars the
abundance derived from this line was very close to that from $\lambda$4161.8 $\rm \AA$,
as shown in Table \ref{abun}. The larger difference is for 
HD 147609 (0.80 dex) followed by HD 12392 (0.45 dex) and HD 188985 (0.30 dex). Lines of 
\ion{Sr}{i} usually result in lower abundances than \ion{Sr}{ii} ones. This effect is
also seen in the solar abundances computed by \citet{g94}. Figure \ref{rsfe1} shows that
the relation [Sr/Fe] vs. [Fe/H] presents a larger dispersion for \ion{Sr}{ii} 
than for \ion{Sr}{i}. Most data are in the range 0.6 $\leq$ [\ion{Sr}{ii}/Fe] $\leq$ 1.40 
and 0.3 $\leq$ [SrI/Fe] $\leq$ 1.2.

{\it Yttrium}: \citet{g94}, \citet{jehin99}, \citet{tl99} and \citet{edv93} found
an increasing trend of [Y/Fe] toward higher metallicities, excluding the peculiar
stars in their samples. In the present work, Y abundance was computed using 
synthesis of 12 lines of \ion{Y}{ii}. In general the results derived from
different lines are in good agreement. A few lines result in different abundances
that can reach 0.7 dex for some stars, as can be seen in Table \ref{abun}. 
The gf-values for the lines of \ion{Y}{ii} were those from \citet{H82},
except for the line $\lambda$6795.4 $\rm \AA$, with log gf from \citet{M94}.
The gf-values from \citet{H82} were also used by \citet{g94} resulting in 
$\log\epsilon_\odot$(Y) = 2.21 $\pm$ 0.02
for the solar abundance, in agreement with \citet{gs98} value of 
$\log\epsilon_\odot$(Y) = 2.24 $\pm$ 0.03. \citet{g94} 
obtained -0.3 $\leq$ [Y/Fe] $\leq$ 0.1 at [Fe/H] $\approx$ -1, and 
[Y/Fe] $\approx$ 0 at [Fe/H] $\approx$ -0.3. \citet{tl99} found -0.2 $\leq$ [Y/Fe] $\leq$ 0
in the range -1 $\leq$ [Fe/H] $\leq$ 0. Figure \ref{rsfe1} shows resulting values
much higher for the present barium stars, in the range 0.50 $\leq$ [Y/Fe] $\leq$ 1.60.

{\it Zirconium}: a combination of results by \citet[][ and references therein]{burr00},
\citet{g94} and \citet{tl99} gives a relation of [Zr/Fe] vs. [Fe/H]
similar to those of Sr and Y. The dispersion increases for [Fe/H] $<$ -1.5, whereas
for higher metallicities, [Zr/Fe] is
in the range -0.2 $\leq$ [Zr/Fe] $\leq$ 0.5. For the sample barium stars 
the dispersion is also present, and
the values are much higher than for normal stars. Table \ref{abun} 
shows that the abundances derived from \ion{Zr}{i} are usually
lower than those from \ion{Zr}{ii} in the range 0.40 $\leq$ [\ion{Zr}{ii}/Fe] $\leq$ 1.60
and -0.20 $\leq$ [\ion{Zr}{i}/Fe] $\leq$ 1.45, the latter with higher dispersion,
as shown in Figure \ref{rsfe1}.

Sr, Y and Zr form the first peak of abundance of the s-process. The reason is
that they have one isotope with a neutron magic number (N=50), $\sp {88}$Sr, $\sp {89}$Y and 
$\sp {90}$Zr. The larger contribution for the abundance of those elements
comes from those isotopes. Figure \ref{rsfe1} shows that 
[Sr,Y,Zr/Fe] vs. [Fe/H] are far higher than [Mo/Fe] vs. [Fe/H]. Some stars
show deficiency in Mo that can reach -0.20.

{\it Molybdenum}:
According to \citet{arlandini99}, the Mo abundance in the solar system has a
contribution of 49.76\% from s-process, 26.18\% from r-process and 24.06\% 
from p-process. For the latter process, Mo is responsible for an abundance peak.

In the present sample, only the line $\lambda$5570.4 $\rm \AA$ was available,
for which the log gf from \citet{B83} was used. For the same line and log gf,
\citet{S00} obtained 1.97 for the solar abundance, close to the value by 
\citet{gs98} of 1.92 $\pm$ 0.05.

Mo is little studied. \citet{S00} obtained 
-0.21 $\leq$ [Mo/Fe] $\leq$ 0.9 for a sample of 10 red giant stars from
the globular cluster $\omega$ Centauri, in the range 
-1.8 $\leq$ [Fe/H] $\leq$ -0.8. The range -0.20 $\leq$ [Mo/Fe] $\leq$ 1.0 
was obtained for the stars of the present sample, showing a much lower 
pattern than the s-elements, as shown in Figure \ref{rsfe1}.

{\it Barium}: \citet{spite78} verified that [Ba/Fe] increases in
the range -3 $<$ [Fe/H] $\leq$ -1.5, that they called as ``halo enrichment''.
For [Fe/H] $>$ -1.5 the Ba enrichment is very slow, or null. This effect
is confirmed by
\citet[][ and references therein]{burr00}, \citet{g94}, \citet{mg01} and
\citet{tl99}, which show [Ba/Fe] increasing to values
close to solar abundance in the range -4 $\leq$ [Fe/H] $\leq$ -2,
and values -0.4 $\leq$ [Ba/Fe] $\leq$ 0.4 for [Fe/H] $>$ -2.

In the present work, all values of [Ba/Fe] are in the range of
0.8 $\leq$ [Ba/Fe] $\leq$ 1.80, showing high Ba overabundance relative to Fe, 
which is a defining characteristics of barium stars. 
Five lines of Ba were used taking into account hyperfine structure.
In general, good agreement on abundances derived from different
lines was found, as shown in Table \ref{abun}. 
Figures \ref{relacsba} and \ref{relacsbafe} and Table \ref{elalbaeu} 
show abundances of $\alpha$- and iron peak elements relative to Ba. The
abundance excesses relative to Ba show a decreasing trend toward increasing [Ba/H].
The decreasing trend of [iron peak/Ba] vs. [Ba/H] could be explained by 
the secondary charater of the s-process, where iron peak elements are the 
seed nuclei. [X/Ba] relative to [Fe/H] show an increasing trend with 
increasing [Fe/H]. This trend is in agreement with the larger efficiency of 
the s-process in producing heavy elements such as Ba, at lower metallicities. 
The s-process site is different from that of Al, Na, $\alpha$- and
iron peak elements, therefore it is not surprising that [X/Ba] vs. [Ba/H] and 
[X/Ba] vs. [Fe/H] are not constant. In both figures \ref{relacsba} and 
\ref{relacsbafe}, the 
halo giant star HD 123396 ([Fe/H] = -1.19) is out of the trend shown by
the other stars.

{\it Lanthanum}: \citet[][ and references therein]{burr00}
and \citet{g94} show a similar behaviour for [La/Fe] and [Ba/Fe]. \citet{g94} 
found a range of -0.4 $\leq$ [La/Fe] $\leq$ 0.05 at -2 $\leq$ [Fe/H] $\leq$ 0,
and \citet{burr00}, in the range -2 $\leq$ [Fe/H] $\leq$ -0.5, found 
0 $\leq$ [La/Fe] $\leq$ 0.5, with 2 stars with [La/Fe] $\leq$ 0.

In the present work, 8 lines of \ion{La}{ii} were used taking into account
the hyperfine structure. For some stars 
abundance differences between lines can reach 0.7 dex, mainly involving the line 
$\lambda$4086.7 $\rm \AA$. The origin of this difference is not related
to the atomic constants since for several stars the agreement is very good.
As an example, the abundance obtained from this line for HD 5424 
was lower than from the line $\lambda$5797.6 $\rm \AA$, while for 
HD 22589 the result is inverted. If the problem were in the atomic
constants, the difference pattern would be always the same. The range 
obtained was 0.6 $\leq$ [La/Fe] $\leq$ 1.70 (see Figure \ref{rsfe1}),
showing the large La overabundance relative to Fe for barium stars, in
contrast to normal stars. 

{\it Cerium}: \citet{g94} and \citet{jehin99} showed
for [Ce/Fe] a behaviour similar to [Ba/Fe] and [La/Fe].
Their results are in the range -0.4 $\leq$ [Ce/Fe] $\leq$ 0.15 
for -2 $\leq$ [Fe/H] $\leq$ 0. In the present work, 8 lines of \ion{Ce}{ii} 
were used for abundance determination. As for La, in some cases, the
abundance result of one line was very different from the others, but
the effect on the average is low, as
can be seen in Table \ref{abun}, \ref{medxfe1} and 
\ref{medxfe2}. Abundance excess relative to Fe is in the
range 0.4 $\leq$ [Ce/Fe] $\leq$ 1.80, with large dispersion among the stars.

{\it Neodymium}: \citet{g94}, \citet{tl99}, 
\citet{jehin99} and \citet[][ and references therein]{burr00} show that
[Nd/Fe] behaves similarly to Ba, La and Ce. According to Figure 5 from \citet{burr00},
a high dispersion in the range -3 $\leq$ [Fe/H] $\leq$ -1.5 is present, with
low values close to [Nd/Fe] $\approx$ -0.6 and high values of $\approx$ +1.5.
For [Fe/H] $>$ -1.5 the results are in the range 
-0.3 $\leq$ [Nd/Fe] $\leq$ +0.3.

In the calculation of Nd abundance for the present sample, the 
hyperfine structure was neglected.
According to \citet{Hartog03}, only the odd isotopes $\sp {143}$Nd and
$\sp {145}$Nd show hyperfine structure, accounting for 20.5\% of the abundance,
hence the effect can be ignored. The abundances derived from 9 lines of 
\ion{Nd}{ii} were computed, and they show a good agreement, with a few
exceptions, similar to previous elements described here.
The range obtained for the abundance excess relative to Fe was 
0.3 $\leq$ [Nd/Fe] $\leq$ 1.70. 

Ba, La, Ce and Nd form the second 
abundance peak of the s-process because of nuclides $\sp {138}$Ba, $\sp {139}$La, 
$\sp {140}$Ce and $\sp {142}$Nd are neutron magic nuclei (N=82).

{\it Lead}: There are a few stars for which the lead abundance was determined.
\citet{vaneck03}, show that some CH stars show
high Pb abundances, and \citet{sivarani04} gathered about 30 halo stars with
high Pb abundances. There are 4 Pb isotopes: $\sp {204}$Pb, $\sp {206}$Pb,
$\sp {207}$Pb and $\sp {208}$Pb. The last is doubly magical (neutron magic
in N=126 and proton magic in Z=82) being the responsible for the third 
abundance peak of the s-process. The best lines for the abundance calculation are
$\lambda$3639.6 $\rm \AA$, $\lambda$3683.5 $\rm \AA$, $\lambda$3739.9 $\rm \AA$, 
$\lambda$4057.8 $\rm \AA$ and $\lambda$7229 $\rm \AA$. Wavelengths
$\lambda$ $<$ 4000 $\rm \AA$ are not reliable with the FEROS spectra and the
$\lambda$7229 $\rm \AA$ line is too weak. For these reasons,
the Pb abundance was determined using the line $\lambda$4057.8 $\rm \AA$,
for which blends have been taken into account. The results are
in the range -0.2 $\leq$ [Pb/Fe] $\leq$ 1.6, showing a large dispersion.
According to \citet{arlandini99}, the s-process is responsible for 46\% of
the Pb abundance with no contribution from the r-process. The missing abundance
is generally attributed to the strong component of the s-process.

\subsection{r-elements}

We included in the r-elements list only those with more than 50\% of
r-process contribution for their abundances, according to \citet{arlandini99}.

{\it Europium}: \citet{g94}, \citet{burr00}, \citet{jehin99}, 
\citet{woolf95} and \citet{mg01} provide an increasing linear relation between
$\log\epsilon$(Eu) and [Fe/H]. On the other hand, [Eu/Fe] decreases toward
higher metallicities. 

In the present work the lines
$\lambda$4129.7 $\rm \AA$, $\lambda$4205 $\rm \AA$,
$\lambda$6437.7 $\rm \AA$ and $\lambda$6645.1 $\rm \AA$ were used for computing Eu
abundances, taking into account the hyperfine structure.
The abundances derived from these 2 first lines are usually lower than
those derived from the 2 last lines (see Table \ref{abun}). 
The 2 first lines show a blend with CN lines in the cooler stars, whereas 
for the hotter stars, this blend is negligible. 

Figure \ref{relacseu} and Table \ref{elalbaeu} show the
relation between [Eu/H] and abundance excess of $\alpha$- and iron peak elements
relative to Eu, showing a larger dispersion than Figure \ref{relacsba}. 
With the exception of O, Mg, Co and Sc, the other $\alpha$- and iron
peak elements show [X/Eu] $\leq$ 0. It seems that there is an 
interval of [Eu/H] where [X/Eu] is constant and, for [Eu/H] $>$ 0 a 
decreasing trend is seen, that could reflect the secondary character of
the r-process with the iron peak elements as the seed nuclei. A
primary r-process \citep[e.g.][]{meyer94} contribution would form Eu
without depletion of other elements. Relative to [Fe/H] 
(Figure \ref{relacseufe}), [X/Eu] seems to be constant, with the
possible exception of [Cr/Eu], [Co/Eu] and [Cu/Eu] which seem to 
increase toward increasing metallicities. The constancy of [X/Fe] could be 
due to the fact that $\alpha$-elements, and probably Na and Al, are
produced in massive stars (mostly in hydrostatic phases), and they are
released in their SNae type II explosions as well as the r-elements.
Thus, if all of them were released through the same events, their 
ratios are expected to be constant.
The dispersion comes from the fact that stars of different masses 
produce different amounts of each element. Regarding the increasing 
or constant behaviour of [iron peak/Eu] vs. [Fe/H], it can be due to 
their production in different amounts in SNae II and SNae Ia.
A larger range of metallicities expanded
toward lower values could illustrate better such behaviour.

{\it Praseodymium}: Little is done on Pr. \citet{g94} determined Pr abundance for
metal-poor stars and obtained results in the range 
-0.2 $\leq$ [Pr/Fe] $\leq$ 0.3, for
-1.5 $\leq$ [Fe/H] $\leq$ 0. Figure \ref{rsfe2} shows that the present results are in
the range 0.2 $\leq$ [Pr/Fe] $\leq$ 1.40, showing the contribution of the
s-process to the Pr abundance. The three lines used were in good internal agreement,
with a few exceptions, as shown in Table \ref{abun}. 

{\it Samarium}: \citet{g94} included Sm in their analysis with results
in the range -0.3 $\leq$ [Sm/Fe] $\leq$ 0.2 for 
-1.5 $\leq$ [Fe/H] $\leq$ 0. In the present work, the results are in the range 
0 $\leq$ [Sm/Fe] $\leq$ 1.40 (Figure \ref{rsfe2}).

The Sm lines are very weak and the hfs of the 5 lines can be
neglected. The abundances derived from several lines are in good agreement, 
as can be seen in Table \ref{abun}.

{\it Gadolinium}: Gadolinium abundances are essentially not found in the 
literature, at least, in
the range of metallicities of the present sample. The lines are very
weak or even invisible in the spectra. Furthermore, they are only
present at $\lambda$ $<$ 5000 $\rm \AA$ with several blends. 

In the present work 3 lines of Gd were used. For most of them there was good
agreement among the abundances derived from different lines (see 
Table \ref{abun}). Figure \ref{rsfe2} shows low dispersion 
in the relation [Gd/Fe] vs. [Fe/H].
 
{\it Dysprosium}: \citet{g94} studied stars in the range of metallicities
-1.5 $\leq$ [Fe/H] $\leq$ 0, and they included Dy in their analysis.
They found a range of -0.2 $\leq$ [Dy/Fe] $\leq$ 0.2 except for one star
for which they found [Dy/Fe] $\approx$ -0.4. The results of the present
sample are in the range of -0.25 $\leq$ [Dy/Fe] $\leq$ 1.65, shown in
Figure \ref{rsfe2}. Two lines were available in the spectra of the sample 
stars, from which a good agreement was obtained.

\subsection{Uncertainties}

The abundance uncertainties were calculated by verifying how much the variation of
each input parameter changes the output value $\log{A_p}$. Table \ref{errab} 
shows the values taken into account in this calculation and the
resulting uncertainties for each element. The procedure was adopted for 2 stars,
HD 5424 of low $\log$ g = 2.0 and HD 150862 of high $\log$ g = 4.6.

The uncertainty on the output value is given by

\begin{equation}
\label{erapinst}
\sigma_{Ap}=\sqrt{(\Delta A_T)^2+(\Delta A_{lg})^2+(\Delta A_v)^2+(\Delta A_m)^2}
\end{equation}
where $\Delta A_T$, $\Delta A_{lg}$, $\Delta A_v$ and $\Delta A_m$ are the differences
in $A_p$ due to variations of 1$\sigma$ in the temperature, log g, microturbulent
velocity and metallicity, respectively.

The average value of $A_p$ ($A_{pm}$) is obtained by averaging individual 
abundances of several lines, and not from several measurements of the same line.
In the latter case, the standard deviation could be used to calculate the 
uncertainty on $A_{pm}$. Considering this, we found more suitable
to apply propagation of errors taking into account the uncertainty 
calculated with equation \ref{erapinst}. Thus, the uncertainty on $A_{pm}$ is
\begin{equation}
\sigma_{Apm}={\sigma_{Ap}\over \sqrt{n}}
\end{equation}
where $n$ is the number of lines. The uncertainty on the logarithm of $A_{pm}$ is
\begin{equation}
\sigma_{\log(Apm)}={\sigma_{Apm}\over A_{pm}\ln{10}}.
\end{equation}

The abundance $\log\epsilon$(X) is related to the output of the synthesis program by
$\log\epsilon$(X) = $\log{A_{pm}}$ + [Fe/H]. Therefore, the uncertainty is

\begin{equation}
\sigma_{\log\epsilon(X)}=\sqrt{\sigma_{\log{(Apm)}}^2+\sigma_{[Fe/H]}^2}.
\end{equation}

The relation between the abundance excess relative to iron [X/Fe] and the
output value of the synthesis program is 
[X/Fe] = $\log{A_{pm}}$ - $\log\epsilon(X)_\odot$, where $\log\epsilon(X)_\odot$ 
is the solar abundance of the element ``X''. The uncertainty is calculated by

\begin{equation}
\sigma_{[X/Fe]}=\sqrt{\sigma_{\log(Apm)}^2+\sigma_{log\epsilon_\odot(X)}^2}
\end{equation}

The uncertainty on [$\alpha$/Fe], which contains the contribution of the 
uncertainty on the abundance of each element taken into account in the calculation
of the $\alpha$'s, is given by

\begin{equation}
\label{eralf}
\sigma_{[\alpha/Fe]} = \sqrt{\sigma_{\log\epsilon(\alpha)}^2+\sigma_{\log\epsilon_\odot(\alpha)}^2+\sigma_{[Fe/H]}^2}
\end{equation}
with

\begin{equation}
\label{eralf2}
\sigma_{\log\epsilon(\alpha)}={1\over n\epsilon(\alpha)}\sqrt{\sum_{i=1}^n (10^{\log\epsilon(X_i)})^2\sigma_{\log\epsilon(X_i)}^2}
\end{equation}
where $n$ is the number of elements ``X'' used in the calculation of $\alpha$,
$\epsilon(\alpha)$ = $10^{\log\epsilon(\alpha)}$. In this work, $n$=5 for
most stars. For 2 stars, HD 87080 and HD 147609, $n$=4 because the oxygen was 
excluded from the calculation. The $\sigma_{\log\epsilon(\alpha)_\odot}$ is 
similar to expression \ref{eralf2}.

For the iron peak elements, the uncertainty is similar to expression \ref{eralf}.

In Figures 4, 5, 6, 7, 8, 9, 10, 11, 12, 13, 17, 18 are shown the maximum
value of uncertainties on each axis. In the corresponding tables
it is possible to check all the values.


\section{Conclusions}

The barium stars have been studied throughout more than five decades,
however several open questions still remain relative to their origin and
characteristics. The aim of the present work was to increase the
knowledge about this class of peculiar stars.

For the first time a detailed study on the behaviour of abundance ratios
for a large number of elements is presented for a relatively large sample
of barium stars.

As the first outcome of this work, the results of the atmospheric parameters 
show that the sample consisted of stars of different luminosity
classes with 4300 $\leq$ T$\sb {eff}$ $\leq$ 6500 K and 
1.4 $\leq$ log g $\leq$ 4.6.

The metallicities obtained are typical of barium stars, in the range
-1.2 $\leq$ [Fe/H] $\leq$ 0.0, most of them with
-0.62$\leq$ [Fe/H] $\leq$ 0.0.

For 7 stars a significant difference was found between the metallicities
resulting from \ion{Fe}{i} and \ion{Fe}{ii} lines, ($\Delta$[Fe/H] $\ge$ 0.2 dex). 
It is very important this to be understood in order
to be possible to determine reliable abundances, given that the errors in
metallicities affect the resulting abundances.
The difference appears when one determines the surface gravities through the
classical equation, which requires the mass values. In the present work,
two methods were adopted. In the first case, the masses were derived from
isochrones and then the surface gravities were determined. In the second case, 
the surface gravities
were derived from ionization equilibrium and then the masses. The difference 
between the surface gravities determined from the two methods can reach 0.6 dex
and in the worst case, over 1 dex for the star HD 147609. This difference
reflects in the differences between the metallicities derived from lines
of \ion{Fe}{i} and \ion{Fe}{ii}. One point to discuss is the fact that the 
masses determined using the log g from ionization equilibrium are very small
for these 7 stars. Another point is that $\Delta$[Fe/H] depends on the 
gf-values adopted (see Sect. 3.5). \ion{Fe}{ii} 
lines are less affected by NLTE effects than \ion{Fe}{i}, and more accurate 
gf-values for \ion{Fe}{ii} lines were used \citep{melendez06}, reducing 
the difference in log g by 0.2 dex.

There is a good agreement between the present results and literature data.
The abundance results obtained for the sample stars show that there are
no correlations with the luminosity classes. The abundances 
found for the $\alpha$-, iron peak, Li, Al and Na are compatible
with the values of [X/Fe] given in the literature for normal disk stars 
in the same range of metallicities, and the s-overabundance is independent
of luminosity class. There are not enough halo stars in the present sample
to identify differences between halo and disk barium stars. The range of 
metallicities is too small to allow a well-defined trend in the 
[X/Fe] vs. [Fe/H] of $\alpha$- and iron peak elements. For heavy
elements, there is a small variation that can be explained by the variable
amount of enriched material that each star received from the more evolved
companion.

The Li abundance decreases toward lower temperatures. This result is consistent 
with the discussion in the literature that the
Li is depleted along the red giant branch evolution.

Less evolved stars show higher C abundances, and [C/O] is approximately constant
with metallicity. Besides being C-rich, the barium stars of the present sample 
are N-rich. It happens
for all stars including the less evolved ones, suggesting that N is also
responsible for the CNO excess in these stars.

For most stars, the excesses of Na, Al, Mg, Si and Ca relative to Fe
are within -0.2 $<$ [X/Fe] $<$ 0.2. O reaches higher values and Ti, lower values.
[Ti/Fe] are approximately similar to those of [V/Fe], [Cr/Fe] and [Ni/Fe],
identifying Ti more as iron peak than an $\alpha$-element. For some stars
the odd-even effect, where Mg and Si are overabundant relative to Na and Al,
can be observed, however, for several stars the abundances of Na and Al are 
higher than those of Si and Mg.

Among iron peak elements, the Sc has the highest abundances. This result
is in agreement with \citet{nissen00}, that identified an ``$\alpha$-element''
behaviour for Sc. For Co, the theoretical prediction is [Co/Fe] $<$ 0 for 
the range of metallicities of
the present sample, however, for most stars [Co/Fe] $>$ 0 was obtained.
For other iron peak elements, V, Cr, Ni and Zn, [X/Fe] show a lower range
of values than [Co/Fe]. Except for 4 stars, [Cu/Fe] is below solar for the
present sample.

The excesses of Na, Al, $\alpha$- and iron peak elements relative to Ba show a 
decreasing trend with [Ba/H], whereas [X/Ba] vs. [Fe/H] show an increasing trend. 
Considering that Ba represents the s-process elements, one can consider that these 
correlations describe the relations between s-process and other 
nucleosynthetic processes.

Regarding the r-process element Eu, 
there is a range of [Eu/H] where [X/Eu] is essentially constant,
[X/Eu] showing a decreasing trend toward higher [Eu/H].
For most stars, [X/Eu] $\leq$ 0, except for O, Mg, Co and Sc. 
[X/Eu] vs. [Fe/H] is constant for Na, Al and $\alpha$-elements as expected.
[X/Ba] and [X/Eu] for the sample stars characterises the abundance behaviour
of different elements relative to the s- and r-processes.

\begin{acknowledgements}
We acknowledge partial financial support from the Brazilian Agencies CNPq
and FAPESP.
DMA acknowledges a FAPESP PhD fellowship n$^{\circ}$ 00/10405-8
and a FAPERJ post-doctoral fellowship n$^{\circ}$ 152.680/2004.
We are grateful to Anita G\'omez for suggesting the analysis of Hipparcos 
dwarf barium stars candidates, to Andrew McWilliam for making available his 
code on hyperfine structure and to the referee, Nils Ryde, for useful comments.
We are also grateful to Licio da Silva, Luciana Pomp\'eia, Paula Coelho and 
Jorge Mel\'endez for carrying out some observations of our sample spectra, 
and to Gustavo Porto de Mello, Wladimir and Graziela, 
for helping us with photometric observations.
This publication makes use of  the SIMBAD database and of data products from 
the Two Micron All Sky Survey,
which is a joint project of the University of Massachusetts and the Infrared 
Processing and Analysis Center/California Institute of Technology, funded by 
the National Aeronautics and Space Administration and the National Science 
Foundation.

\end{acknowledgements}

\end{document}